\documentclass{emulateapj} 
\usepackage{lscape} 
\usepackage{apjfonts}
\usepackage{latexsym}
\usepackage{amsmath} 
\usepackage{subfigure}

\shorttitle{The deepest HST CMD of M32}
\shortauthors{Monachesi et al.}
\begin{document}\title{The deepest HST Color-Magnitude Diagram of M32: Evidence for Intermediate-Age Populations
  \footnotemark[1]}  \footnotetext[1]{Based on observations  made with
  the NASA/ESA Hubble Space Telescope, obtained at the Space Telescope
  Science  Institute,   which  is  operated  by   the  Association  of
  Universities for  Research in  Astronomy, Inc., under  NASA contract
  NAS  5-26555. These  observations  are associated  with GO  proposal
  10572.}

\author{Antonela Monachesi, Scott C. Trager} \affil{Kapteyn
  Astronomical Institute, P.O. Box 800, 9700 AV Groningen, The
  Netherlands} \email{monachesi@astro.rug.nl} \author{Tod R. Lauer}
\affil{National Optical Astronomy Observatory\footnotemark[2],
  P.O.~Box 26732, Tucson, AZ, 85726, USA} \footnotetext[2]{The
  National Optical Astronomy Observatory is operated by AURA, Inc.,
  under cooperative agreement with the National Science Foundation.}
\author{Wendy Freedman, Alan Dressler} \affil{The Observatories of the
  Carnegie Institution of Washington, 813 Santa Barbara Street,
  Pasadena, CA, 91101, USA} \author{Carl Grillmair} \affil{Spitzer
  Science Center, 1200 E. California Blvd., Pasadena, CA 91125, USA}
\and \author{Kenneth J. Mighell} \affil{National Optical Astronomy
  Observatory\footnotemark[2], P.O.~Box 26732, Tucson, AZ, 85726, USA}

\begin{abstract}
  We present the deepest optical color-magnitude diagram (CMD) to date
  of the  local elliptical  galaxy M32. We  have obtained  $F435W$ and
  $F555W$ photometry based  on \textsl{Hubble Space Telescope} ACS/HRC
  images for a  region $110\arcsec$ from the center of  M32 (F1) and a
  background field (F2) about  $320\arcsec$ away from M32 center.  Due
  to  the high  resolution of  our Nyquist-sampled  images,  the small
  photometric errors,  and the depth of our  data (the color-magnitude
  diagram  of  M32  goes  as   deep  as  $F435W  \sim  28.5$  at  50\%
  completeness level) we obtain the most detailed resolved photometric
  study of M32 yet.  Deconvolution of HST images proves to be superior
  than  other  standard  methods   to  derive  stellar  photometry  on
  extremely crowded  HST images, as  its photometric errors  are $\sim
  2\times$ smaller than other methods tried.
 
  The location of the strong red clump in the CMD suggests a mean age
  between 8 and 10 Gyr for $[\mathrm{Fe/H}] = -0.2$ dex in M32.  We
  detect for the first time a red giant branch bump and an asymptotic
  giant branch bump in M32 which, together with the red clump, allow
  us to constrain the age and metallicity of the dominant population
  in this region of M32.  These features indicate that the mean age of
  M32's population at $\sim 2\arcmin$ from its center is between 5 and
  10 Gyr.  We see evidence of an intermediate-age population in M32
  mainly due to the presence of asymptotic giant branch stars rising
  to $\mathrm{M}_{F555W}\sim -2.0$.  Our detection of a blue component
  of stars (blue plume) may indicate for the first time the presence
  of a young stellar population, with ages of the order of 0.5 Gyr, in
  our M32 field.  However, it is likely that the brighter stars of
  this blue plume belong to the disk of M31 rather than to M32.  The
  fainter stars populating the blue plume indicate the presence of
  stars not younger than 1 Gyr and/or blue straggler stars in M32.
  The CMD of M32 displays a wide color distribution of red giant
  branch stars indicating an intrinsic spread in metallicity with a
  peak at $[\mathrm{Fe/H}] \sim -0.2$.  There is not a noticeable
  presence of blue horizontal branch stars, suggesting that an ancient
  population with $[\mathrm{Fe/H}] < -1.3$ does not significantly
  contribute to the light or mass of M32 in our observed fields.
  M32's dominant population of 8--10 Gyr implies a formation redshift
  of $1\la z_f\la2$, precisely when observations of the specific star
  formation rates and models of ``downsizing'' imply galaxies of M32's
  mass ought to be forming their stars.  Our CMD therefore provides a
  ``ground-truth'' of downsizing scenarios at $z=0$.

  Our background field data represent the deepest optical observations
  yet of the inner disk and bulge of M31.  Its CMD exhibits a broad
  color spread of red giant stars indicative of its metallicity range
  with a peak at $[\mathrm{Fe/H}] \sim -0.4$ dex, slightly more
  metal-poor than M32 in our fields.  The observed blue plume consists
  of stars as young as 0.3 Gyr, in agreement with previous works on
  the disk of M31.  The detection of bright AGB stars reveals the
  presence of intermediate-age population in M31, which is however
  less significant than that in M32 at our field's location.  

 \end{abstract}


\section{Introduction}
Elliptical galaxies contain the oldest  stars in the Universe, and the
study of their composition provides  a means of studying the evolution
of the Universe to large look-back times.  Moreover, they represent at
least  50\%  of   the  total  stellar  mass  in   the  local  Universe
\citep{Schechter_Dressler87,  Gallazzi_etal08}.   Understanding  their
formation and evolution is  crucial to understand galaxy formation and
evolution in  general. 

The study of the resolved stellar content in galaxies is a key tool to
reach this goal.  Stars have the imprint of evolutionary parameters
such as age and metallicity and thus provide a fossil record of the
star formation history (SFH) and evolution of a galaxy.  We can derive
the complete SFH of a galaxy by means of deep and accurate
color-magnitude diagrams (CMDs), given that the most direct
information about any stellar population comes from applying stellar
evolution theory to CMDs. Specifically, the direct observation of the
oldest galaxy's main-sequence turnoff (MSTO) is necessary for an
accurate determination of its age and thus its SFH. Thanks to the
capabilities of the \emph{Hubble Space Telescope} (HST), launched in
1990, stellar populations in spirals and dwarf galaxies in the Local
Group can now be resolved with great accuracy, allowing a precise
determination of complete SFHs with an age resolution of $\sim 1$ Gyr
for ages larger than 10 Gyr \citep[see e.g.,][]{Brown_etal06,
  Barker_etal07, Cole_etal07}. Unfortunately, the large distances
  to giant ellipticals, in combination with their high surface
  brightnesses, prevents detection of their intrinsically fainter
  individual stars, limiting knowledge about their stellar populations
  (although there have been studies of the resolved giants near the
  tip of the red giant branch in nearby ellipticals: see, e.g.,
  \citealt{Sakai_etal97, Harris_etal99,Gregg_etal04,
    rejkuba_etal05}). As a consequence, most elliptical galaxies can
  only be studied by the spectra of their integrated light which
possess contributions from all their stars, having a range of
metallicities and ages.  This makes the unambiguous disentanglement of
the age and metallicity of a stellar population difficult, especially
in old populations such as those that dominate the masses of
elliptical galaxies. Stellar population models have been developed to
derive SFH of ellipticals \citep[e.g.][]{Worthey94, rose94} based on
moderate-resolution spectra \citep[e.g.][]{Gonzalez93, coelho_etal09}.
These models have become very sophisticated in disentangling the
non-trivial age and metallicity degeneracy.  However, they still
suffer several uncertainties and there is a pressing need for them to
be tested with direct observations of stars in an elliptical galaxy.

\subsection{M32: A window on the stellar populations of elliptical galaxies}

The Local Group galaxy Messier 32 (M32) is a small satellite of M31
and the nearest elliptical galaxy.  It is classified as a compact
elliptical (cE) galaxy, cE2, due to its low luminosity, compactness
and high surface brightness \citep{Bender_etal92}. M32 is the
prototype of this class of ellipticals, consisting of $\sim 20$
galaxies known so far \citep{davidge91, ziegler_bender98,
  Chilingarian_etal09}, the so-called \emph{M32-like
  galaxies}. Despite the fact that M32 has been extensively observed
and studied, its SFH and therefore its origins are still a matter of
debate.  The proposed models for M32's origins span a wide range of
hypotheses: from a true elliptical galaxy at the lower extreme of the
mass sequence \citep[e.g.,][]{Faber73, Nieto_prugniel87,
  kormendy_etal09} to an early-type spiral galaxy whose concentrated
bulge, unlike its disk, still survives to the tidal stripping process
caused by its interactions with M31 \citep[e.g.,][]{Bekki_etal01,
  Chilingarian_etal09}.

Nevertheless, M32 is \emph {today} an elliptical galaxy and the
nearest system that has properties very similar to the giant
ellipticals: it falls at the lower luminosity end of all of the
structural and spectroscopy scaling relations of giant ellipticals:
the Faber--Jackson relation \citep[e.g.,][]{Faber_jackson76}, the
Kormendy relation \citep[e.g.,][]{Kormendy85}, the
mass--age--metallicity and $[\alpha/\mathrm{Fe}]$--mass relationships
\citep{Trager_etal00b}, and the Mg--$\sigma$ relation and the
Fundamental Plane of early-type galaxies
\citep[e.g.,][]{Bender_etal92}. More recently, \citet{kormendy_etal09}
find that both central and global parameter correlations from recent
accurate photometry of galaxies in the Virgo cluster place M32 as a
normal, low-luminosity elliptical galaxy in all
regards\footnote{\citet{Graham02} has claimed that M32 has a disk,
based on the ability to fit its brightness profile as a bulge plus
exponential disk.  The location of our field F1 would correspond to a
region where both the disk and bulge should equally contribute to the
light under this model.  However, his bulge plus disk fit is not a
unique decomposition. \citet{kormendy_etal09} fit a Sersic profile to
the SB of M32 with $n=2.8$, which places M32 at the low-luminosity end
of normal ellipticals. They interpret the light in the center of M32
that was not fit by their Sersic profile (which was also not fit by
Graham) as a signature of formation in dissipative mergers.  Extra
central light is a general feature of coreless galaxies and is
observed in all the other low-luminosity ellipticals of Kormendy et
al.'s sample.}.
 
Given its proximity, M32 provides a unique window on the stellar
composition of elliptical galaxies, since it can be studied by both
its integrated spectrum and the photometry of its resolved stars.
While we note that the SFH of a low luminosity elliptical such as M32
($r_{\mathrm{eff}}\approx 40\arcsec$, \citealt{choi_etal02,
  kormendy_etal09}) may differ from those of giant ellipticals, it is
a fact that in general models applied to giant ellipticals reach the
same conclusions as those applied to M32 \citep[e.g.,][]{Worthey98}.
M32 is therefore a vital laboratory to test the applicability of the
stellar population models to more distant galaxies.

\subsection{Integrated light studies of M32}

From spectroscopic studies, one of the most important results of
synthetic population models was found by \citet{oconnell80}: models
fail to reproduce M32 with a single old-age and solar-metallicity
population. Various synthetic population models have claimed that M32
underwent a period of significant star formation in the recent past,
i.e.  about 5--8 Gyr ago, \citep[e.g.,][] {oconnell80, Pickles85,
Bica_etal90} based on the presence of enhanced H$\beta$ absorption in
the integrated spectrum of M32, and thus indicate signatures of an
intermediate luminosity-weighted age population
\citep[e.g.,][]{rose94, Trager_etal00b, Worthey04, Schiavon_etal04,
rose_etal05, coelho_etal09}.  \citet{rose_etal05} studied the nuclear
spectrum of M32 and found radial gradients in both the age and
metallicity of the light-weighted mean stellar population of M32: the
population at $1r_{\mathrm{eff}}$ is $\sim 3$ Gyr older and more metal
poor by $\sim -0.25$ dex than the central population, which has a
luminosity-weighted age of $\sim$ 4 Gyr and [Fe/H]$\sim 0.0$.
Extrapolation of the spatially resolved spectroscopy of
\citet{Gonzalez93} results in an average age and metallicity of M32 at
1$\arcmin$ from its center of 8 Gyr old and [Fe/H]$\sim -0.25$
\citep{Trager_etal00b}.  This is consistent with a more recent
estimate by \citet{Worthey04} who found the age of M32 at 1$\arcmin$
to be 10 Gyr old.  The most recent results from stellar population
models are given by \citet{coelho_etal09} who observed high
signal-to-noise spectra at three different radii, from the nucleus of
M32 out to $\sim 2 \arcmin$ from the center of M32. They propose that
an ancient and intermediate-age populations are both present in M32
and that the contribution from the intermediate population is larger
at the nuclear region.  They claim that a young population is present
at all radii \citep[see also e.g.,][]{Trager_etal00b, rose85, rose94,
Schiavon_etal04}, but its origin is unclear.  Moreover, the
determination of ages in integrated spectra is a difficult problem, as
extended horizontal branch morphologies and/or blue stragglers,
unaccounted for in the models, can mimic younger ages
\citep[e.g.,][]{Burstein_etal84, rose85, defreitas_barbuy95,
Maraston_thomas00, Trager_etal05}.  Thus, lacking any direct evidence
for such a young population (ages $< 1$ Gyr), and due to the
uncertainties in the synthesis models, these results should be
considered with some caution.

\subsection{Individual stars studies of M32}

On the other hand, photometric studies of resolved stars have
supported the existence of an intermediate-age population
\citep[e.g.][]{freedman92, davidge_jensen07} by detecting AGB stars
suggestive of a $\sim$ 3 Gyr old population. However, observations by
\citet{davidge_jensen07}, obtained with the NIRI imager on the Gemini
North telescope, do not support spectroscopic studies that find an age
gradient in M32, since they suggest that the AGB stars and their
progenitors are smoothly mixed throughout the main body of the galaxy.
\citet{brown_etal00, brown_etal08}, using ultraviolet observations of
the center of M32, and \citet{Fiorentino_etal10}, using the ACS/HRC
data presented here, have found evidence of an ancient, metal-poor
population by observing blue horizontal branch and RR Lyrae stars,
respectively.  \citet{Worthey_etal04}, using optical observations
obtained with the Wide Field Planetary Camera 2 (WFPC2) on board HST
and presented by \citet{alonsogarcia_etal04}, studied the stellar
populations of the outer regions of M32 and M31 and found that there
is no trace of a main sequence younger than $\sim 1$ Gyr in M32 at a
region $7r_{\mathrm{eff}}$ from its center.  The most extensive study
of the resolved stellar populations of M32 has been carried out by
Grillmair et al.\ (1996, hereafter G96), who resolved individual stars
down to slightly below the level of the HB with the HST WFPC2 in a
region of 1--$2\arcmin$ from the center of the galaxy. Their most
important result is the composite nature of the CMD of M32.  They
concluded that the wide spread in color of the giant stars in their
CMD cannot be explained only by a spread in age but rather by a wide
spread in metallicity.  However, given the age--metallicity degeneracy
on the giant branch, there may well be a mixture of ages present in
their field, but age effects are less important than metallicity on
the giant branch morphology.  For an assumed age of 8.5 Gyr old, the
metallicity distribution function has a peak at $\mathrm{[Fe/H]}\sim
-0.25$, consistent with the extrapolation made from the spatially
resolved spectroscopy of \citet{Gonzalez93}.  The spread in
metallicity found by G96 ranges from roughly solar to below $-1$
dex. This study, as well as those by Brown et al., Alonso-Garcia et al.
and Worthey et al., concluded that the metal-poor population is
insignificant, contrary to the results of \citet{coelho_etal09}.
Finally, a young population of $\la 1$ Gyr claimed by several
population models to be present in the spectrum of M32 has not been
seen by any of the observations of resolved stars.  Overall, the
photometric studies carried out so far only obtained information from
the brighter stars of M32, i.e., the upper CMD.  These studies were
prevented from observing fainter stars by the extreme crowding of M32.
Since upper giant-branch tracks are degenerate in age and metallicity,
much like integrated colors and metallic lines, it is not possible to
derive an age from the upper CMD alone. 

The only  way to  derive the SFH  of M32  and test conclusions  so far
based solely  on integrated colors  and spectral indices is  to obtain
deep CMDs  that reach the MSTO  of M32. Measuring the  position of the
blue turnoff  stars with accurate  photometry is the only  evidence to
test the ages  inferred from population synthesis models.   A deep CMD
and luminosity function  of M32 can be used as  the basis for spectral
synthesis  studies.   An  agreement  between  observed  and  synthetic
indices for M32 would confirm  such indices as simple diagnostic tools
for  constraining stellar  populations  in integrated  light of  other
elliptical galaxies, for which only the integrated light is available,
given their greater distances.  Moreover  the CMD allows for the study
of spreads about mean properties in a way that is currently impossible
with integrated  light.  These  spreads are as  important as  the mean
values in decoding the SFH of the galaxy.

In order to further investigate the stellar content of M32, and with
the primary goal of deriving a complete SFH of this enigmatic galaxy,
we were awarded 64 orbits of the HST to observe the MSTO of M32 with
the High-Resolution Channel (HRC) at the Advanced Camera for Surveys
(ACS).  The proximity of M32, combined with the high resolution of HST
ACS/HRC allows for a remarkable improvement in our study of its
stellar content.  In this paper, we introduce our new observations and
present the deepest optical CMD so far obtained. The CMD presented
here reaches more than 2 magnitudes fainter than the previous optical
CMD by G96 and fully resolves the red giant branch (RGB) and the
asymptotic giant branch (AGB).  We report the discovery of a blue
plume (BP), consisting of young stars and/or blue straggler stars, not
claimed to have been observed before. We also detect for the first
time in M32 a RGB bump and an AGB bump.  By analyzing our CMD we have
achieved the most comprehensive photometric study of the resolved
stellar content of M32.  A follow-up paper will present the
  recent and intermediate SFH of M32 that can be derived from these
  data.  In addition, as discussed above, these data have already been
  analyzed by \citet[][hereafter F10]{Fiorentino_etal10} to study RR
  Lyrae variables in our fields.

The paper is organized as follows.  Section 2 describes our
observations and the reduction of the data.  The photometry performed
and the extensive study of completeness and crowding of the data are
presented in Section 3. Section 4 presents the decontamination of the
M32 field from the light contribution by M31. The analysis of the CMD
of M32 and its luminosity function is presented in Section 5. We
derive the distance to M32 and M31 is Section 6. In Section 7 we
analyze the M31 stellar populations in our background field. We
summarize our findings in Section 8.

\section{Observations and Data reduction}
\subsection{Field Selection and Observational Strategy} 
We obtained deep $B$ and $V$-band imaging of two fields near M32 using
the ACS/HRC instrument on board HST during Cycle 14 (Program GO-10572,
PI: Lauer).  The ACS $F435W$ ($B$) and $F555W$ ($V$) filters were
selected to optimize detection of MSTO stars over the redder and more
luminous stars of the giant branch.  M32 is very compact and is
projected against the disk of M31.  The major challenge was to select
a field that represented the best compromise between the extreme
crowding in M32, which would drive the field to be placed as far away
from the center of the galaxy as possible, versus maximizing the
contrast of M32 against the M31 background populations, which would
push the field back towards the central, bright portions of M32.
Following these constraints, the M32 HRC field (designated F1) was
centered on a location $110''$ south (the anti-M31 direction) of the
M32 nucleus, roughly on the major axis of the galaxy. The $V$-band
surface brightness of M32 near the center of the field is
$\mu_V\approx21.9$ \citep{kormendy_etal09}.  M32 quickly becomes too
crowded to resolve faint stars at radii closer to the center, while
the galaxy rapidly falls below the M31 background at larger radii.

\begin{figure*} \centering
  \includegraphics[width=140mm,clip]{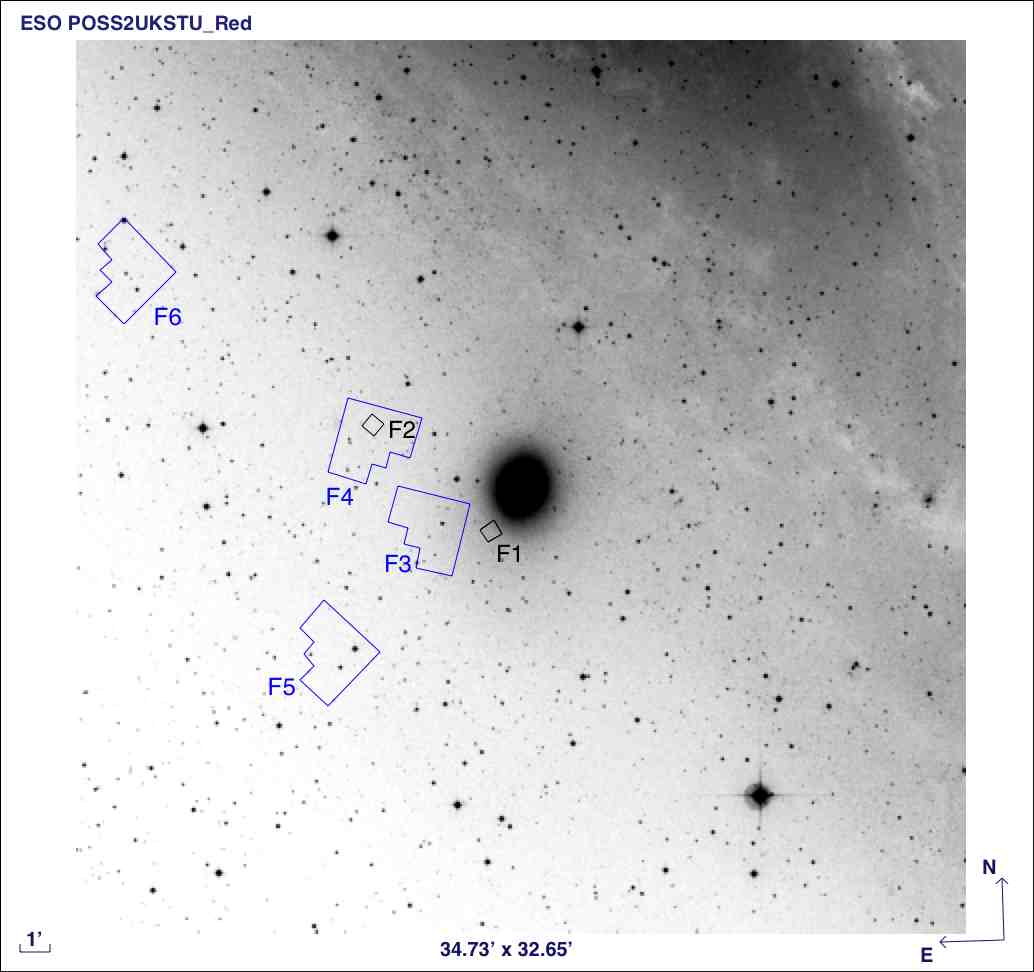}
  \caption{Location of our two HST ACS/HRC pointings: M32 (F1) field
    and M31 background (F2) field, both indicated as black small
    boxes.  Each field covers a region of $26 \times 29
    \,\mathrm{arcsec^{2}}$ on the sky.  The field F1 is located at
    110\arcsec from the nucleus of M32 and represents the best
    compromise between minimizing image crowding and contamination
    from M31.  The F2 field is at the same isophotal level in M31
    corresponding to the location of the F1 field.  Thirty-two
    exposures in each of the $F435W$ ($B$) and $F555W$ ($V$) filters
    were taken for each field. The location of fields F3, F4, F5 and
    F6 is also shown in blue.  They are archival HST/WFPC2 fields near
    M32 that were analyzed in the Appendix to investigate the presence
    of a ``blue plume''.  Information about these observations can be
    found in Table ~\ref{table:wfpc2fieldsdata}.  North is up and East
    is to the left.}
 \label{fig:location}
 \end{figure*}

Even at  the location  of F1, M31  contributes $\sim1/3$ of  the total
light with inner disk and bulge stars (K. Howley, private comm.), thus
it  was  critical  to obtain  a  background  field,  F2, at  the  same
isophotal level  in M31 ($\mu_V\approx22.7$)  to allow for  the strong
M31  contamination to  be  subtracted  from the  analysis  of the  M32
stellar population.  F2, which also contains both inner disk and bulge
M31 stars (K. Howley, private comm.), was located $327''$ from the M32
nucleus at  position angle $65^\circ$.  At this  angular distance M32
has an ellipticity of $\epsilon\approx0.25$ \citep{choi_etal02} and F2
is nearly aligned with M32's  minor axis.  Thus the implied semi-major
axis  of  the  M32  isophote   that  passes  through  F2  is  $435'',$
significantly  larger  than   the  nominal  angular  separation.   The
estimated M32 surface brightness at F2 is $\mu_V\approx27.5,$ based on
a  modest   extrapolation  of  the  $B$-band   surface  photometry  of
\citet{choi_etal02}  and  an assumed  color  of $B-V\approx0.9$.   The
contribution of M32 to F2 thus falls by a factor of $\sim180$ relative
to its surface brightness at F1.  While one might have been tempted to
move F2  even further away from  F1, it clearly serves  as an adequate
background at  the location selected,  while uncertainties in  the M31
background would increase at larger angular offsets.  The locations of
both fields are shown in Figure~\ref{fig:location}.
\begin{deluxetable*}{lcccccc}
\tabletypesize{\scriptsize}    
\tablecaption{Log of observations\label{table:data}} 
\tablewidth{0pt}
\tablehead{\colhead{Field}&\colhead{$\alpha_{J2000.0}$}&\colhead{$
  \delta_{J2000.0}$}  &\colhead{Filter}&\colhead{Exposure time (s)}&
\colhead{Date}  & \colhead{FWHM
  ($\arcsec$)}}
\startdata
 F1 &00 42 47.63 &+40 50 27.40
  &$F435W$ &16$\times$1279 + 16$\times$1320 &Sept 20--22, 2005 &0.04 \\
  F1   &00  42 47.63  &+40  50 27.40  &$F555W$ &16$\times$1279  +
  16$\times$1320 &Sept 22--24, 2005 &0.05 \\ F2  &00 43 07.89 &+40
  54 14.50  &$F435W$ &16$\times$1279  + 16$\times$1320 &Feb  6--8, 2006
  &0.04\\ F2  &00 43 07.89 &+40 54 14.50 &$F555W$ &16$\times$1279
  + 16$\times$1320 &Feb 9--12, 2006 &0.05
\enddata
\end{deluxetable*}

Detection  of  the  MSTO  required  deep exposures  at  F1.   Accurate
treatment  of the  background required  equally deep  exposures  to be
obtained  in  F2.   A summary  of  the  observations  is shown  in  in
Table~\ref{table:data}; briefly, each field was observed for 16 orbits
in each of  the $F435W$ and $F555W$ filters for a  total program of 64
orbits. 

At $B$, and even $V$, HRC undersamples the PSF, despite its
exceptionally fine pixel scale.  All of the images were obtained in a
$0.5\times0.5$ sub-pixel square dither pattern to obtain Nyquist
sampling in the complete data set.  In detail, the sub-pixel dither
pattern was executed across each pair of orbits, with each orbit split
into two sub-exposures.  The telescope was then offset by $0\farcs125$
steps between the orbit pairs in a ``square-spiral'' dither pattern of
maximum extent $\pm5$ pixels to minimize the effects of ``hot
pixels,'' bad columns, and any other fixed-defects in the CCD, on the
photometry at any location.  The data for each filter/field
combination thus comprises 8 slightly different pointings, with
Nyquist-sampling obtained at each location.

In  addition to the  HRC images,  parallel observations  were obtained
with the ACS/WFC channel using  the $F606W$ filter (broad $V$).  Those
images have  been analyzed by \citet{sarajedini_etal09},  who find 324
and 357  RR Lyrae  variables stars in  the parallel  fields associated
with F1 and F2, respectively.

\subsection{Image Reduction, Stacking, and Upsampling}

As  outlined above, the  data set  for each  of the  four filter/field
combinations comprises 32 exposures with non-redundant pointings.  The
images were combined in an  iterative procedure designed to detect and
repair  cosmic-ray  events, hot  pixels,  and  other  defects, with  a
Nyquist-sampled summed image as the final product.

The  first reduction step  was to  interlace the  four images  at each
position within  the larger square-spiral dither pattern  into a rough
Nyquist image  at that position.   In practice, the  sub-pixel dithers
were  accurate  to  $<0\farcs01,$   so  a  simple  interlace  worked
reasonably  well for  the  initial reduction.   For  this first  step,
cosmic  rays  in  one  of   the  four  images  could  be  repaired  by
interpolation among  the three remaining frames.   The resulting eight
Nyquist  images   were  then  shifted  to  a   common  centroid  using
sinc-interpolation  (which does  not smooth  the data),  and  added to
produce an  initial stack.   At this point  the stack  still contained
artifacts from  coincident cosmic-ray events within each  of the eight
subgroups, as well as hot pixels; although both types of artifacts are
reduced in  amplitude by the  averaging implicit in the  larger dither
pattern.

The second  reduction cycle used  the initial Nyquist  summed-image to
then  re-identify and repair  cosmic-ray hits  in each  of the  32 raw
images.  Hot pixels were also identified and repaired at this stage by
finding  coincident   events  in  detector,   rather  than  celestial,
coordinates.  At this point  a higher quality Nyquist summed-image was
generated by  combining all 32  images (trimmed to their  common area)
using  the  Fourier  algorithm  of \citet{lauer99a}.   This  algorithm
produces a  summed image  with double the  native HRC pixel  scale, by
combining the images in the  Fourier domain to eliminate aliased power
from the under-sampled source images.  The algorithm has no adjustable
parameters, approximations,  and so on, and important  for the present
application, induces no smoothing or degradation of the PSF.

The final reduction cycle was just to repeat the second cycle, but
using the output from the second cycle as the input for the detection
and repair of cosmic-ray events and hot pixels.  The final summed
image is thus essentially free of artifacts.  As it eliminates the
undersampling in the HRC, which provides the finest pixel scale of all
HST instruments to begin with, it represents one of the
highest-resolution images obtained with the observatory, given the
blue-bands selected for the observations.

The final image still contains the geometric field distortion inherent
to the HRC.  Since the image is Nyquist-sampled, it can be rectified
using sinc-interpolation, given the STScI two-dimensional polynomial
representations of the distortion, without incurring degradation of
the PSF.  The required correction can be done by multiplying the image
by the appropriate pixel area map (PAM).  We construct a PAM
image\footnote{We have downloaded the script example as well as the
  coefficients files which are needed for the PAM image construction
  from the web page \url{ http://www.stsci.edu/hst/acs/analysis/PAMS}.
  We have executed the script in IRAF.}  for the HRC which has a size
of 2048$\times$2048 pixels, as this is the size of each combined
image. The PAM is expressed in units of the pixel scale corresponding
to our combined image, i.e., half of the native HRC scale, and it has
an approximate value of $\sim$ 1.12 near the image center.  We correct
each combined image as follows:
\begin{equation} \mathrm{CORRECTED_{flux}=COMBINED_{flux}\times PAM}
\end{equation}

\begin{figure*} \centering
\vspace{-0.001cm}
\includegraphics[width=125mm,clip]{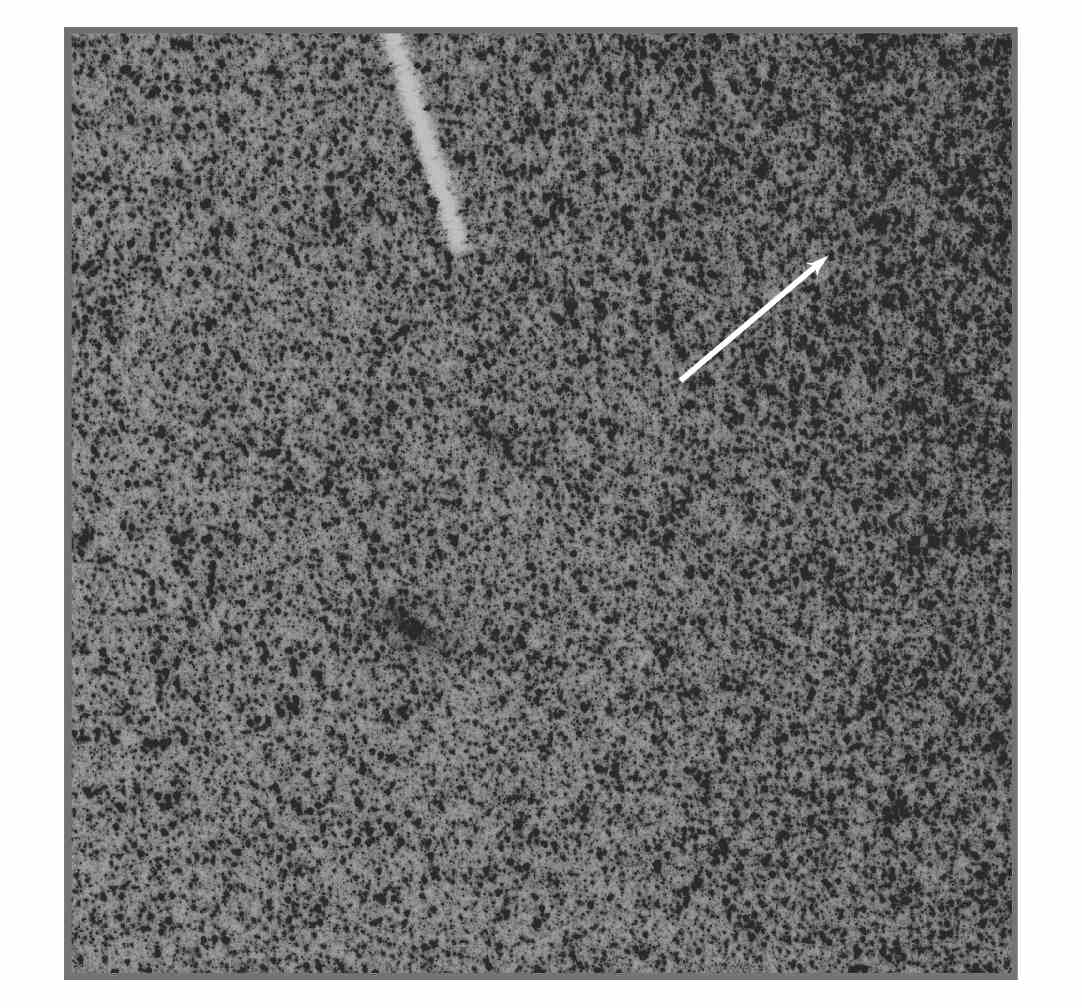}
%\vspace{-0.001cm}
\includegraphics[width=125mm,clip]{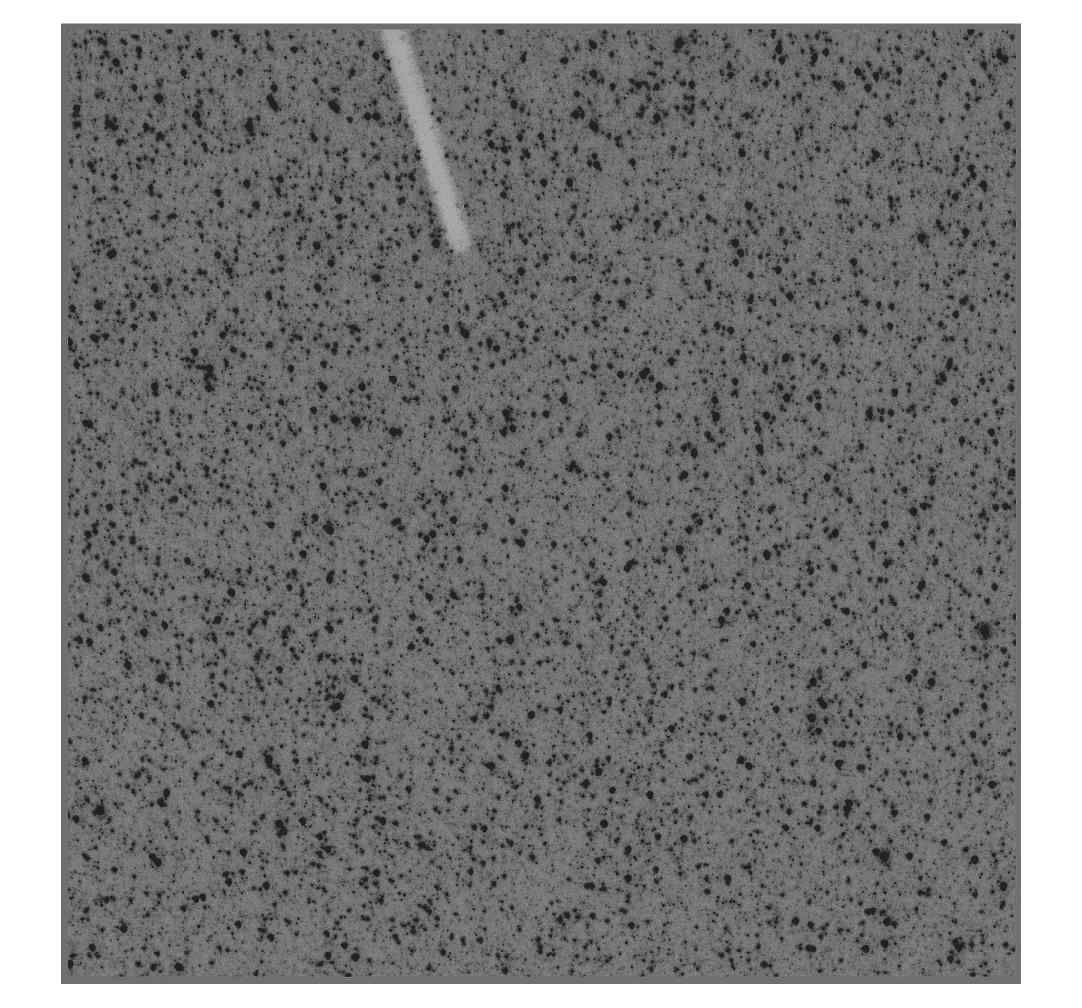}
\caption{Combined images  of the 32  $F555W$ exposures in the  F1 (top
  panel) and F2  (bottom panel) fields displayed with  the same linear
  stretch.  Each  image has a  size of 2048$\times$2048 pixels  with a
  $0\farcs0125$ pixel  scale. There is  a clear difference  in stellar
  density  between the  images.   This difference  indicates that  the
  crowding  is more severe  in F1  than in  F2 field.  We also  note a
  stellar  density gradient  in  the F1  image,  becoming higher  when
  approaching the center  of M32, whose direction is  indicated by the
  arrow in  the top panel.  The long white  spot in the top  center of
  each image is the occulting finger of the ACS/HRC coronagraph.}
\label{fig:fields}
\end{figure*}

The combined images of F1 and  F2 fields used for analysis thus have a
pixel scale of $0\farcs0125$ and  a resolution of $\sim 0\farcs05$ for
point sources. They  are shown in the top (F1)  and bottom (F2) panels
of  Figure~\ref{fig:fields}, in  the  $F555W$ filter,  from which  the
strong crowding in  these fields is clearly seen.   There is however a
difference between the  stellar density in F1 and  F2, as the crowding
is more severe in F1 than in F2.  The arrow in the top panel indicates
the direction towards the center of M32.

\section{Photometry}
The traditional method for extracting stellar photometry in crowded
fields uses standard stellar photometry packages, e.g.  DAOPHOT II
\citep{stetson87}, which are specifically designed for this problem.
This approach is favored over direct deconvolution of the PSF from the
images, as HST images in general are undersampled, and deconvolution
treats the artifacts due to aliasing as genuine sources (see
\citealt{Holtzman_etal91}), resulting in large photometric errors.  In
the present case, however, the Nyquist-sampled and high-S/N summed
images are free from artifacts.  This motivated a re-examination of
using general-purpose PSF deconvolution to mitigate the extreme
crowding in the images, which we did in parallel to reducing the
images with DAOPHOT II.  To our delight, the deconvolved photometry is
superior to that done with DAOPHOT, a conclusion based both on the
sharpness of features in the CMDs derived from the images and
extensive artificial star tests.  We present derivation of stellar
photometry using both methods in this section, showing why we decide
in the end to solely use the deconvolved photometry for the analysis
of the CMD.

\subsection{DECONVOLVED IMAGE photometry}

The  final summed  images  of F1  and  F2 were  deconvolved using  the
Lucy-Richardson  algorithm \citep{Lucy_74,  Richardson_72}.   The PSFs
for the $F435W$ and  $F555W$ images were constructed interactively and
iteratively by summing the  brightest relatively isolated stars in the
images to produce \emph{ad hoc} PSFs, which were then used to clean out
the fainter stars in the wings of the PSF stars, resulting in improved
PSFs, which were  then used to refine the PSFs  further.  In all steps
of the process, sinc-function interpolation was used to shift the PSFs
and their component stars as needed.

The Lucy-Richardson algorithm works iteratively, quickly removing the
``wings'' of the PSFs, but taking considerably longer to enhance
structure on the scale of the central diffraction cores.  In the
present case, we used 640 iterations on the $F555W$ images and 160
iterations on the $F435W$ images.  This heavy level of deconvolution
nearly transforms the images into a set of delta functions, but in
doing so serves to split closely blended stars; pairs of stars as
close as $\sim 0\farcs03$ were separated.  Due to its higher S/N,
stars were identified in the $F555W$, while the $F435W$ image provided
fluxes at the position of the identified stars.  Stars were identified
by a simple peak-finding algorithm.  We had no formally-derived
criterion for the threshold used to identify peaks.  Instead, we
examined faint sources in relatively isolated regions and adopted a
single threshold for a given image that roughly separated what
appeared to be real stars from noise fluctuations. In practice, the
real depth of the photometry was established by the artificial star
tests (ASTs) described after this section. We measured the stellar
fluxes as follows.  After the central pixel of the source was
identified in the $F555W$ deconvolved image, we summed the counts
within a $3\times3$-pixel box centered at this position.  The
positions of the stars identified in the $F555W$ deconvolved image are
used to find the stars in the $F435W$ deconvolved image.  We measured
the fluxes in the $F435W$ deconvolved in the same way, summing the
counts within a $3\times3$-pixel box around the central pixel of the
source in the $F435W$ deconvolved image. All these steps were
performed using algorithms running in the XVISTA
package\footnote{\url{http://ganymede.nmsu.edu/holtz/xvista/}}. The
catalog of each field was cleaned of stars located in the borders, and
in the occulting finger of the image.  The final number of stars
obtained with the deconvolution process is indicated in
Table~\ref{table:detections}.

The deconvolved magnitudes needed to be corrected for a small
non-linearity in the deconvolved flux. This is due to the fact that
the fraction of flux in the box defined to measure the flux varies
slightly with flux.  We generated a correction table from simulated
deconvolutions on a constant-sky level image.  Stars were injected
with appropriate Poisson noise as a function of flux. For each 0.2
step in magnitude we generated 16 stars per simulation and recover
them performing the deconvolution in the same way as was done with the
real stars.  The correction is rather small (less than 0.1 mag) and
only affects the magnitudes of some of the stars. Stars at both the
brighter and fainter end are not affected by this small non-linearity.

In order  to transform the  instrumental magnitudes of the  stars into
apparent magnitudes,  they need to  be corrected for two  effects that
reduce the measured stellar flux: charge transfer efficiency (CTE) and
aperture correction.

\emph{Charge Transfer Efficiency (CTE)}: Charge lost due to imperfect
electron  transfers  from pixel  to  pixel  and  then to  the  readout
amplifier degrades  the photometry. Due to the  gradual degradation of
ACS after its installation in 2002, the effects of CTE are noticeable.
The correction needed for ACS/HRC is given in the ACS Data Handbook:
\begin{multline}   
\displaystyle  
\mathrm{\Delta mag = 10^A\times SKY^B\times FLUX^C\times \frac{Y}{2000}}\times
  \\\frac{(\mathrm{MJD}-52333)}{365}
\end{multline}  
where $\mathrm{A}=-0.44 \pm 0.05$, $\mathrm{B}=-0.15 \pm 0.02$ and
$\mathrm{C}=-0.36 \pm 0.01$ are the most recent coefficients
\citep{Chiaberge_etal09}.  In this formula, SKY is the sky level in
electrons measured near the star, FLUX is the flux of the star in
electrons, Y is the number of transfers which, when the default
amplifier C has been used for readout as in our case, is simply the
$\it{y}$ coordinate of the star.  Finally, the images used to obtain
the stellar magnitudes are constructed using images with very similar
exposure times.  Therefore, the SKY and FLUX values in the formula
should be divided by the number of images used.  The $\mathrm{\Delta
  mag}$ for each star is subtracted from its measured
magnitude. Values of $\mathrm{\Delta mag}$ vary from 0.001 to 0.1.

\emph{Aperture Correction}: The PSFs used for the deconvolution have
limited extent stellar wings in order to avoid as much as possible the
contamination by neighboring stars.  Hence, the contribution of the
flux in the large extent of the stellar wings need to be added to the
measured magnitudes.  The standard procedure to perform this
correction consists of obtaining the flux for a small number of bright
and isolated stars within a $0\farcs5$ aperture radius, after all
resolved stars -- except those to be measured -- have been removed.
The median value of the differences between the magnitudes obtained
from this flux and the one measured is the aperture correction
\citep{stetson_harris88}.  This correction is then applied to all of
the star magnitudes.  After this step, the correction from $0\farcs5$
to ``infinite'' is made using the tables in \citet{sirianni_etal05}.
In our case, such bright and isolated stars are unavailable because
the field is so crowded.  Moreover, even if we could find some bright
isolated stars, we would need to subtract an enormous number of stars
from the image.  The residuals from the PSF-fitting of all those
subtracted stars will remain, adding fluctuations to the image and
therefore significant errors to the photometric measurements.  For
these reasons we have decided to use the encircled energies (EE) which
have been tabulated by \citet{sirianni_etal05} and provide the
fraction of the total source count as a function of the aperture
radius, instead of the usual method.  At each PSF radius, we calculate
the fraction of flux that is missing and add this to the magnitudes.
These values differ in each filter band and they are listed in
Table~\ref{table:detections} as well as the PSF radius used for the
deconvolution, for each filter/field combination.  We are aware that
the aperture corrections calculated with the EE should only be applied
to the photometric data for aperture radii larger than 10 pixels in
the original HRC image, and therefore 20 pixels in our combined
images.  This was only the case for one of the PSFs we have
obtained. However, due to the issues explained above, we have no other
way to correct by aperture correction without introducing significant
errors.

Finally, if $\lambda$ is the bandpass in the ACS/HRC system, the
apparent magnitudes $\mathrm{m_{ap}}$ are transformed into the VEGAmag
system using the zero points 36.73 and 36.80 for $F435W$ and $F555W$
respectively, which were obtained as follows:
\begin{equation} \mathrm{ZP(\lambda) = 2.5\log[t_{exp}(\lambda)] +
ZP_{VEGAmag(\lambda)}}
\end{equation} The values for
$\mathrm{ZP_{VEGAmag(\lambda)}}$ are 25.19 and 25.26 for $F435W$ and
$F555W$
respectively\footnote{\url{http://www.stsci.edu/hst/acs/analysis/zeropoints}}.

\begin{deluxetable}{@{}lcccccc}
  \tabletypesize{\scriptsize}
  \tablecaption{Detections\label{table:detections}} 
  \tablewidth{0pt}
  \tablecolumns{6}           
  \tablehead{&\colhead{Detections\tablenotemark{a}}&
    \colhead{$R^{F435W}_{\rm{PSF}}$\tablenotemark{b}}&
    \colhead{$R^{F555W}_{\rm{PSF}}$\tablenotemark{b}}&
    \colhead{AC$_{F435W}$\tablenotemark{c}}&
    \colhead{AC$_{F555W}$\tablenotemark{c}}}
  \startdata
  \sidehead{Deconvolution}\hline
  F1&58,143&5&5&$-0.25$&$-0.22$ \\
  F2&27,963&6&16&$-0.22$&$-0.10$ \\ \hline
  \sidehead{DAOPHOT II}\hline
  F1&50,583&6&6&$-0.22$&$-0.21$ \\
  F2&19,780&6&6&$-0.22$&$-0.21$
\enddata
\tablenotetext{a}{Final number of stars detected and used to derive
  the CMDs}
\tablenotetext{b}{PSF radius in HRC original pixels}
\tablenotetext{c}{Aperture correction}
\end{deluxetable}

\subsection{DAOPHOT photometry}
We also performed stellar photometry with the standard DAOPHOT II
and ALLSTAR  packages \citep{stetson87, stetson94} to compare with
the photometry obtained from deconvolved images.

We first built a PSF for each combined image from bright and as
isolated as possible stars in each field of view (FoV).  This was done
iteratively using the PSF routine of DAOPHOT II.  The number of stars
that finally remain to construct the PSF is about 50 per image.  After
testing various PSF models, we adopted a ``Penny'' function (a sum of
a Gaussian and a Lorentz function) as the best analytical model for
all the images, according to the Chi value calculated by the PSF
routine.  We adopted a PSF that varies quadratically with position.

We performed PSF-fitting photometry on the images using ALLSTAR. The
procedure was applied to a list of star candidates obtained from a
DAOPHOT routine, from which concentric aperture photometry was
performed to obtain crude apparent magnitude estimates and sky
determination for all the objects found.  Due to the severe crowding,
the procedure of finding star-like objects, concentric aperture
photometry and profile-fitting photometry was performed 3 times in
order to both find the faint stars and improve the photometry on the
bright stars.  After this procedure we had four ALLSTAR output lists,
one for each filter on each field, from which we retain only the
objects having a statistically good photometry. We made use of the Chi
value, sharpness index, and magnitude errors given by ALLSTAR to
eliminate possible false photometric detections, e.g.  background
galaxies, unrecognized blends or cosmic rays.  An extra step was
performed and we cleaned each list of stars located both at the edge
of the image, i.e. those stars having image coordinates X or Y either
$<$ 23 or $>$ 2023, and in the occulting finger of the ACS/HRC CCD,
for which the magnitudes were poorly determined.

We then used  the DAOMATCH and DAOMASTER algorithms  (Stetson 1994) to
correlate the output list from the $F435W$ filter with the output list
from the $F555W$  filter.  This created a combined  star catalog. An
object was  considered to  be a star  if it  is found in  both filters
($F435W$ and $F555W$) within a distance of 2 pixels.  The final number
of   stars  that   we   obtained   for  each   field   is  listed   in
Table~\ref{table:detections}.
 
The last step consisted of applying the CTE and aperture corrections
to obtain the apparent magnitudes in the VEGAmag system.  This was
done in exactly the same way explained in the above section.  The PSF
radius as well as the aperture correction values used for this
photometry are indicated in Table~\ref{table:detections} for each
filter/field combination.

\subsection{Comparison of the two photometric methods}
\begin{figure*}
\hspace{-1.1cm}
 \includegraphics[width=190mm, clip]{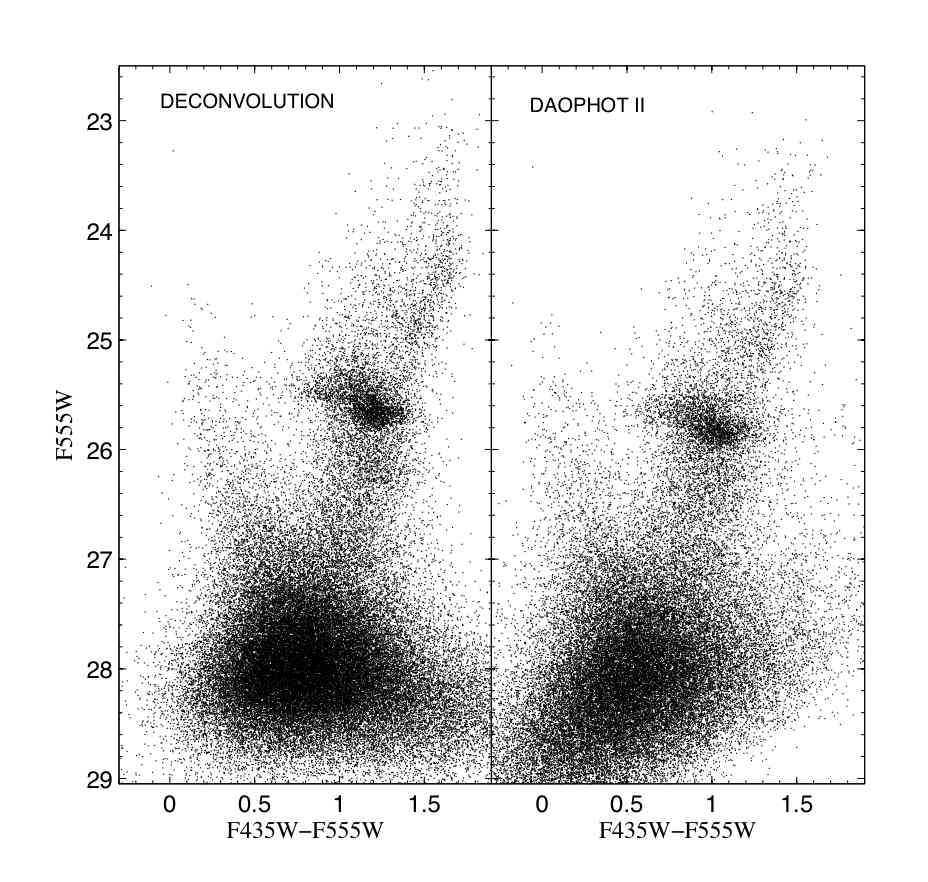}
 \caption{($F435W - F555W$, $F555W$) CMDs of field F1 obtained from
   the deconvolution (left-hand panel) and DAOPHOT (right-hand panel)
   photometry. They contain 58,143 and 50,583 stars respectively, and
   are calibrated onto the VEGAmag HST system.  We can see that
   features in the deconvolved CMD are more clearly delineated than in
   the DAOPHOT CMD, at all luminosities.  All of the features
   described in Section 5 are much sharper and better defined,
   e.g. the RGB and RC. Moreover, the outliers to the red of the RGB
   ($F435W-F555W > 1.5$, $F555W=$ 25--27.5) are greatly reduced in the
   deconvolved CMD when compared with DAOPHOT.  We therefore only use
   the deconvolved CMD for further analysis.  See text for more
   details.}
\label{fig:deconvolvedmcmd}
\end{figure*}

\begin{figure*}
\hspace{-1.1cm} 
\includegraphics[width=191mm, clip]{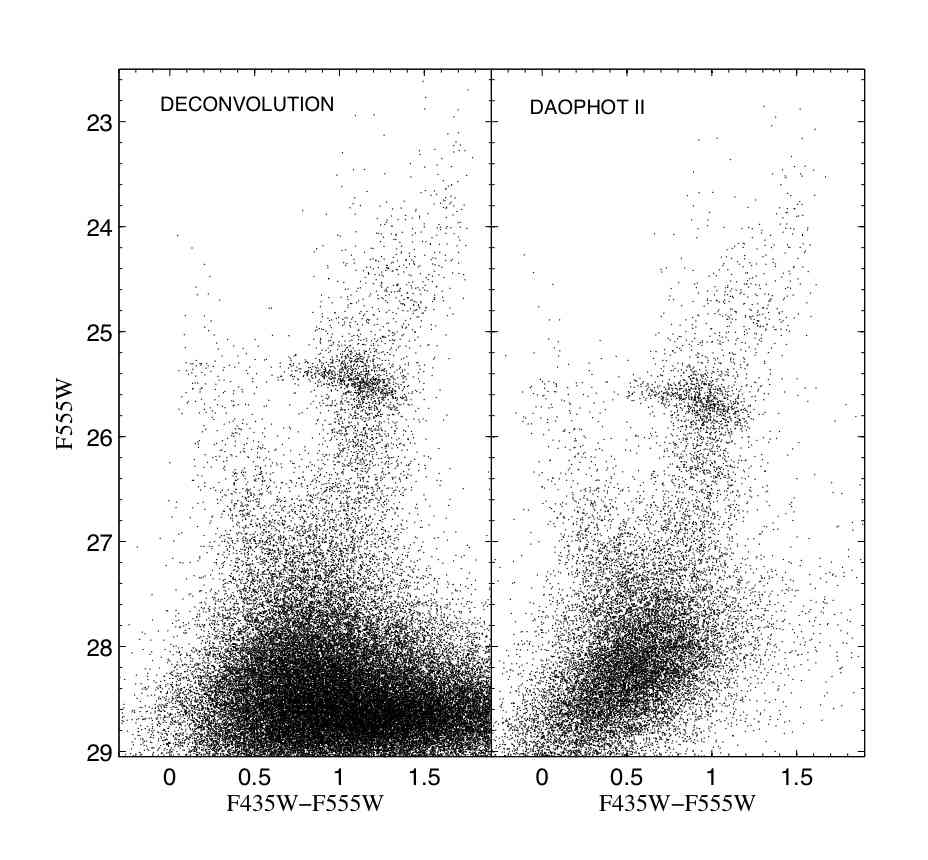}
\caption{Same as Figure~\ref{fig:deconvolvedmcmd} but for field F2.}
\label{fig:calibratedcmds}
\end{figure*}

\begin{figure}
\includegraphics[width=90mm,clip]{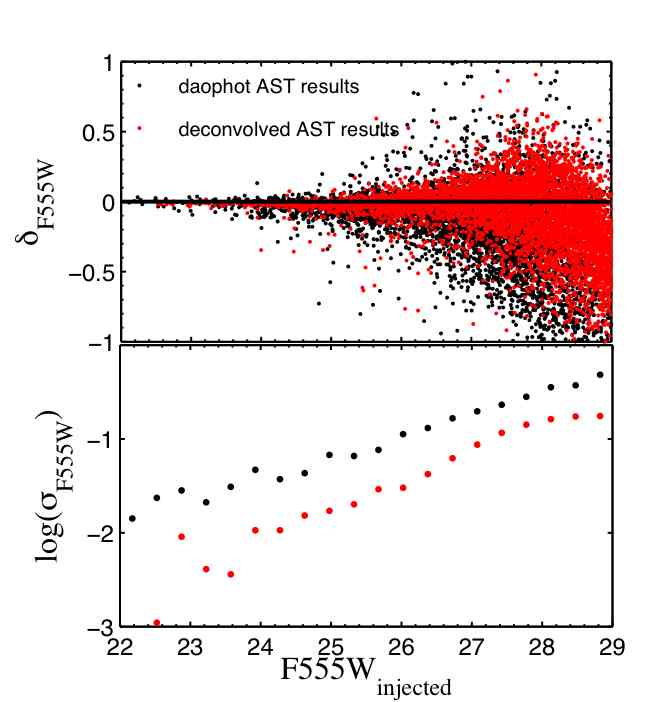}
\caption{Comparison of 10 ASTs results from deconvolved images (red
  dots) and DAOPHOT photometry (black dots). Top panel: (recovered $-$
  injected) magnitudes as a function of injected magnitudes. Note the
  larger spread in the recovered DAOPHOT magnitudes compared with the
  deconvolved magnitudes, especially at the bright end. Bottom panel:
  Mean errors as a function of injected magnitudes.  Errors are
  greatly reduced in the deconvolved photometry when compared with
  DAOPHOT. }
\label{fig:daovsdeconv_ast}
\end{figure}

To compare the two photometric lists, we first directly examine the
CMD obtained from the deconvolved images with that obtained from the
photometry performed using DAOPHOT.  Figure~\ref{fig:deconvolvedmcmd}
shows the deconvolved (left panel) and DAOPHOT (right panel) CMDs
obtained for field F1. The same but for F2 is shown in
Figure~\ref{fig:calibratedcmds}. We see that the deconvolved CMDs look
better, as they produce notably clearer features at all luminosities.
All of the features described in Section 5 are much sharper and better
defined, and the outliers to the red of the RGB ($F435W-F555W > 1.5$,
$F555W=$ 25--27.5) are greatly reduced.  In addition, a visual
inspection of the subtracted images reveals that deconvolution does a
considerably better job in resolving blended stars.  Furthermore, we
compared the results given by the ASTs (see
Section~\ref{ast} for a detailed description) using deconvolved images
with that obtained using DAOPHOT.  Figure~\ref{fig:daovsdeconv_ast}
shows, in the top panel, the differences between the recovered and
injected magnitudes obtained using deconvolved images (red dots) and
DAOPHOT photometry (black dots), as a function of the injected
magnitudes.  The bottom panel shows the mean error of these
differences as a function of injected magnitudes.  For clarity, only
the results of the same 10 ASTs (i.e.  20,000 injected stars analyzed)
are shown.  We can see that there is much more scatter in the
DAOPHOT-recovered magnitudes at all magnitude levels; this is
especially clear at the bright end.  As shown in the bottom panel of
Figure~\ref{fig:daovsdeconv_ast}, the deconvolved photometry results
in smaller errors than DAOPHOT.  All this indicates that the
photometry performed on the deconvolved images is superior to that
obtained using DAOPHOT.

\begin{figure} \centering \subfigure
{\includegraphics[width=88mm,clip]{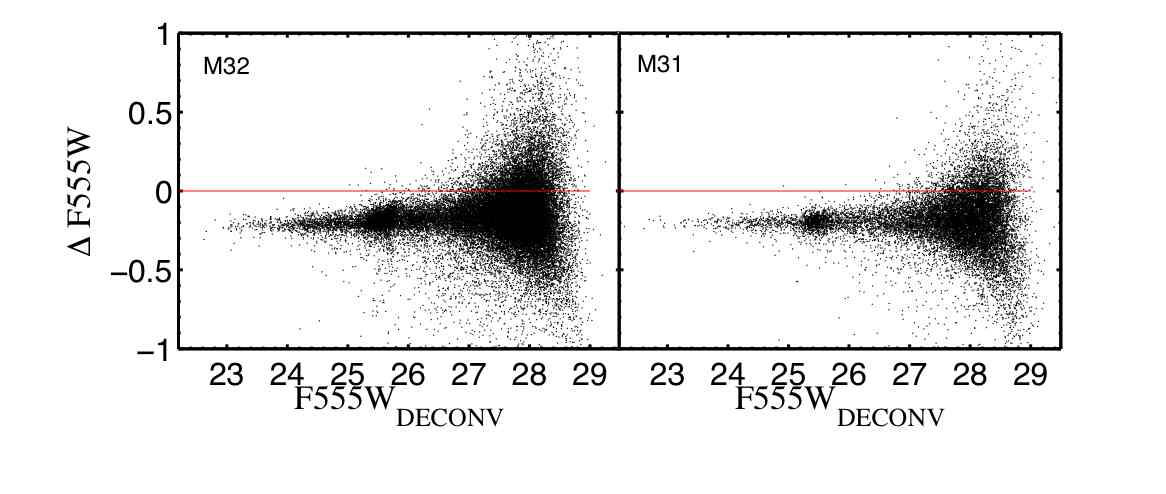}} \subfigure
{\includegraphics[width=88mm,clip]{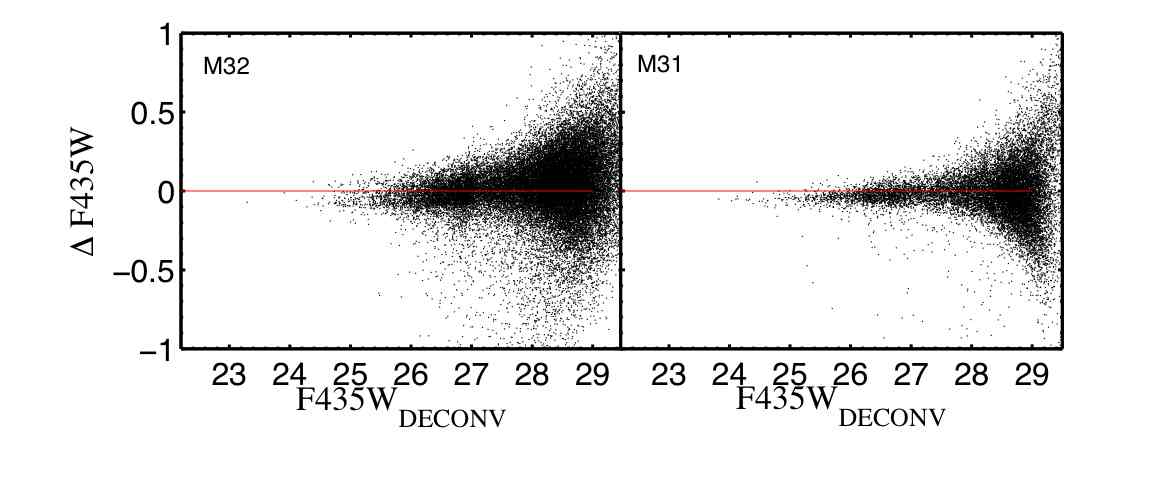}}
\caption{Difference between stars found using deconvolved images and
  the ones found by DAOPHOT II as a function of the deconvolved
  apparent magnitudes. On the two top panels we show the differences
  in the $F555W$ filter (left-hand panel: F1, right-hand panel: F2)
  and on the two bottom panels we show the same but in the $F435W$
  filter (left-hand panel: F1, right-hand panel: F2).  Here
  $\mathrm{\Delta mag= mag_{DECONVOLVED} - mag_{DAOPHOT}}$.  There is
  a constant shift between the deconvolved and DAOPHOT magnitudes that
  we attribute to a difference in the PSFs, considering deconvolution
  to be a more reliable method. Apart from the shift, there is however
  a reasonable good agreement between the two photometric methods for
  magnitudes $\la 27.5$. The photometric results have little to do
  with each other for fainter magnitudes (see text for more details).
  Magnitudes are calibrated onto the VEGAmag photometric system.}
\label{fig:daovsdeconv_V}
\end{figure}

Finally, we have also compared the list of stars obtained from both
methods by cross-correlating them using DAOMATCH and DAOMASTER. The
number of matched stars between deconvolved and DAOPHOT photometry was
$\sim 32,000$ for field F1 and $\sim 17,000$ for field F2.
Figure~\ref{fig:daovsdeconv_V} shows the differences between the
apparent magnitudes in both photometric results as a function of the
deconvolved magnitudes.  We find that there is an almost constant
shift between the magnitudes of the matched stars, especially in the
F555W filter, such that the deconvolved photometry produces
systematically brighter magnitudes. To investigate the cause of this
trend, we compare the PSFs used to deconvolve the images with the ones
obtained using DAOPHOT and we found that the DAOPHOT PSFs have small
wings or none at all.  Due to the severe crowding in our fields, we
could only generate reliable PSFs with very small radii, using DAOPHOT,
which are essentially devoid of wings\footnote{PSFs constructed with
  larger radii produced larger Chi values, given the large effects of
  neighboring stars on determination of the wings.}. We therefore
believe that the PSFs constructed to deconvolve the images are more
reliable and therefore so is the photometry based on the deconvolved
images\footnote{We have also compared the deconvolved and DAOPHOT
  photometry with that obtained using DOLPHOT \citep[a version of
  HSTphot][tailored to work on ACS images]{dolphin00} on our
  individual images. This comparison shows that, at the RC level,
  there is good agreement between DOLPHOT's magnitudes and those
  obtained with the deconvolved images.  This suggests again that the
  photometry obtained from deconvolved images is reliable. The
  photometry performed using DOLPHOT is explained in F10 where it was
  used to search for RR Lyrae variable stars.}.  Nevertheless, if we
correct for the shifts in magnitudes, Figure~\ref{fig:daovsdeconv_V}
shows that both photometric methods agree well for $F555W< 27 $ and
$F435W < 27$.  However at $F555W>27$ and $F435W>27$ the differences
become significant.  Looking at the locations of stars at these faint
magnitudes, it appears that most of the sources are probably products
of blends. Note that the detection limit in the deconvolved images is
determined below to be at $F555W \sim 28$.

\begin{figure*} \centering
\includegraphics[width=190mm]{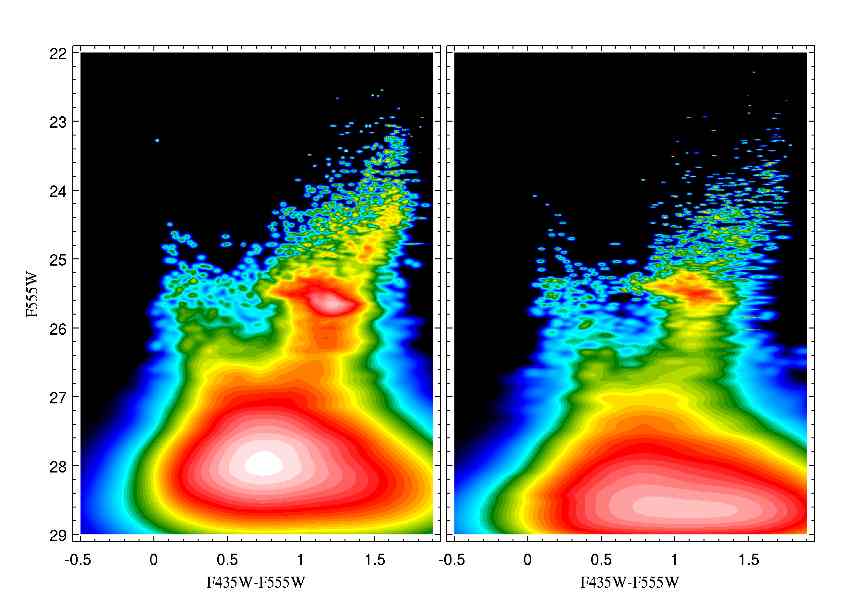}
\caption{($F435W   -  F555W$,  $F555W$)   CMDs  obtained   from  image
  deconvolution  of  F1  (left  panel)  and F2  (right  panel)  in  an
  error-based  Hess representation with  a logarithmic  stretch, where
  features  are  better  highlighted.   The error-based  Hess  diagram
  represents relative  density of stars weighted  by their photometric
  errors.}
\label{fig:hessdeconvolved}
\end{figure*}

Due to all the above reasons, we are convinced that deconvolution,
which has not been previously used to derive stellar photometry, gives
remarkably better results than the standard photometric packages for
these extremely crowded HST images.  We therefore use the CMDs and the
list of stars from the deconvolved images for further analysis.
Figure~\ref{fig:hessdeconvolved} shows the F1 (left panel) and F2
(right panel) deconvolved error-based Hess diagrams with a logarithmic
stretch, where the features are better highlighted.  The error-based
Hess diagrams represent relative density of stars weighted by their
photometric errors as follows.  Each star is represented by an
elliptical gaussian with color and magnitude widths as a function of
its color and magnitude photometric errors given by the ASTs (see next
subsection). The color-magnitude image containing all these elliptical
gaussians has been split into 600$\times$600 bins.  Note that the two
CMDs show a surprisingly similar morphology.  We return to this point
in Section~\ref{sec:m31}.

\subsection{Completeness tests and error analysis}\label{ast}
When analyzing the data and before any detailed interpretation, it is
necessary to have a good understanding of the completeness of the CMD
as a function of both magnitude and color.  The major source of
incompleteness here is crowding, which is the most important
limitation to the analysis of rich stellar fields.  The well-known
method of artificial stars \citep{stetson_harris88} is the best way to
quantify its effects.  Since the crowding effects are different for
the F1 and the F2 fields, due to the differences in stellar density
(Figure~\ref{fig:fields}), such ASTs need to be carried out in each
field separately to reach statistically significant results for both.
The generation of artificial stars was done following the
  prescriptions introduced by \citet{gallart_etal96a} and using the
IAC-STAR code \citep{aparicio_gallart04}.  This code generates
synthetic CMDs by means of interpolation in the metallicity and age
grid of a library of stellar evolutionary tracks. This interpolation
results in a smooth distribution of stars following a given star
formation rate, initial mass function and chemical enrichment law.  In
order to apply the AST method it is important that the magnitudes and
colors assigned to the artificial stars are realistic, covering a
range as wide as the populated one by the real stars to fully sample
the observed colors and magnitudes. It is important that the
  injected stars have realistic colors to test for color effects. We
generated a synthetic CMD of 500,000 artificial stars adopting a
constant star formation rate with ages from 0 to 14 Gyr, and
metallicities $0.0001 < Z < 0.04$ uniformly distributed at all ages.
We have chosen a limiting magnitude for the synthetic stars of about
two magnitudes fainter than the fainter stars observed in our CMD, to
explore the possibility of recovering a very faint, unresolvable
artificial star as if it were much brighter.  The synthetic CMDs given
by IAC-STAR are expressed in a photometric magnitude system different
than the ACS/HRC photometric system. In particular we have chosen the
bolometric correction library from \citet{origlia_leitherer00}, which
has magnitudes in the HST WFPC2 system.  We therefore have transformed
those magnitudes into the ACS/HRC photometric system using the
transformation given by \citet{sirianni_etal05}.  Since the artificial
stars need to be injected into the images, the synthetic CMD needs to
be transformed to instrumental magnitudes.  Hence we have converted
the absolute magnitudes given by the synthetic CMDs into apparent
magnitudes using a distance modulus of $\mu_0= 24.43$
\citep{ajhar_etal96}, a reddening of $E(B-V) = 0.08$
\citep{burstein_heiles82}, and an extinction of $A_{F555W} = 0.25$
\citep{sirianni_etal05}.  We have then applied the photometric
corrections in reverse order to obtain the artificial stars onto the
raw magnitude system.  The artificial stars are placed into the images
with random pixel locations using the PHOTONS routine in XVISTA.  The
number of artificial stars injected per experiment should not be
larger than about 5\% of the real stars found in the images \citep[see
e.g.][]{Grillmair_etal96a, Fuentes-Carrera_etal08} to avoid a
significant increase of the (already extreme) crowding in the images.
We have used 2,000 artificial stars per AST.  Once the artificial
stars are placed into the images, photometry was performed using
deconvolved images as before, since it is the deconvolved photometry
that we use for the CMD analysis.  The reduction of the original and
artificial images must be carried out identically for the comparison
to be valid. A second requirement for a valid AST is that the
reductions should be performed without knowing which stars in the
synthetic frame are added and which are real. For each AST we obtain a
deconvolved photometry list as output file.  Then, we append the file
containing the injected positions and magnitudes of the artificial
stars to the output file from the original photometry of real stars.
Next we match this appended file with the star list of the artificial
image.  This provides us with a list of recovered stars, i.e.,
injected stars that are paired with stars from the artificial stars
subset. The matching is done with DAOMATCH and DAOMASTER using the two
star lists as two lists to be matched. Stars within 1 pixel of radius
distance are considered to be recovered.  This process has been
repeated as many times as necessary in \emph{each} image in order to
achieve a very large number of artificial stars analyzed and to
recover a number of stars similar to the one of real stars in the
image. In this way, we can statistically sample the whole CMD diagram,
both in magnitude and color indices. In total $\sim 5\times 10^5$
stars have been used to perform $\sim$ 250 ASTs on \emph{each image}.
The procedure is applied to both filters of each field, i.e. F1 and
F2.

\begin{figure*} \centering %\vspace{-3cm}
\includegraphics[width=87mm,clip]{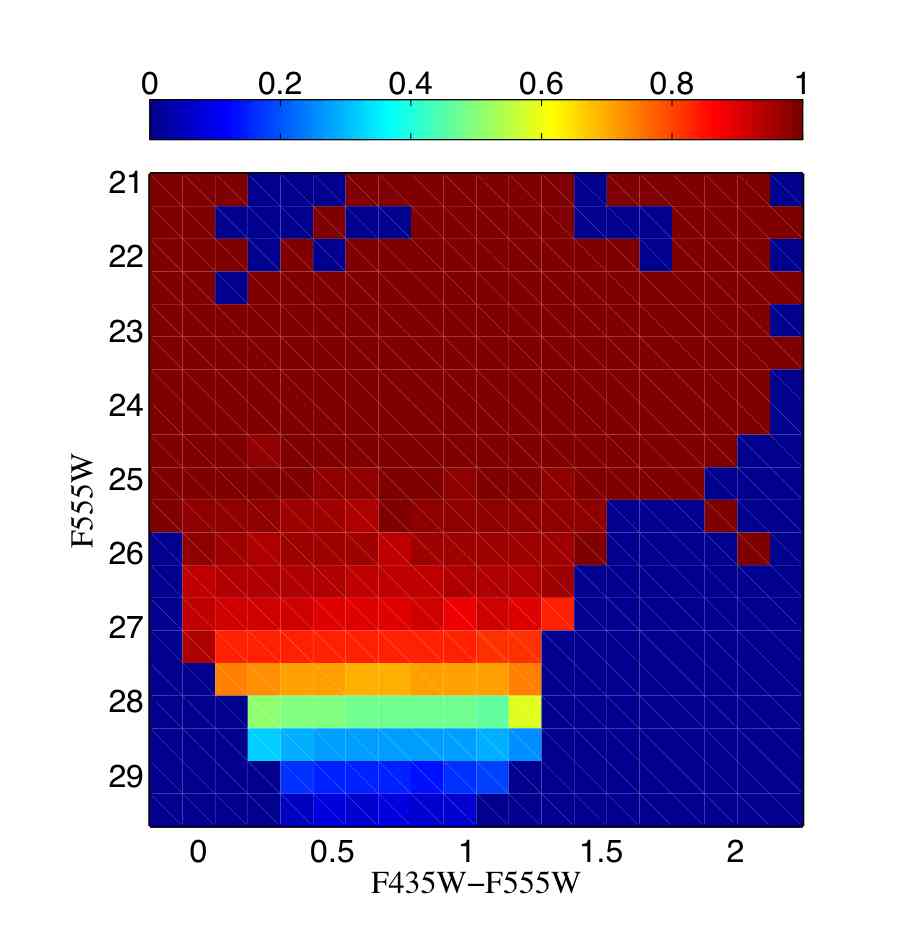}
\hspace{0.001cm} \includegraphics[width=87mm,
clip]{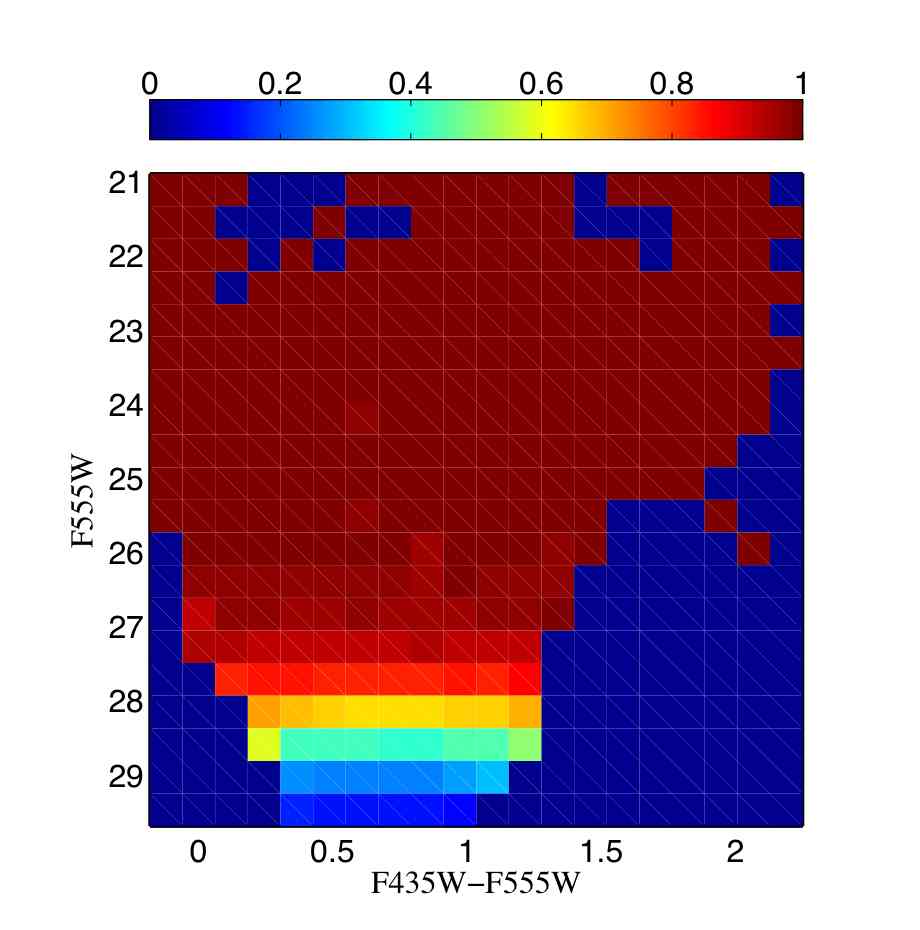}
\caption{Completeness fractions for fields F1 (\emph{left-hand panel})
  and F2  (\emph{right-hand panel}) as a function  of apparent $F555W$
  magnitudes  and  $(F435W  -  F555W)$  colors.   The  values  of  the
  completeness  fraction are  determined for  a given  bin  $i$,$j$ of
  magnitude $i$ and color $j$,  indicated by the color bar.  For field
  F1 the  50\% completeness level  is at $F555W  \sim 28$.  In  the F2
  field the  50\% completeness level is  at $F555W \sim  28.5$, half a
  magnitude deeper, indicating a less-crowded field than F1. }
\label{fig:grid}
\end{figure*}

A comparison between the number of injected and recovered artificial
stars provides information about the crowding effects on the
photometry and gives us the completeness level of our data.  We define
the completeness factor as the fraction of recovered stars in a given
color-magnitude bin as follows:
\begin{equation} \Delta_{i,j}=\frac{N^{rc}_{i,j}}{N^{in}_{i,j}}
\end{equation} where $N^{rc}_{i,j}$ is the number of
stars whose recovered color indexes and magnitudes lie in the $i,j$
color magnitude bin, and $N^{in}_{i,j}$ is the number of stars whose
injected color indexes and magnitudes lie in the $i,j$ color magnitude
bin \citep{gallart_etal96a}.  The data were divided for this
calculation into 20 bins in both magnitude and color.
Figure~\ref{fig:grid} shows the completeness fractions obtained for
both the F1 and F2 fields.  The color bar in these figures indicates
the values of the completeness fraction level, increasing from 0 to 1,
where 1 represents 100\% completeness.  The magnitudes in these
figures are expressed in the VEGAmag system. The 50\% completeness
level for F1 is located at $F555W \sim$ 28 ($F435W \sim$ 28.5)
independent of color. Our completeness factor falls rapidly to zero
below $F555W \sim 28$, suggesting a limiting magnitude $F555W_{lim}
\sim 28$.  The 50\% completeness level for F2 is half a magnitude
deeper, at $F555W \sim$ 28.5, indicating again that this field is
slightly less crowded.

\begin{figure*} \centering \subfigure
  {\includegraphics[width=80mm,clip]{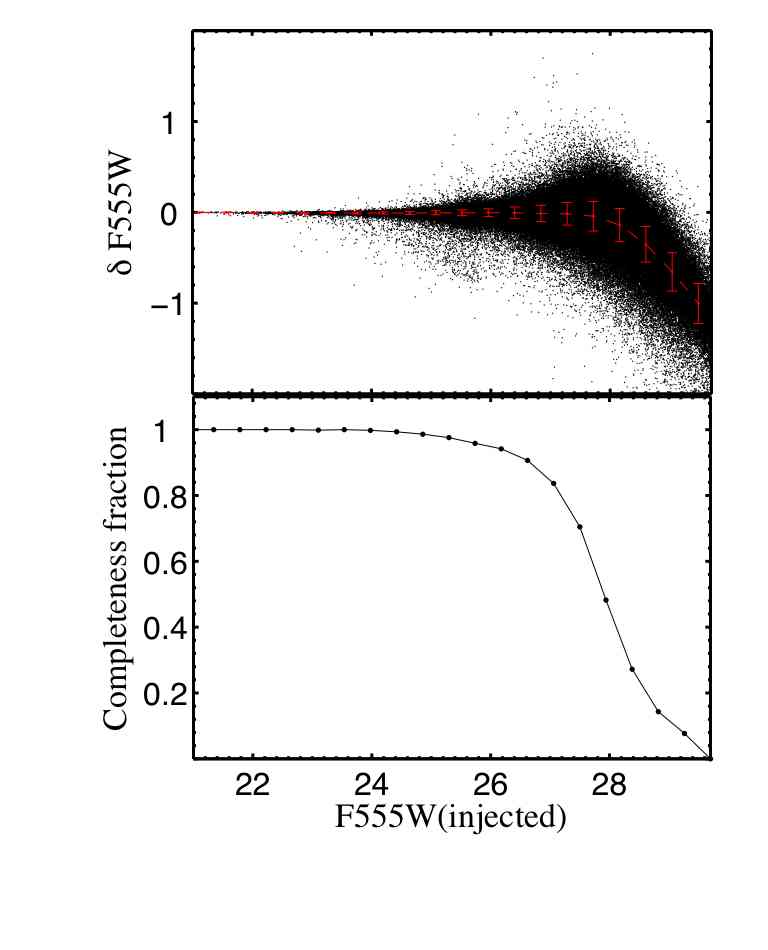}}
  \subfigure
  {\includegraphics[width=80mm,clip]{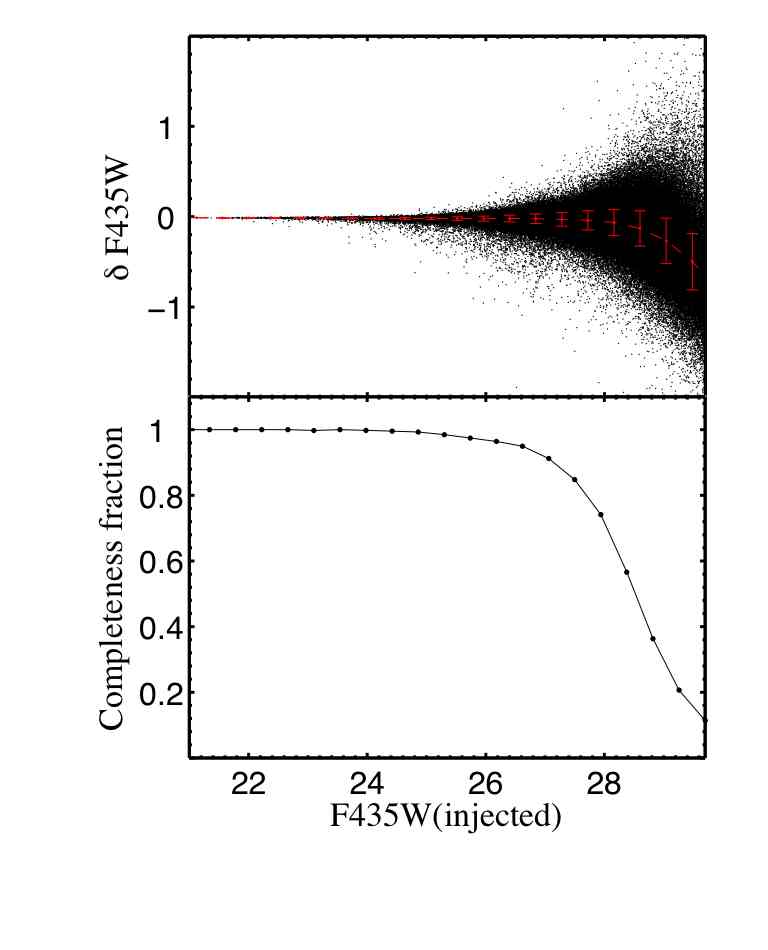}}
  \caption{ Results  from ASTs for field F1.  (Recovered $-$ Injected)
    $F555W$  (\emph{left-hand  panel})  and $F435W$  (\emph{right-hand
      panel}) as  a function of  the injected magnitudes are  shown in
    the top  panels. The  1-d completeness fraction  as a  function of
    magnitude  is  given in  the  bottom  panels  and it  follows  the
    theoretical  color-magnitude  locus of  the  M32  stars. The  mean
    photometric errors as a  function of magnitude are indicated.  The
    median  of these  differences and  the error  of such  medians are
    illustrated as red curves  and red error bars respectively.  Stars
    whose   recovered   magnitudes   differ  significantly   ($|\delta
    \rm{mag}|  \ge 1$)  from their  input magnitudes  are  products of
    blends and we do not consider them as recovered.}
\label{fig:diffmaganderrors}
\end{figure*}

\begin{figure} \centering
\includegraphics[width=90mm,clip]{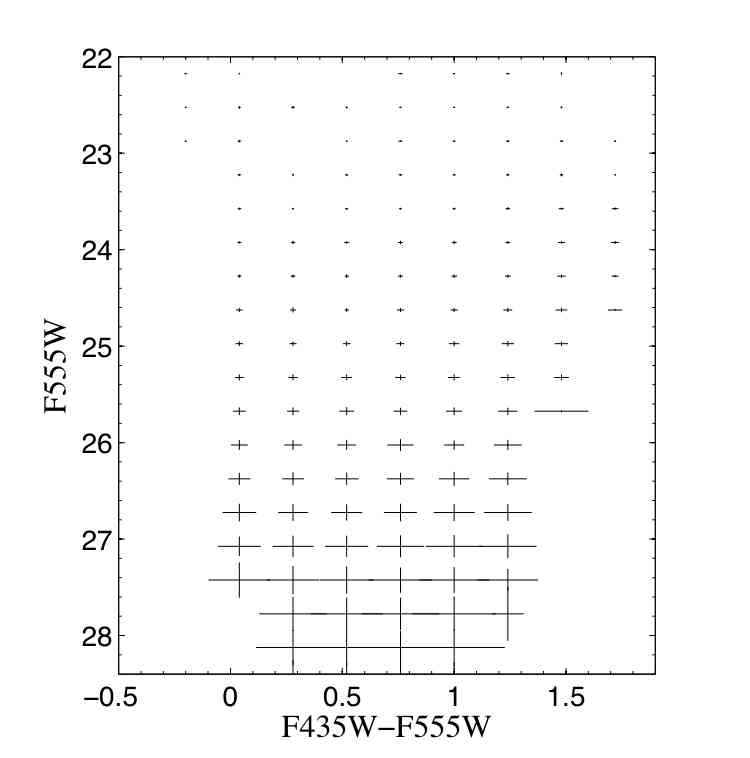}
\caption{Photometric color  and magnitude errors  for the CMD  of M32,
  estimated from the ASTs.  The errors are negligible
  ($< 0.05$ mag) for stars brighter than  $F555W = 26$ and start to become
  significant ($> 0.1$ mag) for stars fainter than $F555W \sim 27$.}
\label{fig:photerrorscmd}
\end{figure}

The  actual photometric  errors are  quantified as  the  difference in
magnitude  between   the  recovered  and  injected   stars,  shown  in
Figure~\ref{fig:diffmaganderrors}   for  F1.    In  this   figure  the
difference $\mathrm{\delta mag = mag_{recovered} - mag_{injected}}$ is
plotted  as a  function of  the  injected magnitudes  in both  $F555W$
(left-hand  panels)  and  $F435W$  (right-hand panels)  filters.   The
completeness fractions as  a function of magnitude are  plotted in the
bottom  panels  of the  same  figure.   The  median of  the  magnitude
differences  is  negligible  for  magnitudes  having  more  than  70\%
completeness.  We see  from these figures that the  magnitudes of some
recovered  stars  differ significantly  from  their input  magnitudes,
having  significantly  negative   $\mathrm{\delta  mag}  <  -1$  (i.e.
recovered brighter).  The appearance  of aberrant stars coincides with
the dramatic drop in the completeness at $F555W \sim 28$.  These might
be stars  that were  only detected because  they are located  at noise
spikes. We  therefore do not  consider stars with magnitudes  that lie
below our 50\% completeness level as recovered.
Photometric errors are defined for a given bin of magnitude $i$ and
color $j$ as the errors in the median of the magnitude differences in
that bin.  We show in Figure~\ref{fig:photerrorscmd} the amplitude of
photometric errors throughout the theoretical CMD locus of the M32
stars.  Our photometry shows excellent accuracy for magnitudes $F555W
< 26.5$ and the mean errors in magnitude and color are $\sim 0.18$ mag
and $\sim 0.23$ mag, respectively at the 50\% completeness level.

We  want  to emphasize  that  a  same  AST analysis  was  additionally
reproduced        using       DAOPHOT.        As        shown       in
Figure~\ref{fig:daovsdeconv_ast}, we  have obtained larger photometric
errors  with DAOPHOT  photometry and  much more  scatter  in recovered
magnitudes.   Thus, the ASTs  were also  used to  prove quantitatively
that the deconvolved photometry is superior than DAOPHOT.

\section{M32 field decontamination}

\subsection{M31 contamination}  M31 is clearly the  dominant source of
contamination in our M32 field  (F1). M32 lies at a projected distance
of 5.3 kpc south of the center of M31 and therefore contamination from
its disk and bulge is  significant. Moreover, at the position in which
the F1  field was  taken, the  closest possible to  the center  of M32
without being overwhelmed by crowding  effects, one third of the light
is coming  from M31. To correct  for this statistically,  we have also
obtained images  of a comparison  field located at the  same isophotal
level within M31. As explained in previous sections, those images were
processed in the same way as the F1 images.

Since  both fields are  located at  the same  isophotal level  in M31,
correcting  for M31  stars  would require  that  for each  F2 star  we
subtract  the closest  one in  color and  magnitude from  the  F1 star
list. However, as  we have already addressed in  previous section, the
crowding  differs between  the  two fields.   Image  crowding is  more
important  for  F1  than for  F2  so  there  are different  levels  of
completeness  in the  images that  should be  taken into  account.  We
therefore cannot  simply subtract the  F2 stars from  the F1 CMD  in a
``one-by-one''  way. Instead,  the  number of  stars  removed from  F1
depends on  the completeness  fractions computed at  F1 and  F2 fields
\citep[e.g.,][]{gallart_etal96a}.  Assuming that the population of M31
stars  is statistically  the same  in both  the F1  and F2  fields, we
corrected the F1 CMD as follows: For each F2 star of magnitude $i$ and
color $j$  an ellipse was defined in  the F1 CMD centered  at $i$, $j$
and with semi  axes $err_i$ and $err_j$.  The  semi-axes correspond to
the magnitude  and color photometric errors, estimated  from the ASTs,
affecting  the given  region  of the  F1  CMD.  We  also consider  the
photometric errors in magnitude  and color affecting the corresponding
region of the F2 CMD when generating the semi-axes of the ellipse. For
a given  ellipse in the contaminated  F1 CMD, a number  $F_n$ of stars
was subtracted randomly from the F1 list, where
\begin{equation} F_n= \frac{\Lambda_{i,j}^{\mathrm{F1}}}{\Lambda_{i,j}^{\mathrm{F2}}},
\end{equation}  and $\Lambda_{i,j}^{\mathrm{F1}}$ and $\Lambda_{i,j}^{\mathrm{F2}}$ are
the corresponding completeness factors for F1 and F2 in the $i$,$j$
bin of the CMD \citep{gallart_etal96a}, as previously calculated.  The
closest integer to $F_n$ is chosen as the number of stars to be
subtracted.  If the number of stars in a given ellipse is smaller than
the number of stars expected to be removed, we enlarge the semi-axes of
the ellipse by a factor of two. The remaining stars are deleted from
this larger ellipse in order of proximity to the color and magnitude
of the F2 field star considered. This happens most often in regions of
the CMD where the density of stars is low.
The advantage of this process \citep{gallart_etal96a} is that
the region in magnitude and color from where the stars are removed
varies along the CMD, i.e. the interval in magnitude and colors from
which stars are removed is changing size depending on the photometric
errors. Thus, brighter stars -- the ones with smaller
photometric errors -- are subtracted from a small region around the
field star in the CMD, whereas the fainter stars, and therefore the
ones with larger photometric errors, are allowed to be removed from a
larger region of the F1 CMD with a size controlled by the error.

\begin{figure*}\hspace{-0.001mm}
\includegraphics[width=92mm,clip]{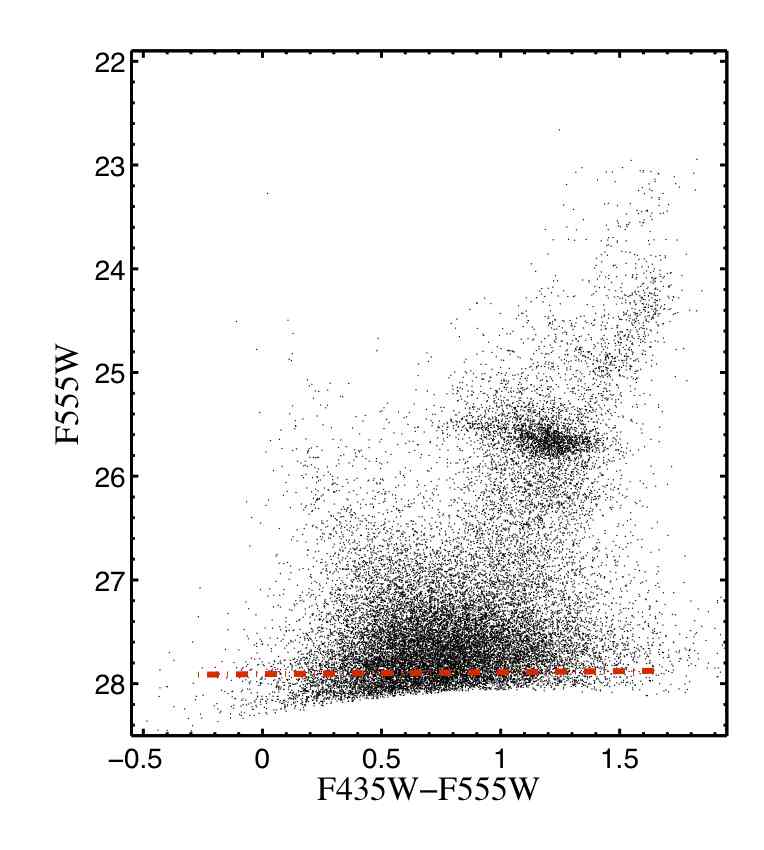}
%\hspace{-0.001mm}
\includegraphics[width=92mm,clip]{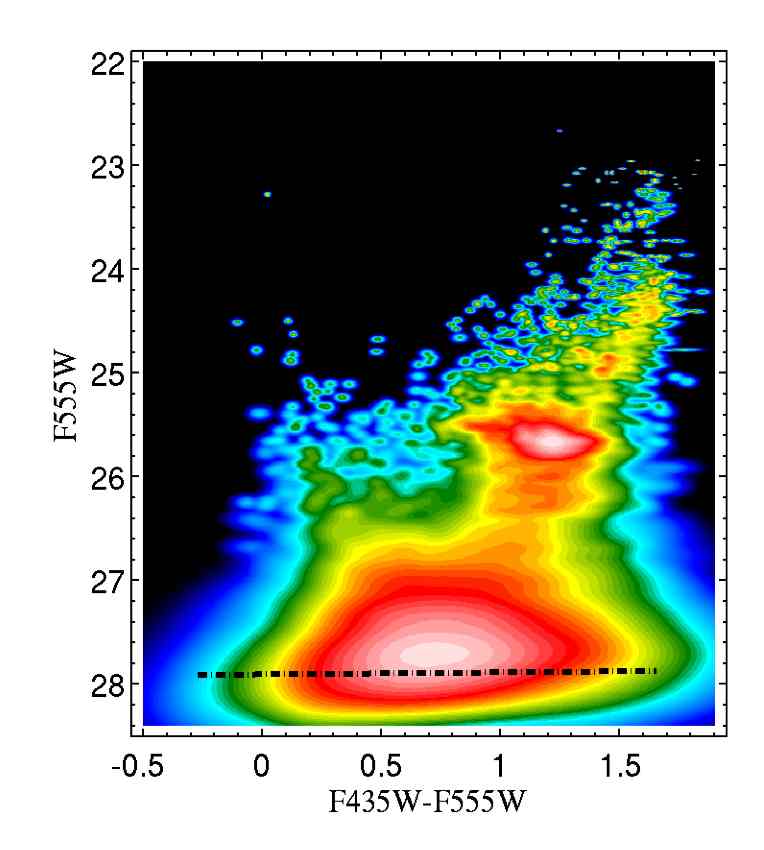}
\caption{Left panel: CMD of M32, corrected for contamination by the
  M31 background stars. The different crowding levels between the F1
  images and the F2 images were taken into account to statistically
  perform the decontamination (see text for more description). Right
  panel: Error-based Hess representation of the decontaminated CMD of
  M32, where the features are better highlighted. To construct this
  Hess diagram, the stars were replaced by elliptical gaussians with
  color and magnitude photometric errors as color and magnitude
  gaussian widths, respectively. The CMD was then divided into
  600$\times$600 bins.  In both panels, the dashed line indicates the
  50\% completeness level of our data.  Apparent magnitudes are
  calibrated onto the VEGAmag ACS/HRC system.}
\label{fig:decontaminatedm32}
\end{figure*}

We   show    in   Figure~\ref{fig:decontaminatedm32}   the    F1   CMD
decontaminated from  the M31  background stars. The  50\% completeness
level is also shown in this Figure.
The number of M32 stars remaining after this decontamination process
is $\sim26000$ of the 58143 originally detected in F1
(Table~\ref{table:detections}).

\subsection{Galactic foreground  stars} Our  field F1 is  quite small,
and  the  contamination  by  the  Galactic foreground  stars  is  very
small. We  have however still estimated  it from star  count data. The
Besan\c con group  model of stellar population synthesis  of the Galaxy
\citep{robin_etal03} predicts the amount of stars in a given magnitude
interval  for a  given  location. This  model  predicts 14  foreground
Galactic  stars  in  the  range  of  $V  =  22-30$  in  our  $\sim  29
\,\mathrm{arcsec^{2}}$  ACS/HRC field.  This  is of  course negligible
compared  with  the  thousands  of  stars  we  have  obtained  in  our
photometric  catalog, and  therefore we  do not  consider foreground
stars further in our analysis.
\\
\section{The Stellar Populations of M32}
\begin{figure*}
\centering
\includegraphics[width=160mm]{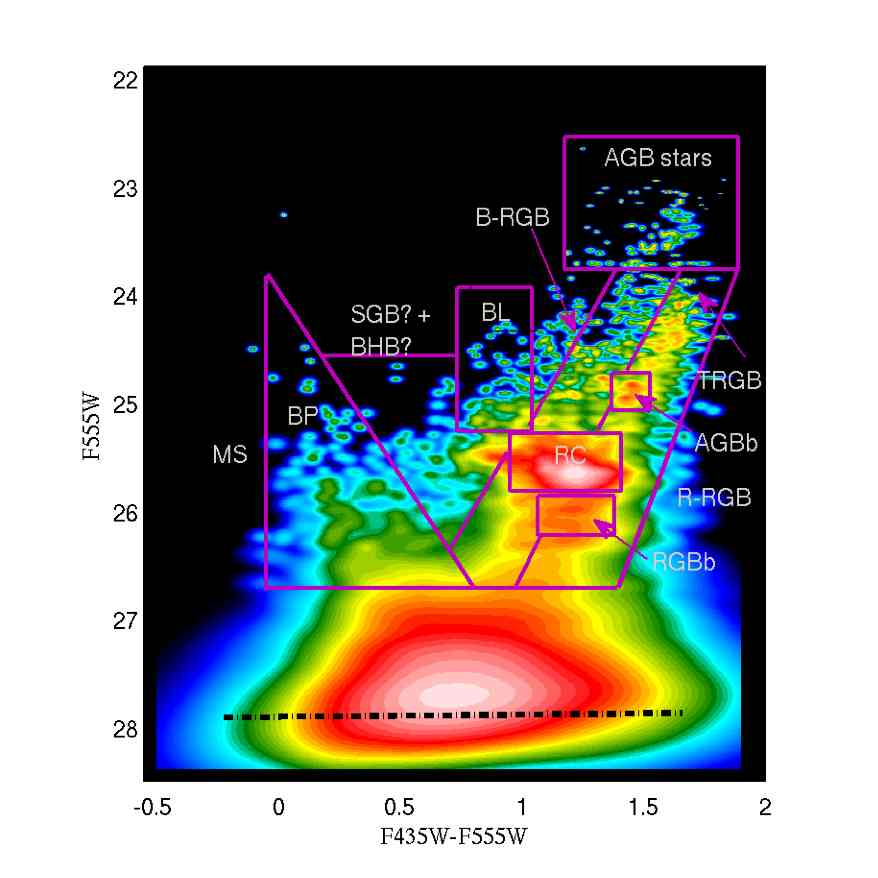}
\caption{Error-based Hess diagram for M32, corrected for contamination
  by  the M31 background  stars. The  boxes indicate  various features
  that represent different stellar populations. MS: Main Sequence; BP:
  Blue Plume;  SGB: Subgiant branch; BHB: Blue  Horizontal Branch; BL:
  Blue  Loop; RC:  Red  Clump;  RGBb: Red  Giant  Branch bump;  R-RGB:
  Red-Red Giant Branch; B-RGB: Blue-Red Giant Branch; TRGB: Tip of the
  Red Giant Branch; AGB: Asymptotic Giant Branch; and AGBb: Asymptotic
  Giant  Branch  bump. The  dotted-dashed  line  indicates  the 50\%
  completeness level  of our data. Magnitudes are  calibrated onto the
  VEGAmag system.}
\label{fig:hessm32_features}
\end{figure*}

The CMD we have obtained is deep enough to allow a comprehensive study
of the stellar populations of M32, and we can gain some insights into
them by comparing our CMD with theoretical isochrones.  In what
follows we present the most detailed resolved photometric study of M32
carried out so far.  Figure~\ref{fig:hessm32_features} shows the CMD
of M32 with boxes highlighting features that reveal the different
stellar populations.  We see evidence for an intermediate-age and old
population -- ages between 2 and 10 Gyr -- due to the presence of a
strong red clump, an extended and bright asymptotic giant branch, a
prominent red giant branch, and the red giant branch bump as well as
the asymptotic giant branch bump.  We also see possibly evidence of a
young population -- ages younger than 2 Gyr -- due to the presence of
stars occupying the blue plume producing an extended main sequence,
blue loop stars, and a possible bright subgiant branch.  Evidence of
an ancient -- older than 10 Gyr -- population could be represented by
blue horizontal branch stars together with a well-populated red giant
branch.  Note that a well-defined blue horizontal branch in our CMD is
not present, but we have observed RR Lyrae stars in F1 (F10) and there
are stars in the region where we would expect to see blue horizontal
branch stars.  We emphasize here that all these features are above the
80\% completeness level where the photometric errors are very small.
Hence what we see in the CMD at this level represents the intrinsic
properties of the stars.  Note in Figure~\ref{fig:hessm32_features}
that, although we have the highest resolution and deepest data for M32
yet obtained, the severe crowding of our fields makes it impossible to
reach the oldest main-sequence turn-offs.  This unfortunately will
remain a challenge beyond existing telescopes and even near-future
space-telescopes such as JWST.

In the following, we discuss the different stellar populations of M32
in detail.  We assume a distance modulus (DM) of $\mu_0= 24.53$ (this
paper, below), Galactic reddening $E(B-V)=0.08$
\citep{burstein_heiles82}, and extinction $A_{F555W}=0.25$
\citep{sirianni_etal05}. Note that we have only considered Galactic
reddening, on the assumption that M32 is dust-free. We note that no
dust features are seen in the surface photometry residual maps of the
M32 center \citep{Lauer_etal98} or envelope \citep{choi_etal02}.  F10
tested for internal extinction in F1 and F2 due to M31 and/or M32 by
using the intrinsic properties of the RR Lyrae variables in the
fields.  They found that the mean reddening values obtained in both
fields agree within the errors and are further consistent with the
assumed Galactic reddening.  We also tested for differential
extinction over the HRC field by comparing the RGB colors of CMDs
constructed at different quadrants of the image and found no
difference. We use theoretical isochrones from the Padova library
\citep{marigo_etal08, girardi_etal02, girardi_etal08} as they are
available in the HST ACS/HRC photometric system at different ages and
metallicities.  The metallicity in these models assumes
$\mathrm{[M/H]=\log}(Z/Z_{\odot})$, $Z_{\odot}=0.019$, and
$[\alpha/\mathrm{Fe}]=0$. Although our photometry could be transformed
onto traditional magnitude systems (e.g., Johnson--Cousins) for
comparison to other theoretical isochrones, such transformations
always introduce significant systematic errors \citep{sirianni_etal05}
and we prefer to stay as much as possible in the original photometric
system of the data.

\subsection{Intermediate-age (2 $\le$ Age $\le$ 8 Gyr) and old (8 $<$
  Age $\le$ 10 Gyr) populations}\label{sec:iapop}

\subsubsection{The Red Clump (RC)}

The most prominent feature in our CMD is the RC formed mostly by the
reddest low-mass stars burning helium in their cores\footnote {The
  bottom part of the blue loop (core-He-burning intermediate-mass
  stars, see below) also contributes to this RC.}.  A strong RC, as we
see here, indicates the presence of intermediate-age/old metal-rich
stars.  Models of core-helium burning stars predict that the RC
luminosity depends on both age and metallicity \citep{cole98,
  girardi_etal98}. For a given metallicity, old stars form a fainter
RC than young stars whereas for a given age, lower metallicity stars
form a brighter RC. For a population of known age and metallicity, the
RC is at a fairly constant color and luminosity, hence these stars can
serve as good standard candles to derive distances both within our own
Galaxy and to nearby galaxies and globular clusters
\citep{percival_salaris03}.  We make use of this fact to derive the
distance to M32 in Section~\ref{sec:distance}.

We now attempt to estimate a mean age and metallicity of M32, based on
the constraints that the presence of this feature impose. Constraints
on age and metallicity of these populations can be obtained from the
analysis of the locus and width of the red giant branch (RGB) together
with the position of the RC \citep{ferguson_johnson01,
  rejkuba_etal05}.

\begin{deluxetable}{@{}lccccc}
  \tabletypesize{\scriptsize}  
  \tablecaption{M32 RC, RGB$\mathrm{b}$ and AGB$\mathrm{b}$ magnitudes
    and colors.\label{table:magandcolors}}              
  \tablewidth{0pt}
  \tablecolumns{5}            
  \tablehead{ &\colhead{$F555W$}&\colhead{$(F435W-F555W)$}&
    \colhead{V\tablenotemark{a}}&\colhead{ $\Delta V
      \rm{(bump-RC)}$\tablenotemark{b}}} 
  \startdata
  RC  &$25.66\pm0.08$&$1.20\pm0.08$&$25.33\pm0.09$& \nodata\\ 
  RGBb&$26.21\pm0.10$&$1.15\pm0.18$&$25.89\pm0.10$&$0.56\pm0.13$\\
  AGBb&$24.87\pm0.09$&$1.44\pm0.03$&$24.52\pm0.09$&$-0.81\pm0.13$
  \enddata
  \tablecomments{Errors are $1\sigma$ deviations.}
  \tablenotetext{a}{Transformed onto the Johnson-Cousins system
    following \citet{sirianni_etal05}.} 
  \tablenotetext{b}{Difference between the RGBb (AGBb) and the RC V mean
  magnitudes, used to estimate a mean age and metallicity of M32.}
\end{deluxetable}

\begin{figure}                  
  \centering                  
  \subfigure{\includegraphics[width=70mm,clip]{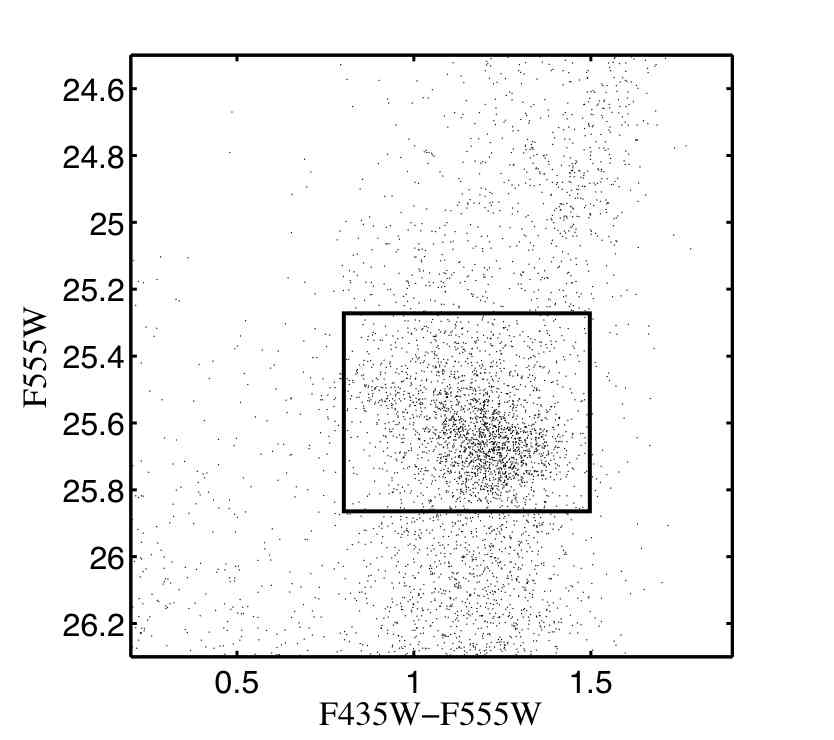}} 
  \subfigure{\includegraphics[width=40mm, clip]{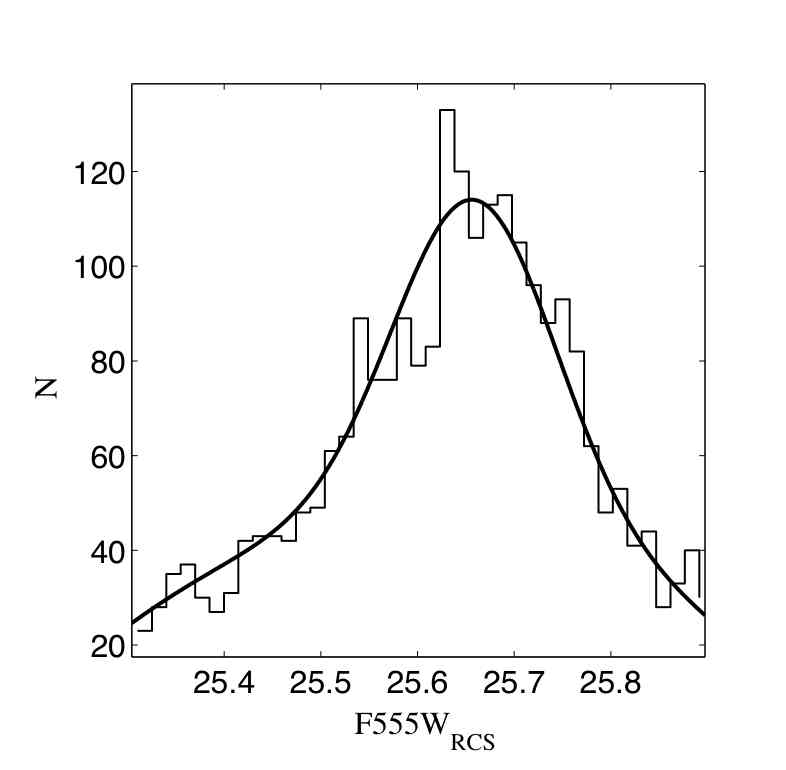}}
  \subfigure{\includegraphics[width=40mm, clip]{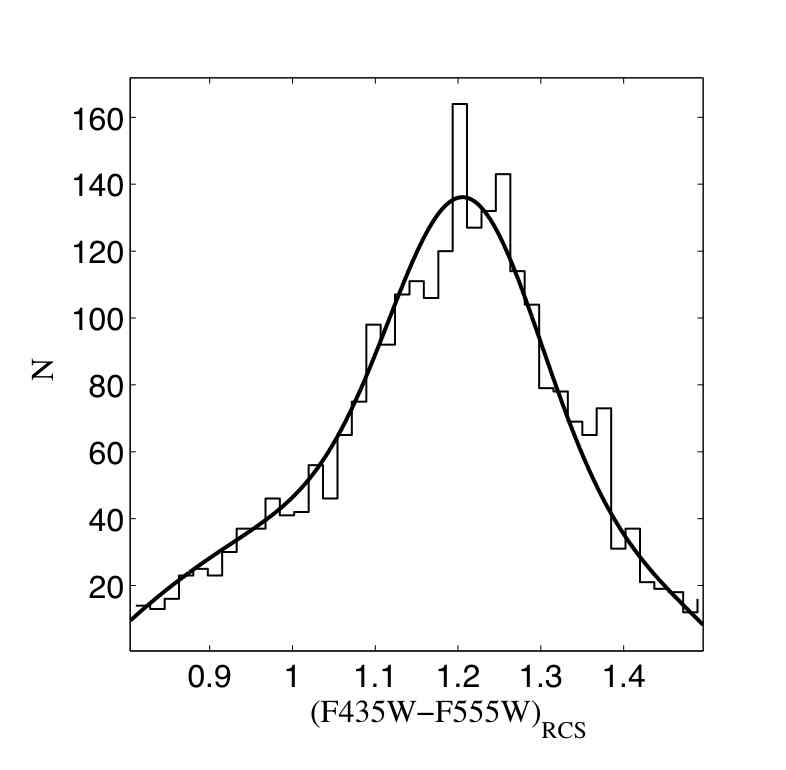}}
  \caption{\emph{Top panel}: Close-up of the M32 CMD corrected for M31
    background contamination in the region of the RC.  Note the
    complex morphology of the RC.  We observe a bluer and brighter end
    of the RC, at $F435W - F555W \sim 0.75$, which indicates the
    presence of lower metallicity stars in combination with young
    stars at the bottom of the blue loop. The red end of the RC, at
    $F435W - F555W \sim 1.30$, is fainter, which indicates the
    presence of both older ages and higher metallicities stars. Stars
    inside the box $25.30 \le F555W \le 25.90$ and $0.80 \le (F435W -
    F555W)\le 1.50$ are selected to determine the mean apparent
    magnitude of the RC in our field.  In total 2525 stars lie inside
    this box.  \emph{Bottom panel}: A non-linear least-squares fit of
    the function $N(F555W)$, Equation~\ref{eq:rc}
    \citep{paczynski_stanek98}, to the histogram of stars in the clump
    region is shown in the left-hand panel.  $N(F555W)$ is a Gaussian
    representing the RC population plus a second term representing the
    RG stars that contaminate the RC selection.  The coefficients
    found for this fit with 95\% confidence bounds are $F555W_{m}$ =
    25.66 and $\sigma = 0.082$.  In the right-hand panel we show the
    fit to the histogram of the color distribution of RC stars. The
    same formalism was used in this case and we obtain $(F435W -
    F555W)_m= 1.20$ and $\sigma = 0.088$.}
\label{fig:rcs}
\end{figure}

We begin by measuring the mean luminosity and color of the RC. We
consider a rectangle in the CM plane with $25.30 < F555W < 25.90$ and
$0.80< (F435W - F555W) < 1.50$ defined in such a way that all the RC
stars remain inside it (Figure~\ref{fig:rcs}).  We find 2525 stars in
this rectangle; note that some of these will be RGB stars.  Note in
Figure~\ref{fig:rcs} the complex morphology of the RC in the CMD of
M32.  As stated above, the RC's morphology depends not only on the
metallicity but also on the age of the stellar system.  In this case
we observe a bluer and brighter end of the RC, at $F435W - F555W \sim
0.75$, which indicates the presence of lower metallicity stars and
young intermediate-mass stars at the bottom of the blue loop (see next
subsection).  The red end of the RC, at $F435W - F555W \sim 1.30$, is
fainter, indicating the presence of both older ages and higher
metallicities stars.  We will quantitatively study the complex
morphology of the RC when deriving the SFH of M32 in a follow-up
paper.  We make a histogram of the luminosity of these stars and
measure the peak magnitude of the RC by fitting the $F555W$-band
luminosity function with the following function from
\citet{paczynski_stanek98},
\begin{multline}\label{eq:rc} 
\displaystyle N(F555W)= a + b(F555W - F555W_{m}) + \\ 
c(F555W - F555W_{m})^2 + \frac{N_{RC}}{d}e^{-\left[\frac{(F555W -
      F555W_{m})^2}{2\sigma^2_{RC}}\right]},
\end{multline} 
a Gaussian representing the RC population plus terms representing the
contamination due to RGB stars.  Here, $F555W_m$ is the mean apparent
magnitude, $\sigma_{RC}$ is the width of the RC and $N_{RC}$ is the
number of stars selected to determine the apparent magnitude of the
RC.  A non-linear least-squares fit of this function to the histogram
of stars in the clump region provides $F555W_m$ and $\sigma_{RC}$.  We
show in the top panel of Figure~\ref{fig:rcs} the stars inside the box
that were selected to determine the mean apparent magnitude of the RC
in our field. In the lower left-hand panel of the same figure we show
the histogram of the magnitudes of those RC stars together with the
fit.  We can see that the data are well fit by this function.
The mean color of the RC is calculated using the same formalism (lower
right-hand panel of Figure~\ref{fig:rcs}) and its value, as well as
the mean magnitude of the RC, is listed in
Table~\ref{table:magandcolors}.  Models by \citet{marigo_etal08} and
\citet{girardi_etal08} suggest, for the observed mean magnitude and
color of the RC, a mean age of M32 of 8--10 Gyr for a metallicity
$Z=0.012$ ([Fe/H]$\sim -0.2$ dex, see Section~\ref{sec:rgb})
consistent with the bulk of the stellar population being old.

There are uncertainties in this estimate: The mean magnitude and color
of the RC  are also well fit  with a mean stellar population  of 5 Gyr
and  solar  metallicity   $Z=0.019$,  reflecting  the  age-metallicity
degeneracy that occurs in the  RGB. Moreover, the uncertainties in the
distance modulus  obtained (see below) could  modify these parameters,
possibly changing the mean age  by $\pm2$ Gyr. 

\subsubsection{The RGB bump (RGBb) and the AGB bump (AGBb)}

We detect for the first time in M32 a feature in the RGB that we
identify as the RGB bump (RGBb) located at $F555W\sim$ 26.10 (see
Figure~\ref{fig:hessm32_features}).  This feature is the consequence
of the following process that occurs at the beginning of the RGB
phase.  During evolution along the RGB, the H-burning shell moves away
from the core of the star, which is increasing in luminosity at almost
constant temperature.  As the shell moves out to regions of 'fresh'
hydrogen it encounters the chemical discontinuity left behind by the
maximum penetration of the convective envelope.  When this happens,
the rate at which the star climbs the RGB drops for a short period and
even reverses for a while, until the shell adapts to the new
environment, and then the star again increases its luminosity, burning
in a regime of constant H content. As a result, the star crosses the
same small portion of the RGB evolutionary path ---the same luminosity
interval--- three times, producing a peak in the luminosity function
\citep{iben68, sweigart_gross78, king_etal85,
renzini_fusipecci88}. The time that a low mass star spends during the
RGBb phase is a considerable fraction ($\sim$ 20\%) of the total RGB
lifetime, and the RGBb can be easily observed in an intermediate-age
or old stellar system provided there is a large number of stars.
Stellar evolution models predict that the brightness of this feature
depends on both the age and metallicity of the system. For a given
metallicity old stars have a lower RGBb luminosity than young stars
\citep[][hereafter AS99]{alves_sarajedini99}. 

\begin{figure}                                
  \subfigure{\includegraphics[width=42mm,clip]{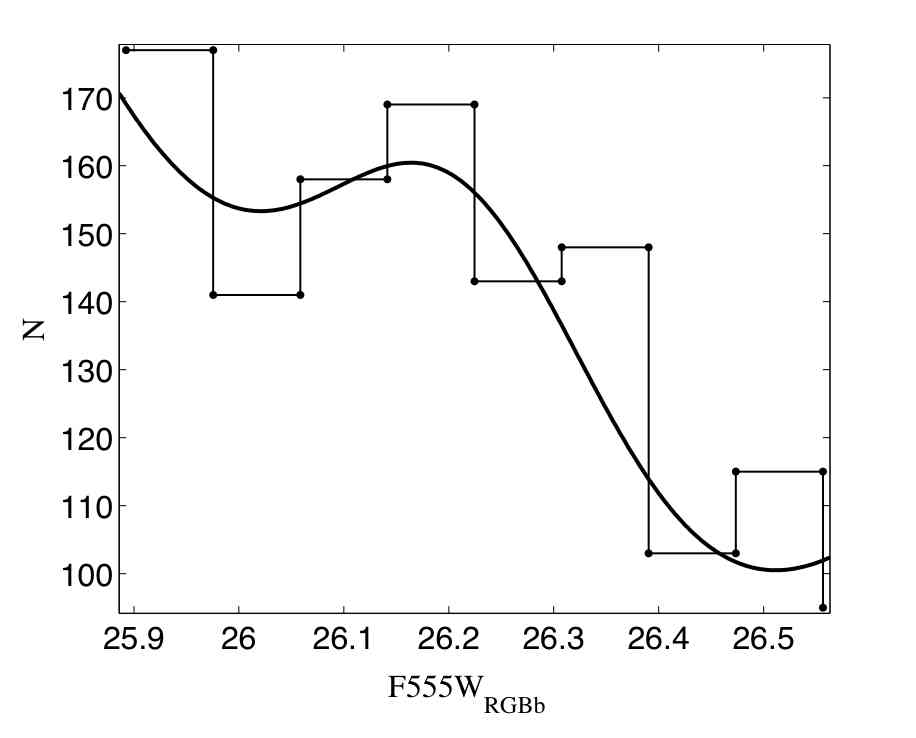}} 
  \subfigure{\includegraphics[width=42mm, clip]{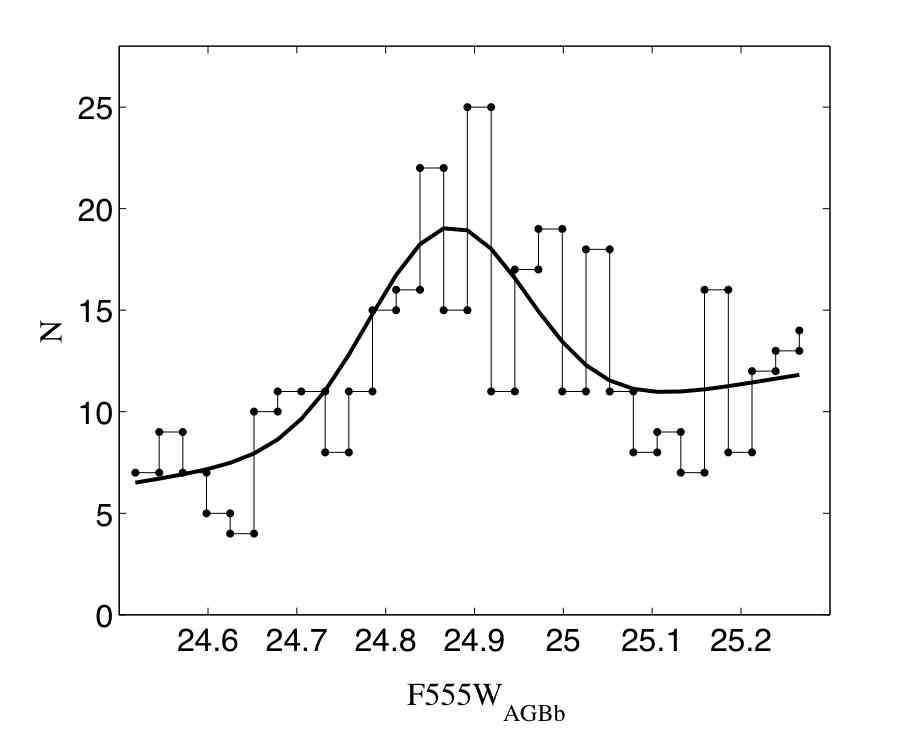}}
  \caption{\emph{Left-hand panel}: A Gaussian plus a quadratic fit to
    the $F555W$-band luminosity around the RGB bump. A number of 1370
    stars lie in this region. The peak is at $F555W$ = 26.21 and
    $\sigma = 0.10$.  The bin size of the luminosity function is 0.1
    mag, since this is the approximate mean value of the photometric
    errors in that region of the CMD. From this fit, we find that the
    RGB bump is a $5$--$8\sigma$ detection. \emph{Right-hand panel}: A
    Gaussian plus a straight line fit to the histogram of stars in the
    AGB bump region. The peak is at $F555W$ = 24.87 and $\sigma =
    0.09$. A number of 372 stars lie in the region. Note here that the
    bin size of the histogram is 0.03 mag, given that the photometric
    errors are negligible at these color and magnitude levels. The ABG
    bump is a $4$--$7\sigma$ detection.}
\label{fig:bumps}
\end{figure}

We calculate the mean magnitude of the RGBb by fitting a Gaussian plus
a quadratic function to the $F555W$-band luminosity function around
the bump (Figure~\ref{fig:bumps}). The bin size of the distribution is
0.1 mag, since this is the approximate mean photometric error in this
region of the CMD. The mean color of the RGBb was obtained by fitting
a Gaussian to the color distribution around the bump; the inferred
value and its standard deviation are listed in
Table~\ref{table:magandcolors}. We can estimate the significance of
this bump by comparing the number of RGB background stars (i.e. the
quadratic component of the fit) with the number of stars that are in
the bump (the Gaussian component of the fit). There are $1127\pm34$
RGB background stars, where the error is simply Poisson error. The
number of remaining stars that are in the RGB bump is
$219\pm51$. Thus, the RGB bump is a $5$--$8\sigma$ detection.

We also detect for the first time in M32 the asymptotic giant branch
bump (AGBb), a bump in the Hess diagram at the beginning of the AGB
phase.  Here a process analogous to the one at the beginning of the
RGB phase occurs related to the formation of the He-burning shell
\citep{caputo_etal89, fusipecci_etal90,
  sarajedini_forrester95,Gallart_98,ferraro_etal99}.  As a consequence
a feature similar to the RGBb is seen. In this case, the He-exhausted
core contracts rapidly and heats up, and the H-rich envelope expands
(the luminosity increases) and cools so effectively that the H-burning
shell extinguishes, causing the base of the convective envelope to
penetrate inward again.  Eventually, the expansion of the envelope is
stopped by its own cooling and it re-contracts.  Therefore the
luminosity decreases and the matter at the base of the convective
envelope heats up. When the H-burning shell reignites, the envelope
convection moves outward in radius ahead of the H-burning shell, and
the luminosity increases again.  As a consequence of this process the
star will cross the same luminosity interval three times, and an
increase of star counts in this luminosity interval is therefore
predicted.  There is, like for the RGBb, a good probability of
observing this feature in intermediate-age or old systems, provided
that they are well-populated enough to detect such a fluctuation. AS99
and \citet{cassisi_etal01} have shown that the luminosity of the AGBb
is a function of the mean age and metallicity of the stellar
populations generating this feature.  We identify the clump of stars
seen at $F555W \sim 24.80$ with the AGBb.  To obtain the mean
luminosity value of this bump, we fit a Gaussian plus a straight line
to the $F555W$-band luminosity function around it
(Figure~\ref{fig:bumps}). Note that, in this case, the bin size of the
distribution (0.03 mag) is smaller than the one used for the
luminosity function around the RGBb. This is consistent with the fact
that the photometric errors are negligible in that region of the
CMD. A Gaussian was fit to the color distribution around the bump to
obtain its mean color. The mean luminosity and color inferred from
these fits are listed in Table~\ref{table:magandcolors}. We can
estimate the significance of the AGB bump in the same way as we did
for the RGB bump: we find that the number of stars in the background
is $265\pm 16$ and the number of stars in the bump is $85\pm25$. This
implies that the AGB bump is detected at the $4$--$7\sigma$ level.

AS99 presented values of the magnitude difference between the RGBb and
RC ([$\Delta V$(RGBb-RC)]) as well as between the AGBb and the RC
([$\Delta V$(AGBb-RC)]) for four Galactic globular clusters: M5, NGC
1261, NGC 2808, and 47 Tuc.  They showed that predictions of
theoretical models are in good agreement with the ages, metallicities
and [$\Delta V$(AGBb/RGBb-RC)] values of these clusters.  These
predictions can be used as a consistency check on our age and
metallicity determinations.  It is important to note here that the
measurement of [$\Delta V$(AGBb/RGBb-RC)] is distance independent.

Using the predictions presented in AS99, the magnitude of the RC of
M32 in F1, located between the brighter AGBb and the fainter RGBb,
strongly indicates that populations older than 2.5 Gyr and
metallicities higher than about $-0.7$ dex dominate in our field (see
Table 2 and Figure 4 in AS99).  To obtain more quantitative
information, we use of the mean luminosities of the AGBb and RGBb
listed in Table~\ref{table:magandcolors}. We then transform these
magnitudes onto the Johnson-Cousins photometric system using
\citet{sirianni_etal05} calibrations, as the models by AS99 are given
on that photometric system, and we calculate the differences between
the AGBb and RGBb mean magnitudes and the mean magnitude of the RC.
These values are also indicated in Table~\ref{table:magandcolors}.
Given the AGBb-RC magnitude difference, Figure 6 of AS99 suggests that
M32 metallicity is likely to be higher than $\mathrm{[Fe/H]} \sim
-0.4\,\mathrm{dex}$ regardless of age. However since their models do
not extend to more metal-rich regimes, it is difficult to obtain a
tighter constraint.  Nevertheless we confirm the metal-rich nature of
the stellar population in M32.  On the other hand, if we assume such a
metal-rich population, the RGBb-RC magnitude difference suggests (see
Figure 6 of AS99) that the mean age of M32 is likely to be in between
5 and 10 Gyr, consistent with the value found above from the RC alone.

\subsubsection{The    RGB:    The    metallicity    distribution    of
  stars in M32}\label{sec:rgb}

The Red Giant Branch (RGB) in our CMD at $F555W \la 26.75$ and $0.75
\la (F435W - F555W) \la 1.50$ is the evolutionary phase where stars
are burning H in a shell while He has not yet been ignited in their
cores. The lifetime of a star on the RGB is a decreasing function of
its initial mass, hence the probability of observing low mass stars in
this phase is very high.  The color and morphology of the RGB for a
stellar system strongly depend on its metallicity.  On the other hand,
for a given metallicity, the RGB moves to the red as a stellar
population ages. Although the age dependence of the RGB color is not
as strong as its metallicity dependence, an age--metallicity
degeneracy certainly exists on the RGB.

\begin{figure}
  \centering 
  \includegraphics[width=95mm]{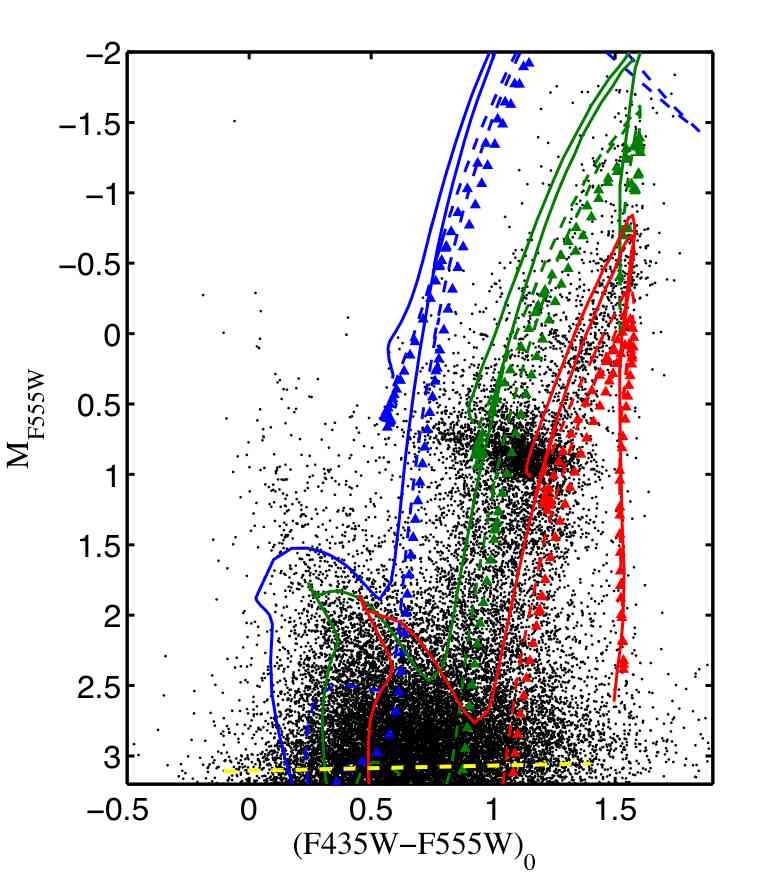}
  \caption{CMD of M32 corrected for contamination by M31 background
    stars, reddening \citep[$E(B-V)=0.08$,][]{burstein_heiles82},
    extinction \citep[$A_{F555W}=0.25$,][]{sirianni_etal05} and
    distance $\mu_0 = 24.53$, which was obtained using the RCS method
    (see text for description).  Isochrones from \citet{marigo_etal08}
    and \citet{girardi_etal08} are superimposed with metallicities of
    $Z$ = 0.0008,0.008, and 0.030 for ages of 2 (solid), 5 (dashed)
    and 9 (triangles) Gyr.  Note a good match with the features that
    represent an intermediate-age population. Note also that the width
    of the RGB cannot be explained with a single metallicity and, even
    though a single age with a spread in metallicity could reproduce
    this, we cannot exclude some age spread.}
\label{fig:ia_isochronesm32}
\end{figure}

Figure~\ref{fig:hessm32_features} shows that the RGB has a rather wide
spread in color.  Given the small (almost negligible) photometric
errors at these magnitude and color levels, this cannot be explained
by a single-age and -metallicity population but rather by an intrinsic
large spread in the metallicity distribution.  We show in
Figure~\ref{fig:ia_isochronesm32} that a population with a single age
and a range of metallicities can adequately reproduce the width of the
RGB.  In spite of this, some age spread cannot be excluded. This is in
agreement with G96 who showed that the spread in color of the M32 CMD
is indicative of its metallicity range.

Figure~\ref{fig:ia_isochronesm32} shows isochrones superimposed on the
CMD of M32 corrected for reddening
\citep[$E(B-V)=0.08$,][]{burstein_heiles82}, extinction
\citep[$A_{F555W}=0.25$,][]{sirianni_etal05} and distance $\mu_0 =
24.53$ (this paper below), that represent populations of 2 (solid
lines), 5 (dashed lines) and 9 (triangles) Gyr with metallicities of
$Z =$ 0.0008 (bluest), 0.008, and 0.03 (reddest). We can see that
these isochrones cover the entire RGB and match the features we just
discussed. However it is clear that not all of them match our data
well.  For example, the most metal-poor isochrones are too blue
compared with our data, thus suggesting that very metal-poor stars are
unlikely to be present. On the contrary, metal-rich isochrones do a
better job in matching both the bright and faint end of the RGB. As
stated earlier, an age-metallicity degeneracy is present in this
region of the CMD, and therefore differences in ages cannot be
distinguished if we look solely at the RGB.

\begin{figure}
  \centering 
  \subfigure{\includegraphics[width=90mm,clip]{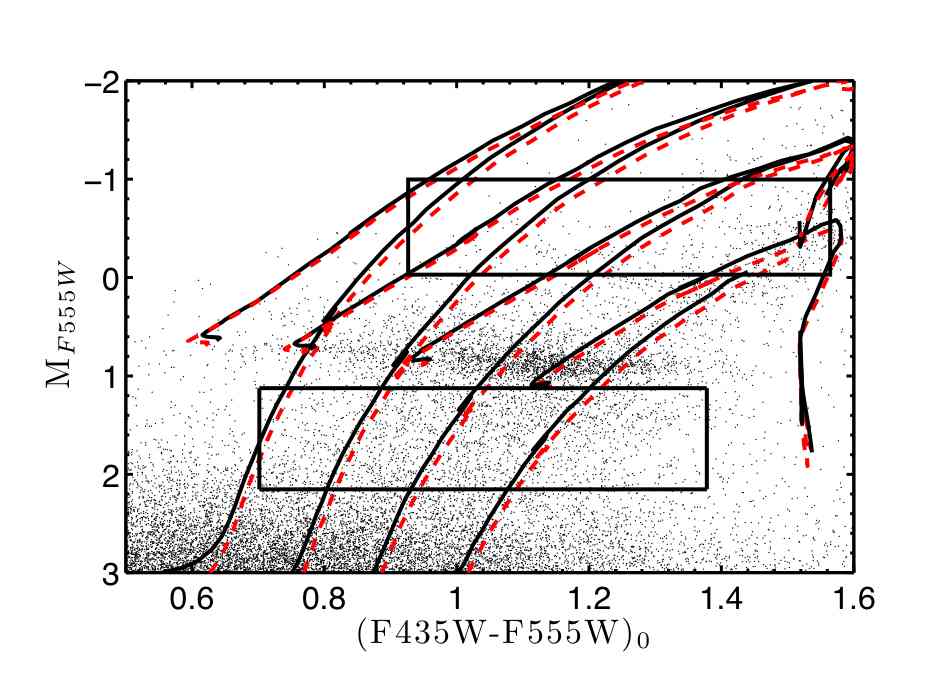}}
  \subfigure{\includegraphics[width=90mm,clip]{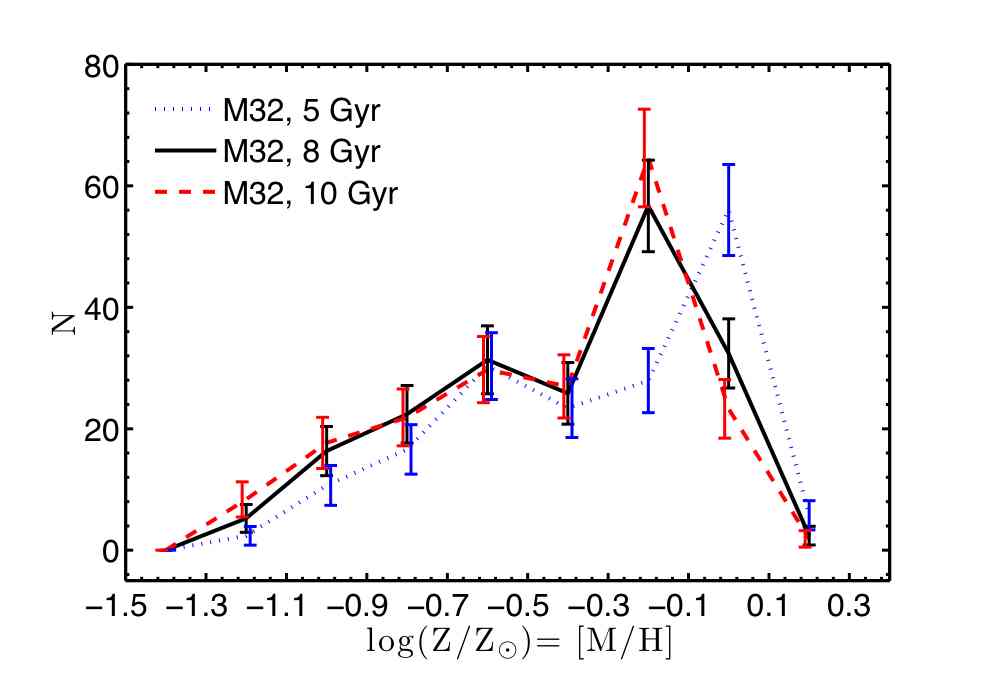}}
  \subfigure{\includegraphics[width=90mm,clip]{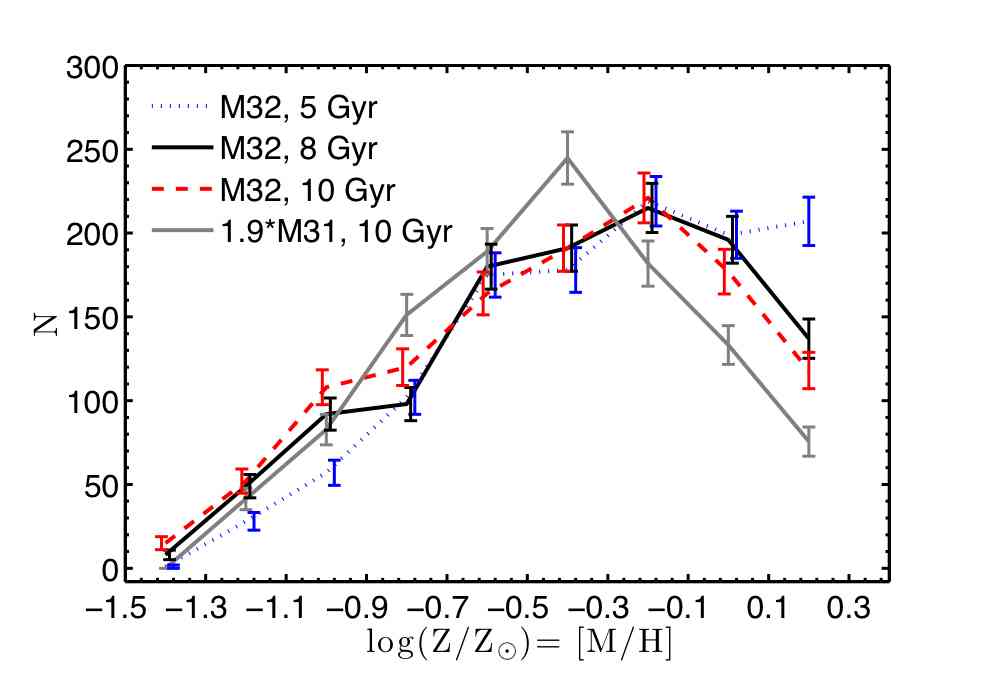}}
  \caption{\emph{Top panel} Close-up of the RGB region of the
    decontaminated CMD of M32 (the magnitudes are de-reddened and
    corrected for distance). Stars inside the two boxes were used to
    compute the two MDFs.  Black and red--dashed curves are 8 and 10
    Gyr-old \citet{girardi_etal08} and \citet{marigo_etal08}
    isochrones covering a wide range in metallicity, $Z=$ 0.0012,
    0.003, 0.008, 0.03.  We counted stars between the isochrones of
    different metallicities to compute the MDF.  Errors are simply
    Poisson errors. \emph{Middle panel} MDF of M32 derived using stars
    above the RC, defined as the number of stars per bin size in
    $\log(Z/Z_{\odot})=$[Fe/H].  The counting has been corrected for
    contamination by AGB stars in this region.  Blue, black and red
    lines represent 5, 8 and 10 Gyr-old MDFs, respectively.  The peak
    of the distribution is at $\mathrm{[Fe/H]} \sim0.0$ for 5 Gyr and
    is slightly more metal-poor, at $\mathrm{[Fe/H]} \sim -0.2$, for 8
    and 10 Gyr-old populations.  \emph{Bottom panel} Same as the
    middle panel but considering stars below the RC, which are
    affected by photometric errors but not by AGB stars.  Here the
    peak of the distribution is at $\mathrm{[Fe/H]}\sim -0.2$ for all
    ages.  The MDF of M31 is also computed for a 10 Gyr-old population
    and is illustrated by the gray solid line.  The M31 MDF has been
    normalized to the M32 MDF, and its peak is given at
    $\mathrm{[Fe/H] \sim -0.4}$, indicating that our background field
    contains more metal-poor stars. Note that the M31 metal abundance
    distribution looks very similar to that of M32.}
\label{fig:mdf}
\end{figure}

To obtain the metallicity distribution function (MDF) of M32, shown in
Figure~\ref{fig:mdf}, we have used isochrones from the model grid of
the Padova library \citep{girardi_etal02, marigo_etal08,
  girardi_etal08} for ages of 5, 8 and 10 Gyr and $\log(Z/Z_{\odot})=
\mathrm{[M/H]}$ from $-1.2$ to $0.3 \, \mathrm{dex}$ with a
metallicity step (bin size) of $\mathrm{[M/H]} = 0.2 \, \mathrm{dex}$.
Although we do not see the 5 Gyr MSTO\footnote{We would only see a 5
  Gyr MSTO for a very metal poor population which is unlikely to
  contribute significantly to the M32 population in our fields: see
  Figure~\ref{fig:ia_isochronesm32}.}, it is possible that M32
contains such a population due to the observation of bright AGB stars
that confirm the presence of an intermediate-age population
(Sec.~\ref{sec:agb}). For the metallicity distribution, we have first
considered only RGB stars located below the RC and above the 80\%
completeness level, to avoid contamination by AGB stars that ascend
from the RC.  We selected a box containing 1166 stars of absolute
magnitudes $1.2 < M_{F555W} < 1.8$ and dereddened colors $0.7 < (F435W
-F555W)_0 < 1.4$ to compute the MDF below the RC.  Selecting these
stars guarantees an unambiguous metallicity assignment but, even
though stars in this region of the CMD have small photometric errors,
they are not negligible. Figure~\ref{fig:photerrorscmd} shows that at
the level of apparent magnitudes $F555W \sim$ 26.5, which corresponds
to an absolute magnitude of $M_{F555W} \sim 1.7$, and color $\sim 1$
the photometric errors in colors in our selected box are in between
$\sim 0.05$ and $\sim 0.1$ mag. The width of the RGB at those
magnitudes is $\sim 0.7$ mag and thus the small photometric errors
cannot explain the observed spread in color. We also note that
variations in ages, from 2 to 9 Gyr, on the RGB can only account for a
$\sim 0.15$ mag variation in color (see
Figure~\ref{fig:ia_isochronesm32}). On the other hand, if we use stars
above the RC, for which the photometric errors are truly negligible,
implying that the color variation of this region is only due to an
intrinsic metallicity distribution in the populations\footnote{There
  is a spread in age based on the appearance of bright AGB stars, but,
  again, such a variation can account only for a modest spread in the
  RGB and not for the width that is observed (G96).}, we need to
correct for the contamination by AGB stars. We selected a box
containing 500 stars of absolute magnitudes $-0.9< M_{F555W} < 0.0$
and dereddened colors $0.9 < (F435W -F555W)_0 < 1.65$ for the MDF
above the RC.

We have derived a MDF for these two groups of stars.  The top panel of
Figure~\ref{fig:mdf} shows the selected stars (boxes) and
representative isochrones considered for the MDF calculation. Solid
and dashed curves are the 8 and 10 Gyr isochrones, respectively.  The
middle and bottom panels of Figure~\ref{fig:mdf} show the resulting
MDF of M32 above and below the RC, respectively, defined as the number
of stars per bin size in $\log(Z/Z_{\odot})$. We have counted RGB
stars in the CMD between fixed-age isochrones (either 5, 8 or 10 Gyr
old) covering the range of metallicities stated above. We have
attempted to correct the MDF above the RC for AGB contamination by
taking into account the theoretical ratio of AGB to RGB stars at
different ages and metallicities. We calculated the ratio of AGB to
RGB stars occupying the isochrone section considered, i.e. for a given
age and metallicity, using the ``int-IMF''\footnote{The ``int-IMF'' is
  the integral of the IMF under consideration (as selected in the
  form, in number of stars, and normalized to a total mass of 1
  $M_{\odot}$) from 0 up to the current initial stellar mass: see
  \url{http://stev.oapd.inaf.it/}.} column of Padova's isochrones.  We
subtracted the corresponding number of AGB stars from the RGB counting
between the isochrones considered to derive the MDF.  Overall, we
obtain a rather smooth distribution with many more metal-rich stars
than metal-poor ones. The general peak of this distribution is given
at $\mathrm{[Fe/H]} \sim -0.2 \, \mathrm{dex}$ with the exception of
the 5 Gyr MDF above the RC that has its peak at $\mathrm{[Fe/H]} \sim
0.0 \, \mathrm{dex}$. This peak agrees with previous results
\citep{rose85, rose94, grillmair_etal96b, Trager_etal00b,
  coelho_etal09}. We note that the peak in the MDF above the RC is
more pronounced compared to that in the MDF below the RC. We believe
that this arises from the cooler giant stars going back towards the
blue at the TRGB due to the strong opacity present in the $V$ band.

Note that there are very few stars with metallicities $\mathrm{[Fe/H]}
< -1.2$, which implies that the enrichment process largely avoided the
metal poor stage \citep{Worthey_etal96}. Moreover, it is possible that
some of the B-RGB is due to stars with ages $< 2$ Gyr (see
Figure~\ref{fig:isochronesm32} below), hence the number of metal-poor
stars is likely to be even smaller. Note also that a few biases
  should have been taken into account when deriving the MDF, such as
  e.g. the different RGB lifetimes at different metallicities
  \citep{Rood72} or the rate at which stars leave the main sequence
  \citep{Renzini_buzzoni86}. These biases, however, mostly affect the
  metal-poor tail of the metallicity distribution \citep[see][]
  {Zoccali_etal03}, which implies that our (very weak) metal-poor tail
  is actually an upper limit. The shape and peak of the MDF agrees
very well with the photometric MDF of G96.  We even obtain the same
peak value, which however disagrees with the synthetic population
results by \citet{coelho_etal09}, who found a significant amount of
metal-poor stars at a location that samples the positions of both G96
and our field. \citet{coelho_etal09} claim that they do not understand
this difference, although they say that one might not expect an MDF
derived from photometry to match an MDF derived from spectroscopic
data \citep[although see][]{Trager_etal00b}.  Nevertheless, the
difference is significant.

\subsubsection{The tip of the red giant branch (TRGB)}

Another important CMD feature that confirms the results obtained so
far is the tip of the RGB (TRGB). This corresponds to the He-burning
ignition through the He flash marking the end of the RGB phase.  For
very metal rich systems, such as the globular clusters NGC 6553 and
NGC 6528, the TRGB in the $B$, $V$ and $R$ filters is fainter than in
metal-poor systems due to the strong molecular opacities of TiO bands,
which become very deep in the cool giants. This effect is so strong in
the $V$ band that the TRGB in a $V$--$(B-V)$ color--magnitude diagram
is accompanied by a vertical sequence of stars extending to fainter
magnitudes and almost merging with the hotter giants
\citep{Ortolani_etal92}. The location of what we identify as the TRGB
in the CMD of M32, at the apparent magnitude of $F555W \sim 24$, (see
Figure~\ref{fig:hessm32_features}) corresponds to the theoretical
predictions of a system as old as $\sim$8.5 Gyr with [Fe/H] $\sim
-0.2$. This again confirms previous (e.g., G96) and our own results in
this section.

\subsubsection{AGB stars}
\label{sec:agb}

\begin{figure}
  \centering 
  \includegraphics[width=95mm]{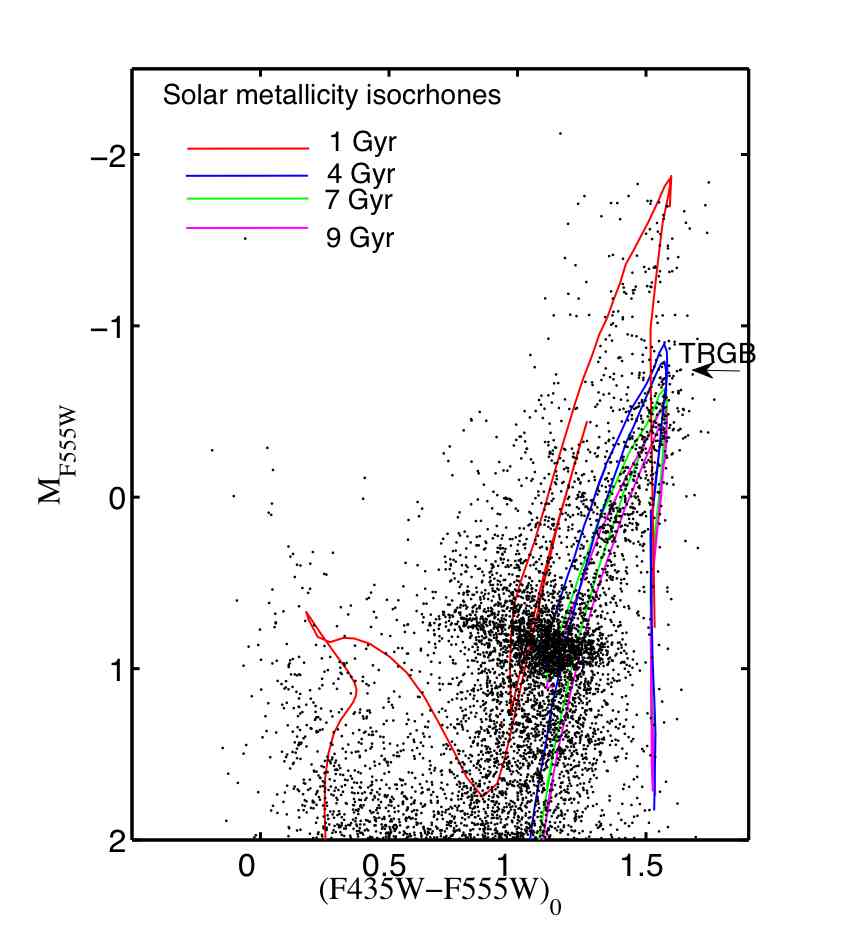}
  \caption{CMD of M32 corrected for contamination by M31 background
    stars, reddening \citep[$E(B-V)=0.08$,][]{burstein_heiles82},
    extinction \citep[$A_{F555W}=0.25$,][]{sirianni_etal05} and
    distance $\mu_0 = 24.53$, obtained using the RCS method (see text
    for description).  Isochrones from \citet{marigo_etal08} and
    \citet{girardi_etal08} are superimposed with ages of 1 (red), 4
    (blue), 7 (green) and 9 (magenta) Gyr at solar metallicity. Bright
    AGB stars, located above the TRGB, are clearly present at ages
    younger than 7 Gyr.}
\label{fig:agb_isochronesm32}
\end{figure}

Finally, the bright extension of AGB stars seen above the first-ascent
red giant branch (or TRGB) is a signature of an intermediate-age
population \citep[see e.g.][]{freedman92, gallart_etal05}.  We thus
identify the bright stars seen in Figure~\ref{fig:hessm32_features} at
apparent magnitudes $F555W < 24$ and colors $1.0 \la (F435W - F555W)
\la 1.6$ as intermediate-age AGB stars. We find $\sim 130$ of these
stars in the CMD of M32. To test whether blends of fainter stars could
mimic bright AGBs we make use of the AST results. From AST, we
considered all the recovered bright AGB stars and we looked at their
injected counterpart stars, i.e. we looked at where these bright stars
came from theoretically. We obtained that $\sim$ 97\% correspond to
the injected bright AGB. We are confident hence that \emph{the
  detected AGB stars in our photometry are not artifacts of crowding}
since they cannot be generated by blends of fainter stars. The
existence of these stars confirms the results of previous studies
\citep[e.g.,][]{freedman92, elston_silva92, davidge_jensen07}, and
strongly supports the presence of an intermediate-age population in
M32. Figure~\ref{fig:agb_isochronesm32} shows a decontaminated CMD of
M32, corrected for distance and extinction with Padova solar
metallicity isochrones superimposed, for ages of 1, 4, 7 and 9
Gyr. Clearly, bright AGB stars above the TRGB can be present between 1
and 7 Gyr. 
%For the 9 Gyr isochrone, we cannot see bright AGB stars above the
%TRGB.
These bright AGB stars thus represent the evolved population resulting
from star formation that occurred less than 7 Gyr ago in M32.

To summarize, the RC suggests a mean age of 8--10 Gyr for a
metallicity of $\mathrm{[Fe/H]} = -0.2\,\mathrm{dex}$, consistent
with the position of the TRGB. The RC, RGBb and AGBb suggest a
dominant population of stars with a mean age of 5--10 Gyr and a mean
metallicity of $\mathrm{[Fe/H]} \gtrsim -0.4\,\mathrm{dex}$. In
addition, stars younger than 7 Gyr are present as bright extended-AGB
stars.

\subsection{Young  populations?:  Ages  $<$ 2  Gyr}  \label{sec:young}

\begin{figure}
  \centering
  \includegraphics[width=95mm]{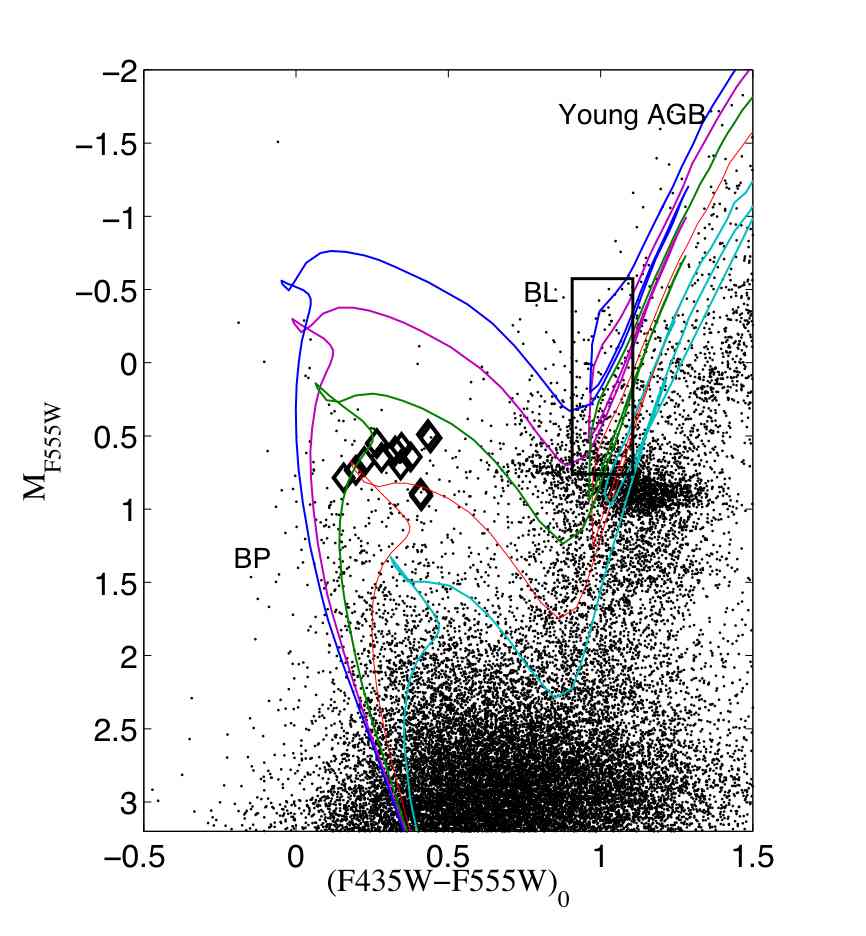}
  \caption{CMD of M32 corrected for contamination by the M31
    background stars, reddening
    \citep[$E(B-V)=0.08$,][]{burstein_heiles82} and extinction
    \citep[$A_{F555W}=0.25$,][]{sirianni_etal05}.  The distance
    modulus $\mu_0 = 24.53 \pm 0.12$ was obtained using the RCS method
    (see text for more description).  Solar metallicity isochrones
    from \citet{marigo_etal08} are superimposed, with ages of
    0.4,0.5,0.7,1, and 1.5 Gyr.  Note the good fit to the extended MS
    of the CMD, and the suggestion of the presence of subgiant
    stars. On the other hand, the diamonds show the location of the RR
    Lyrae found in this M32 field (F10).  These isochrones also fit
    the BL region (delineated by the black box) well, suggesting the
    presence of young AGB stars with masses of $2.5$--$2.8\,
    M_{\odot}$.}
\label{fig:isochronesm32}
\end{figure}

Young populations are mostly represented by stars occupying the blue
plume (BP) seen in the CMD of Figure~\ref{fig:hessm32_features} at
$F555W \sim 24.5$ from $F555W \sim 27$ and $(F435W - F555W) < 0.5$,
suggesting the presence of an extended main sequence (MS).
We note that the presence of a BP in M32 has either not been claimed
or not observed in previous photometric works \citep[e.g., G96;
][]{Worthey_etal04}.  Therefore, we begin this section by addressing
whether this feature is real or not.  Blends of fainter blue stars
could appear in our photometry as brighter blue stars, generating a BP
which is actually artifact of crowding.  We have tested this using the
results from the ASTs.  We estimated the fraction of recovered BP
stars that were actually injected as red or fainter stars; $\sim$ 14\%
of recovered BP stars are blends, thus indicating that $\sim$ 86\% are
genuine blue stars. We are hence confident that the detected BP is
indeed real.  Moreover, we have investigated archival observations of
fields near M32, which also seem to indicate the presence of a BP in
M32. This analysis is shown in the Appendix.

Figure~\ref{fig:isochronesm32} shows Padova isochrones
\citep{marigo_etal08, girardi_etal08} for young ages superimposed on
the decontaminated, de-reddened and corrected for distance M32 CMD.
The isochrones have solar metallicity $\rm{Z}_{\odot} = 0.019$ and a
range of ages of 0.4, 0.5, 0.7, 1, and 1.5 Gyr.  This suggests that
M32 at F1's location may contain stars as young as $\sim 0.5-1.5$ Gyr;
other higher metallicities, for Z ranging from Z$_{\odot}$ to 0.03
($\mathrm{[Fe/H]} \sim +0.2$), show similar results.  However, lower
metallicities isochrones are too blue compared with the observed data.
On the other hand, we can see that any significant presence of
populations younger than 0.4 Gyr is ruled out in this field, although
stars as young as 0.5 Gyr may be present.  We even appear to find a
possible subgiant branch (SGB), with a region occupied by stars as
soon as they leave the MS. These SGB stars are consistent with
isochrones of ages from 0.5 Gyr to 1.5 Gyr for stars with masses
$1.6-2.5$ $M_{\odot}$ and a range of metallicities $Z=0.019 - 0.03$
($\mathrm{[Fe/H]} \sim 0$ to $+0.2$).  We also show in
Figure~\ref{fig:isochronesm32} the position of known RR Lyrae
variables in field F1 (F10), indicating the presence of some blue
horizontal branch stars in this region of the CMD.  It is worth noting
that the number of RR Lyrae discovered in this field (17, of which
$7^{+4}_{-3}$ belong to M32 ) cannot account for all of the stars in
this region ($\sim 100$).  Nevertheless, the SGB region of the CMD is
the most affected by blends: stars either bluer or redder are likely
to blend and occupy this region. We tested this using the results from
ASTs and found that only a $\sim 38\%$ of stars recovered in this
region were actually injected there.  Most of the stars are therefore
blends and we are not able to assume that they are all actual SGB
stars.

The presence of a blue loop (BL) is another sign of young populations
and has never been detected before in M32. Stars in this region are
intermediate-mass stars with ages between $\sim 0.4$ and $\sim 1$ Gyr
burning helium in their cores \citep[see, e.g.,][and references
therein]{Sweigart87, Xu_li04} .  Theoretical models predict that their
positions strongly depend on metallicity
\citep[e.g.,][]{girardi_etal00}.  The box in
Figure~\ref{fig:isochronesm32} shows the location of BL stars.  We
considered the fact that these stars could also be artifacts of
crowding and we analyzed this again using the results from the ASTs.
We found that $\sim 80\%$ of the recovered stars in the BL region were
actually injected there. We are certain that most of the stars in the
BL are real and not products of blends.  We can see from
Figure~\ref{fig:isochronesm32} that their magnitudes and colors are
consistent with the assumed solar metallicity and ages of 0.4--0.9 Gyr
and they have masses in the range $\sim 2.25$---$3\, M_{\odot}$.
Isochrones with lower metallicities predict the location of the BL
bluer than observed.  Note also that the isochrones in
Figure~\ref{fig:isochronesm32} suggest the presence of young AGB stars
as well, of ages between 0.7--1 Gyr, with masses of 2.5--$2.8\,
M_{\odot}$. However this region may be also occupied by older RGB and
AGB stars with lower metallicities (see the previous section) and much
lower masses, of the order of $0.9$--$1\,M_{\odot}$.

\begin{figure}
\hspace{-0.5cm}
%\subfigure {
\includegraphics[width=45mm,clip]{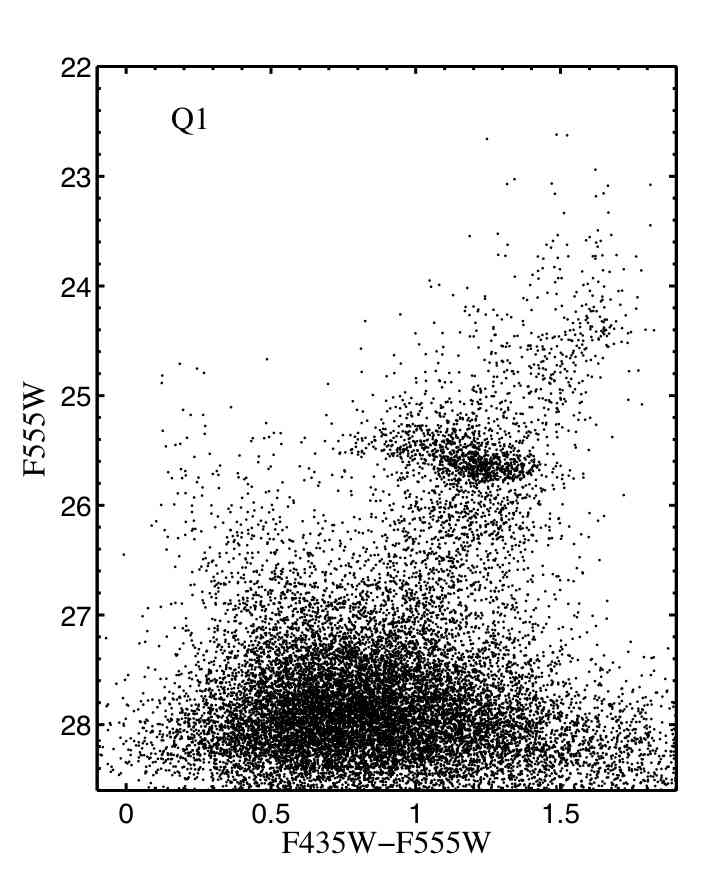}%}
%\hspace{-0.5cm}
%\subfigure {
\includegraphics[width=45mm,clip]{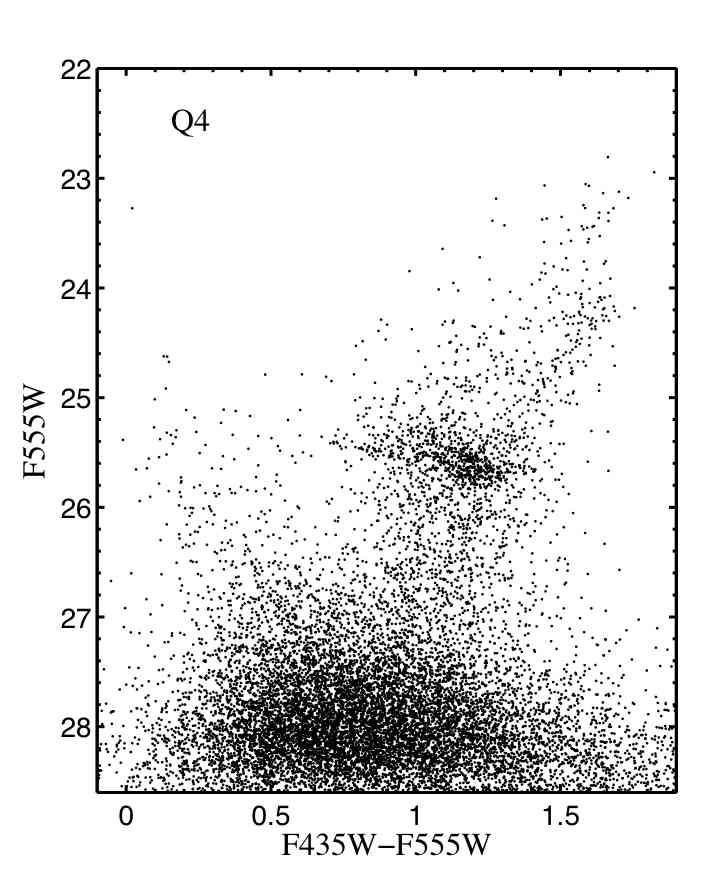}%}
%\subfigure {
%  \includegraphics[width=60mm,clip]{hess_sub_quadrantsdraft.eps}%}
\caption{\emph{Left  panel}: CMD  of the quadrant  Q1 closest  to the
  nucleus of M32  in F1.  \emph{Right panel}: CMD  of the quadrant Q4
  furthest away  from the nucleus of  M32. We compare the  BP in these
  two CMDs in order to test whether it belongs to M32 or it represents
  a  constant M31  background in  F1.  It  seems that  the  fainter BP
  $F555W >  26$ is  indeed stronger in  Q1, which indicates  that this
  portion of the  BP comes from M32.  However,  the brighter BP stars,
  which assured  the presence of a  young population, do  not show any
  significant difference between the two CMDs.  They could represent a
  constant   M31    disk   background    in   F1.    See    text   and
  Table~\ref{table:starcounts} for more information.}
\label{fig:quadrants}
\end{figure}
\begin{deluxetable*}{@{}cccccccc}
  \tablecaption{Star counts  in different  boxes of the  CMDs. Regions
    are                 indicated                 as                in
    Figure~\ref{fig:hessm32_features}\label{table:starcounts}}
  \tablewidth{0pt}  \tablecolumns{7}  \tablehead{\colhead{CMD  Region}
    &\colhead{F1                        \tablenotemark{a}}&\colhead{F2}
    &\colhead{Q1$_{\rm{F1}}$            \tablenotemark{b}}           &
    \colhead{Q4$_{\rm{F1}}$\tablenotemark{c}}  &  \colhead{F1$-$F2}  &
    \colhead{Q1$_{\rm{F1}}-$    Q4$_{\rm{F1}}$}&   \colhead{   F1/F2}}
  \startdata BP stars& &% &
  & & & &\\
  $24.0<F555W<25.5$&$\phn \phn 93~\pm~10$ & $\phn \phn 64
  ~\pm~\phn8$% &34
  & $\phn \phn 22~\pm~\phn 5$ & $\phn \phn 27 ~\pm~\phn5$ & $\phn\phn29 ~\pm~13$ & $\phn\phn-5 ~\pm~\phn7$ & $1.45~\pm~0.24$\\
  $25.5<F555W<26.0$&$\phn169~ \pm~13$ & $\phn\phn80~\pm~\phn9$% &89
  & $\phn\phn48~\pm~\phn7$& $\phn\phn45 ~\pm~\phn7$ &$\phn\phn89 ~\pm~16$ & $\phn\phn\phn3 ~\pm~10$ & $2.11 ~\pm~0.29$\\
  $26.0<F555W<27.0$&$1477 ~\pm~38$ & $\phn534 ~\pm~23$ % &967
  & $\phn  451 ~\pm~21$&  $\phn 336 ~\pm~18$&$\phn  943 ~\pm~44$&$\phn
  115 ~\pm~27$ & $2.76 ~\pm~0.14$ \\\hline BL stars& &% &
  & & & &\\
  $23.5<F555W<25.2$&$\phn263 ~\pm~16$ & $\phn 127 ~\pm~11$% &142
  & $\phn \phn  70 ~\pm~\phn8$ & $\phn \phn  64 ~\pm~\phn8$& $\phn 136
  ~\pm~20$  &  $\phn\phn\phn  6~\pm~11$  & $2.07  ~\pm~0.22$  \\\hline
  Bright AGB & &% &
  & & & &\\
  $22.0<F555W<24.0$&$\phn 211 ~\pm~14$& $\phn 101~\pm~10$% &129
  & $\phn\phn 64 ~\pm~\phn8$& $\phn \phn 55 ~\pm~\phn7$& $\phn\phn 110
  ~\pm~17$ & $\phn\phn\phn 9 ~\pm~10$ &$2.09 ~\pm~0.25$ \\\hline RGB &
  & &% &
  & & &\\
  $24.0<F555W<25.2$&$1205 ~\pm~35$& $\phn372~ \pm~19$% &830
  & $\phn 358~ \pm~19$& $\phn 268~\pm~16$& $\phn 833 ~\pm~40$ & $\phn \phn 90 ~ \pm~25$ & $3.24 ~\pm~0.19$\\
  $25.2<F555W<26.0$&$3998 ~\pm~63$& $1243~ \pm~35$% &2741
  & $1174 ~\pm~34$& $\phn 894 ~\pm~30$& $2755 ~\pm~72$ & $\phn 280 ~ \pm~45$ & $3.22 ~\pm~0.10$ \\
  $26.0<F555W<26.8$&$2857 ~\pm~53$& $\phn 953 ~\pm~31$ % &1915
  & $\phn 876  ~\pm~30$ & $\phn 631 ~\pm~25$ &  $1904 ~\pm~61$ & $\phn
  245~\pm~40$ &$3.01 ~\pm~0.11$
\enddata
\tablenotetext{a}{Before  decontamination  for  M31
      stars.}
\tablenotetext{b}{Quadrant  closest to the  center of  M32, where
      the  stellar  density  in  F1   is  higher.  See  Figure  2.} 
\tablenotetext{c}{Quadrant furthest away  from the center of
      M32. See Figure 2.} 
\end{deluxetable*}

The bright stars composing the BP and BL features remain in the CMD
even after the statistical decontamination for M31 stars.  This would
indicate that they belong to M32.  However, while we took care to
characterize the M31 disk ground by devoting equal time to the F2
images, the surface density of the M31 disk may (unfortunately) be
higher at F1 than it is at F2.  \citet{Arp66} noted that M32 (Arp 168)
appears to have a diffuse ``plume'' of emission to the south of its
nucleus, and thus in the direction of F1.  \citet{choi_etal02}
attempted to isolate this feature through a variety of surface
photometry models, and suggested that it is due to the tidal
interaction of M31 with M32.  The obvious inference is that the plume
comprises stars that have been stripped from M32; however, if the
plume instead represents material drawn for the M31 disk, then this
may enhance the contribution of disk stars to F1 over what we would
have inferred from F2.  To test whether the young population (ages $<$
1 Gyr) comes from M32 or M31, we have first compared the CMDs obtained
from closest and furthest quadrants from the nucleus of M32 (hereafter
Q1 and Q4, respectively) in F1.  Since, given the rapid decline of the
M32 surface brightness, we observe a gradient in the stellar density
of M32 over F1 (Figure 2), we should therefore detect a larger number
of bright BP and BL stars as we approach the nucleus of M32 if they
belong to M32.  Figure~\ref{fig:quadrants} shows the CMDs in apparent
magnitudes obtained in Q1 (left panel) and in Q4 (right panel).  The
number of BL stars as well as BP stars brighter than $F555W \sim 26$
seems to be roughly the same in both CMDs.  However, the number of
fainter BP stars ($F555W > 26$) is significantly larger in Q1.  To
quantify this, we define eight boxes in different regions of the CMDs
and we count stars in them.  The different regions are the BP, BL, AGB
and RGB (see Figure~\ref{fig:hessm32_features}) and the corresponding
boxes are indicated in Table~\ref{table:starcounts}.  This analysis
was done not only in the Q1 and the Q4 CMDs but also in the F1 (before
decontamination for F2) and F2 CMDs.  Assuming Poisson statistics we
can infer whether (F1$-$F2) and/or (Q1$-$Q4) is significantly positive
for each of the boxes.  The results of (Q1$-$Q4) suggest that there is
a constant background of bright blue stars over F1, since there is no
significant difference between the number of bright BP and BL stars.
The ratio between the number of stars in F1 to F2 is also shown in
Table ~\ref{table:starcounts} for each of the regions.  This fraction
is expected to be $\sim$ 3 according to the surface brightness
estimates at F1's location.  Note that this is clearly the case for
the RGB stars and the fainter BP stars ($26.0 < F555W < 27.0$).
However, the ratio is lower than 3 for the brighter portion of the BP,
the BL and bright AGB stars which indicates that the M31's background
is relatively enhanced in F1 compared to what F2 represents.  As
stated above, a plausible explanation for an enhanced contribution of
disk stars to F1 is that the ``diffuse plume'' observed in M32 at F1's
location \citep{Arp66} represents material drawn for the M31
disk. Given the null difference between stars in Q1 and Q4, it is
likely that all of the bright BP stars belong to M31.  It is likely
then that the very young population that we see in the decontaminated
CMD of M32 (i.e.  stars with ages $\sim 0.5$ Gyr) belongs to the disk
of M31 rather than to M32.  Instead, stars in the BP at magnitudes
$F555W > 26$ \emph{do} belong to M32 and represent a population of
ages between 1 and 2 Gyr (see Figure~\ref{fig:isochronesm32}).
Further investigation is required to conclusively determine whether a
young population of stars with ages $\sim 0.5$ Gyr do or do not belong
to M32.  Wide field images from fields closer to the nucleus of M32,
as could be obtained with HST/WFPC3, would reveal this.  Note that the
Q1$-$Q4 difference in the bright AGB region is positive, and the F1/F2
ratio for these stars is significantly higher than for the brightest
BP stars\footnote{The F1/F2 ratio of BL stars is also higher than for
  the brightest BP stars. However, BL stars are more contaminated by
  blends than AGB stars, which are nearly no contaminated at
  all. Moreover, the Q1$-$Q4 difference for the BL stars is not
  significantly positive.}.  This implies that, even though there is
an enhancement of M31 AGB stars over what F2 represents, it is likely
that a fraction of them still belong to M32.

\paragraph {Young population  vs. Blue straggler stars}
Blue straggler stars  (BSSs) are stars hotter, bluer  and brighter than
the MSTOs in a CMD and thus  generate a blue plume, similar to the one
we see in the CMD.  As they  can mimic a young population (ages $<~ $2
Gyr),  it is  crucial  for the  derivation  of the  SFH to  understand
whether the BP we find in  M32 is \emph{entirely} populated by old BSSs
instead of a genuinely young population.

BSSs have been seen in globular and open clusters
\citep[e.g.,][]{ferraro_etal04, mapelli_etal04, mapelli_etal06,
  piotto_etal04, demarchi_etal06} where there is no recent star
formation or the spread in stellar age is small enough such that their
identification as old BSSs (instead of as a young population) is very
clear.  In a field population it is difficult to unambiguously prove
the nature of the BP as old BSSs \citep[see,
e.g.,][]{mateo_etal95,hurleykeller_etal99, aparicio_etal01,
  carrera_etal02, mapelli_etal09}.  Only \citet{momany_etal07} in
their work indicated that their BSS candidates in dSphs may be real
BSSs. They found a statistically-significant anti-correlation between
the specific frequency of BSS candidates with the HB stars and the
absolute magnitude of their dSph sample, similar to what has been
observed in both globular clusters \citep{piotto_etal04} and open
clusters \citep{demarchi_etal06}.  \citet{momany_etal07} claimed that
this anti-correlation can be used as a classification tool such that
galaxies following the anti-correlation are more likely to have real
BSSs rather than young MS stars (the anti-correlation would be hard to
explain if BSS candidates were young stars). It is worth of mentioning
that they selected their sample in such a way that dwarf
spheroidals/irregulars in which there is current or recent ($\leq 500$
Myr) star formation were not considered.  Therefore, an ideal test for
our data would be to calculate the frequency of BSS candidates in M32
and compare our results to the proposed relation of Figure 2 of
\citet{momany_etal07}.  However, the box considered to include the
candidates of old BSSs goes from the magnitudes of the oldest MSTOs up
to where there are blue stars observed. We do not reach the oldest
MSTOs; the stars at the appropriate magnitude level are dominantly
products of blends, which will contaminate the BSS candidates in our
data.  We are therefore unable to test, at least in this way, whether
the BP consists of old BSSs.  Nevertheless, we favor the idea that the
BP cannot be entirely populated by old BSSs but instead contains some
genuine young stars due to the following: Brown et al.  (2006) argue
that, even though the formation mechanism of BSS is not completely
understood, BSSs in an old population will be limited to $M < 2
M_{\sun}$, whereas the masses required to explain stars as bright as,
and even exceeding, the HB luminosity level that we see in our CMD are
higher than that value ($2<\rm{M}<3 \rm{M_{\odot}}$).  This implies
that at least the brightest BP stars observed are truly young
stars. However, as explained above, they are likely to belong to the
disk of M31 rather than to M32. The fainter BP, $F555W > 26$,
associated to M32 is not assured to be only young stars and it could
even be composed only of BSSs.

\subsection{Ancient population: Ages $>$ 10 Gyr}
We believe that old and metal-poor stars should be present in M32,
since previous generations of stars are required to produce the
metal-rich population we observe. It is clear from our CMD that there
are not many of these stars since there is no noticeable presence of
blue horizontal branch (BHB) stars. Therefore a well-developed old,
metal poor population is not present.  Nonetheless, RR Lyrae variable
stars were found in our M32 field with our data (F10), revealing the
presence of BHB stars.  The pioneering work of \citet{brown_etal00,
  brown_etal08} showed evidence of extreme horizontal branch (EHB)
stars in the central region of M32 from Space Telescope Imaging
Spectrograph (STIS) UV observations.  By detecting RR Lyrae stars in
our data, we find a different manifestation of a consistent picture.

The RR Lyrae stars found with the data from this study have a mean
metallicity of $\mathrm{\langle[Fe/H]\rangle} \sim-1.5$ (F10). Taking
into account the metallicity distribution we have obtained (see above)
and assuming that these metal-poor stars are ancient stars, we can
estimate the fraction of these stars in our data. We find that, when
assuming a 10 Gyr-old population, metal-poor stars with
$\mathrm{[Fe/H]}\leq -1$ represent \emph{at most} $\sim$ 5.7\% of the
total $V$-band light in our M32 field (recall that the number of stars
in metal-poor tail of the MDF is an upper limit).  \citet{Worthey94}
models can be used to obtain the stellar M/L ratio for different
metallicities at a given assumed age.  We estimate that a metal-poor
10 Gyr-old population contributes $\sim$ 4.5\% of the total mass in
our observed field. Note however that there is little constraint on
ancient, metal-rich populations.  The fact that the hot HB stars do
not significantly contribute to the optical $V$-band light of M32
suggests that the strong Balmer lines found in spectroscopic studies
\citep{oconnell80, Pickles85, Bica_etal90,rose94} actually represent a
young population or BSSs.  \citet{Trager_etal00a} estimated that the
BHBs needed to account for the strong Balmer lines of M32
($\mathrm{H}\beta = 1.9$~{\AA}, corresponding to an integrated light
age of $\sim$ 8 Gyr) at F1's location would contribute to the $\sim$
5\% of the $V$-band light.  This would imply that $\sim$ 25\% of the
\emph{total} light would have to come from metal-poor stars, which is
$\sim$5 times more than the amount of metal-poor $V$-band light that
we have found.  On the other hand, to explain the
high-$\mathrm{H}\beta$ strength of M32 arising from a population of
BSSs would require that $\ga 15\%$ of the $V$-band light come from
BSSs \citep{Trager_etal00b}. This will be explored in a follow-up
paper where the derived SFH will allow us to quantify the light
contribution of the BP in F1.

\subsection{The Luminosity function of  M32}

\begin{figure}
\centering \includegraphics[width=98mm,clip]{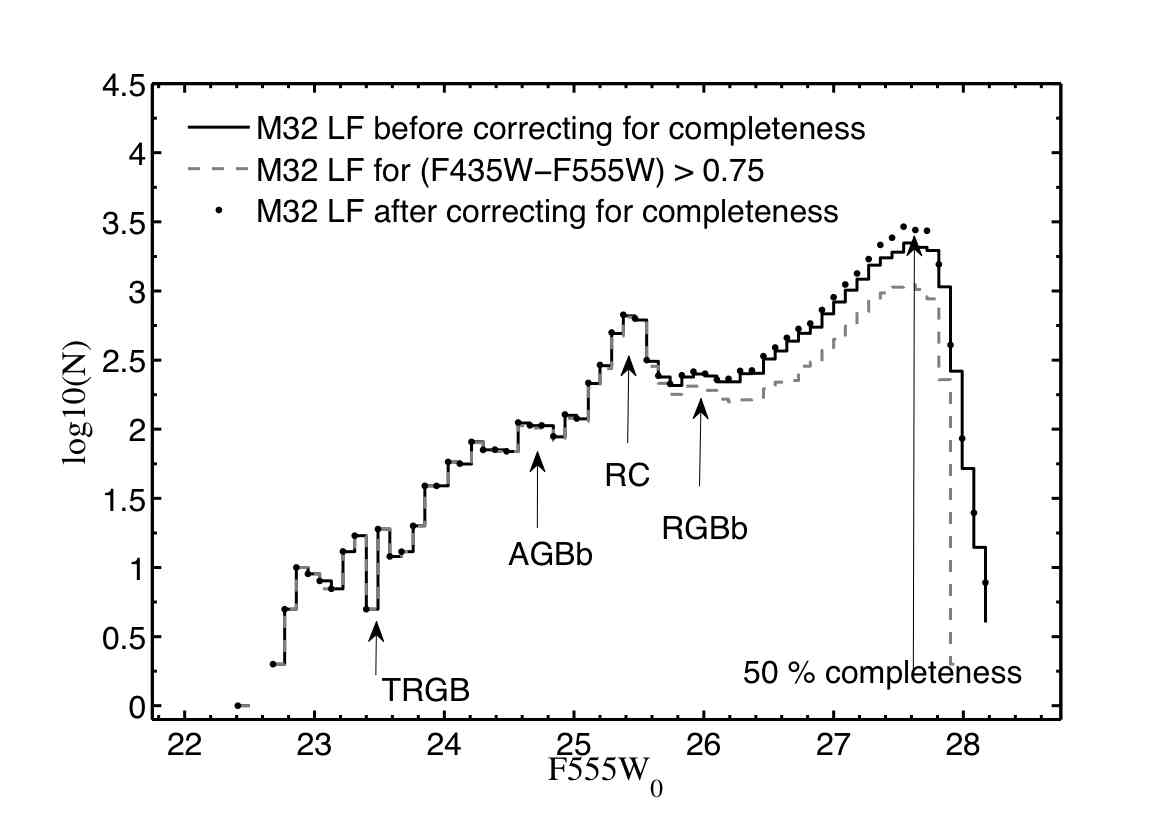}
\caption{$F555W_0$-band Luminosity Function (LF) for M32, corrected
  for extinction and decontaminated for M31 background stars, is shown
  in the solid histogram.  The completeness-corrected LF is shown as
  black dots. The gray-dashed line shows the LF for stars redder than
  $(F435W-F555W)\sim 0.75$ to avoid contamination from, for example,
  the BP.  The 50\% completeness level is indicated at a magnitude of
  $F555W_0 \sim 27.75$. The main features of this LF are indicated by
  arrows.  The TRGB lies at $F555W_0 \sim 23.7$, the AGB bump at
  $F555W_0 \sim 24.7$, the RC at $F555W_0\sim 25.4$, and the RGB bump
  at $F555W_0\sim 25.9$.}
\label{fig:lf}
\end{figure}
\begin{deluxetable*}{@{}cccc|cccc|cccc}
  \tablecaption{M32    Luminosity    Function\label{table:luminosity}}
  \tablewidth{0pt}  \tablecolumns{12} \tablehead{  \colhead{ $F555W_0$
      \tablenotemark{a}} &  \colhead{$N$ \tablenotemark{b}} &\colhead{
      $N_{corr}$    \tablenotemark{c}}   &   \colhead{$N_{(col>0.75)}$
      \tablenotemark{d}}   &   \colhead{$F555W_0$}   &   \colhead{$N$}
    &\colhead{    $N_{corr}$}     &\colhead{    $N_{(col>0.75)}$}    &
    \colhead{$F555W_0  $}  &\colhead{  $N$  }&  \colhead{  $N_{corr}$}
    &\colhead{ $N_{(col>0.75)}$}}  \startdata 22.45  & 1 &  1.00 &  1 &
  24.44 & 73 & 73.64 & 51 & 26.43 & 257 & 276.43 & 135 \\ %\hline
  22.53 &  0 & 0.00  & 0 &  24.52 & 77  & 77.55 & 66  & 26.51 &  329 &
  349.93 & 138 \\ %\hline
  22.62 & 0  & 0.00 & 0  & 24.60 & 107 &  107.77 & 61 & 26.59  & 366 &
  389.89 & 143 \\ %\hline
  22.70 &  2 & 2.00  & 2 &  24.69 & 91  & 91.94 & 86  & 26.67 &  418 &
  453.06 & 150 \\ %\hline
  22.78 &  9 & 9.00  & 2 &  24.77 & 96  & 97.35 & 91  & 26.76 &  480 &
  526.73 & 196 \\ %\hline
  22.87 &  6 & 6.00 &  11 & 24.85 &  79 & 79.40 &  89 & 26.84  & 517 &
  568.22 & 192 \\ %\hline
  22.95 & 8  & 8.00 & 7  & 24.94 & 119 &  120.55 & 86 & 26.92  & 652 &
  721.45 & 188 \\ %\hline
  23.03 & 8  & 8.00 & 8  & 25.02 & 110 &  111.61 & 77 & 27.01  & 751 &
  852.73 & 251 \\ %\hline
  23.11 & 8  & 8.00 & 3 & 25.10 &  191 & 194.20 & 103 &  27.09 & 929 &
  1052.41& 267 \\ %\hline
  23.20 & 12 & 12.08 & 10 & 25.18  & 257 & 261.86 & 100 & 27.17 & 1106
  & 1277.55& 312 \\ %\hline
  23.28 & 10 & 10.00 & 10 & 25.27  & 400 & 409.63 & 170 & 27.25 & 1356
  & 1598.82 & 356 \\ %\hline
  23.36 & 13 & 13.00 & 16 & 25.35  & 534 & 544.64 & 210 & 27.34 & 1586
  & 1896.35 & 433 \\ %\hline
  23.45 & 14 & 14.00 & 6 & 25.43 & 628 & 641.55 & 339 & 27.42 & 1671 &
  2072.59 & 535 \\ %\hline
  23.53 & 5 & 5.00 & 15 & 25.52  & 433 & 445.35 & 456 & 27.50 & 1936 &
  2462.01 & 608 \\ %\hline
  23.61 & 16 & 16.00 & 4 & 25.60 & 235 & 242.03 & 601 & 27.59 & 2009 &
  2633.19 & 775 \\ %\hline
  23.69 & 15 & 15.00 & 17 & 25.68  & 203 & 211.08 & 466 & 27.67 & 1885
  & 2519.17 & 844 \\ %\hline
  23.78 & 21 & 21.00 & 14 & 25.76  & 201 & 208.43 & 232 & 27.75 & 1626
  & 2260.72 & 907 \\ %\hline
  23.86 & 33 & 33.00 & 18 & 25.85 & 220 & 228.30 & 190 & 27.83 & 740 &
  1075.84 & 977 \\ %\hline
  23.94 & 36 & 36.11 & 33 & 25.93 & 235 & 247.53 & 146 & 27.92 & 179 &
  276.44 & 891 \\ %\hline
  24.02 & 56 & 56.39 & 32 & 26.01  & 229 & 239.49 & 173 & 28.00 & 44 &
  72.50 & 872 \\ %\hline
  24.11 & 50 & 50.00 & 53 & 26.09  & 200 & 209.56 & 183 & 28.08 & 12 &
  21.28 & 623 \\ %\hline
  24.19 & 68 & 68.00  & 46 & 26.18 & 196 & 205.74 &  171 & 28.16 & 4 &
  7.77 & 113 \\ %\hline
  24.27 & 67 & 67.12 & 57 & 26.26 & 230 & 243.67 & 170 & & & &
  \\ %\hline
  24.36 & 64 & 64.11 & 70 & 26.34 & 234 & 247.77 & 139 & & & &
\enddata
\tablenotetext{a} {Dereddened magnitude of the bin center.}
\tablenotetext{b}{Raw counts in a given magnitude bin for the
  decontaminated data.}  \tablenotetext{c} {Completeness-corrected
  counts for the same decontaminated data.}  \tablenotetext{d}{Raw
  counts in a given magnitude bin for stars with $(F435W-F555W) >
  0.75$}
\end{deluxetable*}

The extinction-corrected $F555W_0$ luminosity function (LF) for M32 is
given in Table~\ref{table:luminosity}.  The LF is measured by dividing
the $F555W_0$ magnitudes into 65 bins with a bin width of $\Delta
F555W_0 = 0.1$ mag.  The LF is shown as the solid histogram in
Figure~\ref{fig:lf}.  We have also calculated the
completeness-corrected LF (black dots).  The long vertical arrow marks
the 50\% completeness level at $F555W_0 \sim 27.75$.  The main
distinct features are indicated by arrows: the TRGB at $F555W_0 \sim
23.7$, the AGBb at $F555W_0 \sim 24.7$, the RC at $F555W_0 \sim 25.4$
and the RGBb at $F555W_0 \sim 25.9$.  These features are consistent
with the CMD analysis above.  We also show the LF for stars redder
than $(F435W-F555W) = 0.75$, as gray-dashed histogram.  We can see
that the features that we find on the RGB region, in particular the
RGBb, are better identified when we plot only the redder stars thus
avoiding contamination from, for example, the BP.

\smallskip To summarize this section, M32 is dominated by a 8--10 Gyr
old, metal-rich ($[\mathrm{Fe/H}] = -0.20 \, \mathrm{dex}$)
population, due to the RC location and width of the RGB.  This result
is supported by the locations of the RGBb and the AGBb.
Intermediate-age stars of $4\pm3$ Gyr are also present in M32, as
revealed by the bright AGB stars observed and the morphology of the
strong RC.  We see evidence for 1--2 Gyr MSTO stars and/or BSSs due to
the presence of the BP.  An ancient metal-poor population does not
contribute much to the light of M32 in F1. There is, however, little
constraint on its metal-rich counterpart.
 
\section{The Distance to M32 and M31}\label{sec:distance}

We determine the distance to M32 using the Red Clump Stars
(RCS)\footnote{A widely-used standard candle for determining the
  distance to a stellar system is the TRGB method \citep{lee_etal93},
  which determines distances using the TRGB discontinuity in the
  $I$-band.  However, an essential property of the TRGB distance
  indicator in the original method \citep{lee_etal93} is that the
  absolute magnitude in the $I$ band is less sensitive to changes on
  metallicity than the $B$ or $V$ magnitudes (note in
  Figure~\ref{fig:ia_isochronesm32} how the position of the TRGB on
  the $F555W$ band varies with metallicity at a fixed age).  Thus, the
  TRGB method cannot be applied here to obtain the distance to M32
  because we do not have the correct magnitude bands to do so.}.
\citet{udalski98} has stressed that this method has many advantages
with respect to other widely used standard candles such as Cepheid and
RR Lyrae: for example, RCS are easy to recognize in a CMD and large
samples are usually present in galaxies, which is certainly true in
our case.  Another important advantage is the fact that the mean
absolute magnitude of RCS in the Solar neighborhood has been
calibrated by \emph{Hipparcos} \citep{ESA_1997, Perryman_etal97}
parallaxes with an accuracy of 10\%.  The disadvantage however is that
its properties are not the same for all galaxies, given the
age-metallicity dependence of the RC which affects the value of its
absolute magnitude.  Since models of core-helium-burning stars predict
that the RC luminosity depends on both age and metallicity
\citep{cole98, girardi_etal98, girardi_salaris01}, ``population
corrections'' to the RC absolute magnitude obtained using
\emph{Hipparcos} parallaxes need to be made before it can be used as
an accurate extragalactic distance indicator.
\citet{percival_salaris03} have calculated how the absolute magnitudes
of RCS differ for a given age and metallicity when compared with the
Solar-neighborhood RC absolute magnitude, $\Delta
M^{\mathrm{RC}}_{\lambda}$.  Taking these variations into account, we
can still use the RCS to estimate the distance of M32. The true
distance modulus of the galaxy will be given by
\begin{equation}\label{eq:mu} \displaystyle \mu_0=(m-M)_0 =
m^{\mathrm{RC}}_{\lambda} - M^{\mathrm{RC}}_{\lambda, \mathrm{local}}
- A_{m_{\lambda}} + \Delta M^{\mathrm{RC}}_{\lambda},
\end{equation} where
\begin{equation}   \Delta   M^{\mathrm{RC}}_{\lambda}=  M^{\mathrm{RC,
      theory}}_{\lambda,      \mathrm{local}}     -     M^{\mathrm{RC,
      theory}}_{\lambda, \mathrm{galaxy}}.
\end{equation} 
The mean apparent magnitude of RC calculated in section 5.1.1 above is
$F555W= 25.66 \pm 0.082$. This value is then de-reddened assuming
$E(B-V) = 0.08$ \citep{burstein_heiles82} and an extinction of
$A_{F555W} = 0.25$ \citep{sirianni_etal05}.  On the other hand, the
\emph{Hipparcos} absolute magnitude and the theoretical population
corrections to the Red Clump absolute magnitudes were calibrated and
calculated on the UBV system.  We need therefore to transform our data
from the ACS/HRC VEGA system $F435W$ and $F555W$ to the ground-based
$B$ and $V$ magnitudes.  We use the \citet{sirianni_etal05}
transformations for this.  The $M_V^{\mathrm{RC}}$ from Hipparcos for
the solar neighborhood is calibrated as $M^{\mathrm{RC}}_{V,
  \mathrm{local}}= 0.73\pm 0.03$ \citep{Alves_etal02}.  The population
correction $\Delta M^{\mathrm{RC}}_{\lambda}$ is calculated by
\citet{percival_salaris03} and we refer to that paper for a detailed
explanation. Briefly, to study the age and metallicity dependence of
the RC brightness, they compare their empirically derived population
corrections with the theoretical models of \citet[see Figure 2 in
Percival \& Salaris 2003]{girardi_etal00}.  In our case, assuming an
average $\langle [\mathrm{Fe/H}]\rangle = -0.2 \, \mathrm{dex}$ and
an age of 8 Gyr for our RCS (G96, this paper), $\Delta
M^{\mathrm{RC}}_{V} = -0.1$. \citet{percival_salaris03} calculate the
residuals of $\Delta M^{\mathrm{RC}}_{\lambda}$ for their \emph{model
  -- observed} fits taking into account both the metallicity and age
residuals. The mean residual in $\Delta M^{\mathrm{RC}}_{V}$ is $+0.03
\pm 0.07$, where the error is the $1\sigma$ error.  The true distance
modulus of M32 from Equation~\ref{eq:mu} is therefore
\begin{equation}   \mu_0=  (25.33   \pm  0.088)-(0.73   \pm   0.03)  +
  (-0.10+0.03 \pm 0.07)
\end{equation} or
\begin{equation} \mu_0(\rm{M32})=  24.53 \pm 0.12
\end{equation}
in    agreement    with    previous   results,    e.g.    $24.2\pm0.3$
\citep{freedman92a}, $24.55 \pm 0.08$ \citep{tonry_etal01}, $24.39 \pm
0.08$ \citep{jensen_etal03}, and $24.53 \pm 0.21$ (F10).

We can also obtain the distance to M31 from our background field using
the RCS of F2.  This will also give us a relative distance measurement
between these two galaxies.  To this end, we selected 1107 stars from
field F2 having apparent magnitudes $25.0 < F555W < 26.0$ and colors
$0.95 < (F435W-F555W) < 1.32$.  We then fit the histogram of the
luminosities and colors of these stars with Equation~\ref{eq:rc}.  We
find $F555W_{m}= 25.49 \pm 0.06$ and $(F435W-F555W)(RC)= 1.14 \pm 0.10
$.  After we de-redden ($E(B-V)=0.08$) and transform these values onto
the UBV system \citep{sirianni_etal05}, we consider the differences
between the mean RC population in M31 at our field location and the
RCS in our solar neighborhood.  We consider the metallicity
$[\mathrm{Fe/H}] = -0.40\,\mathrm{dex}$ and an age of 8 Gyr for our
M31 field, which corresponds to a population correction $\Delta
M^{\mathrm{RC}}_{V} = +0.02$ \citep{percival_salaris03}.  We obtain a
true distance modulus of
\begin{equation} \mu_0(\rm{M31})=  24.45 \pm 0.14
\end{equation}
for M31, in agreement with previous estimates of its true distance
modulus, e.g.  $24.44\pm0.11$ \citep{freedman_madore90}, $24.50 \pm
0.10$ \citep{brown_etal04}, and $24.47 \pm 0.07$
\citep{mcconnachie_etal05}.  The most up-to-date values have been
obtained using Cepheids \citep{saha_etal06} and RR Lyrae
\citep{sarajedini_etal09, Fiorentino_etal10}, $\mu_0 = 24.54 \pm
0.07$, $24.46 \pm 0.11$, and $24.49 \pm 0.19$, respectively. Clearly
the obtained values here for M32 and M31 distance moduli are
comparable, and, within the errors, it is not possible to determine
with our data whether M32 is projected in front of or behind M31. We
are therefore unable to confirm the results of \citet{Ford_etal78},
who found evidence of M32 being in front of M31.

\section{M31 stellar populations in F2}\label{sec:m31}

Field F2 serves not only  as a background, allowing the M32 photometry
to  be  decontaminated for  M31,  but is  also  very  important as  it
contains information  about the stellar populations in  the inner disk
and  bulge of  M31.   In this  section  we first  compare the  stellar
populations in F2 with the ones in F1. We then analyze the M31 stellar
population from the F2 CMD and compare it with what it is known in the
literature.

\subsection{Differentiating the F1 and F2 CMDs}

The CMDs of F1 and F2 are shown in Figure~\ref{fig:hessdeconvolved}.
We note that, at first glance, they are very similar.  They seem to
share the same morphology: a wide RGB, a noticeable RC, and a BP.
Furthermore, when we statistically subtract field F2 from F1, we
notice that many features remain in the CMD of M32.  Thus, due to the
similarities in both CMDs, we cannot be certain there is not M32
contamination in field F2.  Is it possible that F2 field is
contaminated with tidal debris from M32?  If there were M32
contamination in the background field, we have overcorrected the M32
CMD and we are thus missing information about our primary field F1.
Note that this is unlikely: the predicted contribution of M32 to the
F2 field is very small, representing a surface brightness of
$\mu_V\approx27.5$, while M31 has $\mu_V\approx22.7$ in this field.
On the other hand, we want to be sure that the contamination from M31
was completely removed from F1 (statistically speaking), and not
under-subtracted.

\begin{figure}
  \centering
  \hspace{-1.cm}
  \includegraphics[width=90mm,clip]{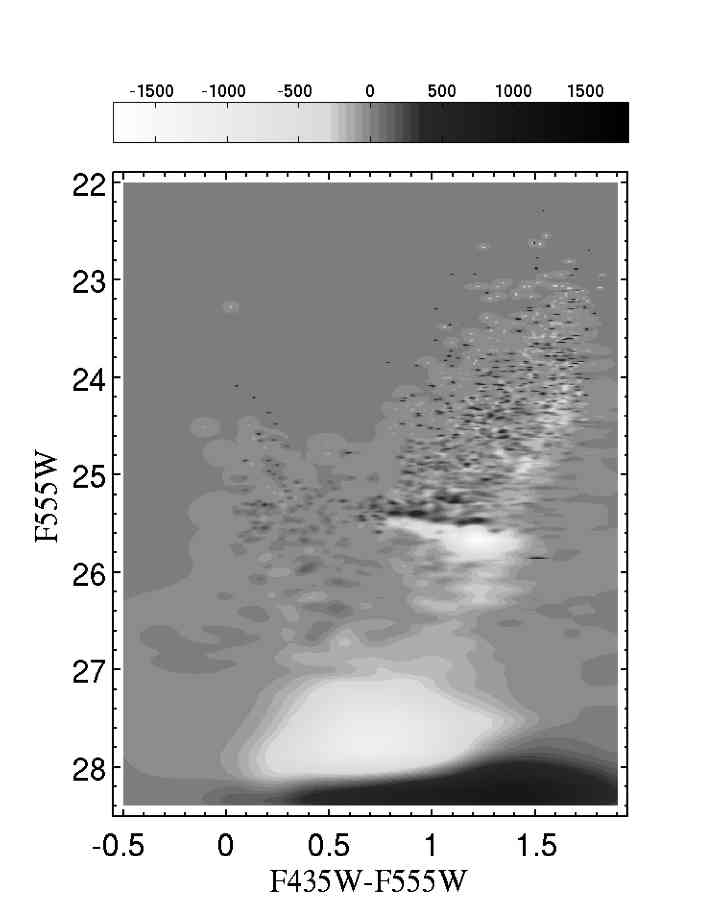} 
  \caption{Subtraction of the normalized F1 error-based Hess diagram
    from the F2 one. The normalization is such that the number of
    stars in F1 and F2 are the same.  The normalization factor is
    0.32. Negative values imply over-subtraction and therefore more
    M32 stars, and positive ones represent under-subtraction and
    therefore more M31 stars.  The subtraction reveals significant
    morphological differences between the M32 and M31 background
    CMDs.}
\label{fig:m31minusm32}
\end{figure}

To  investigate whether  there are  differences in  the  morphology of
these two  CMDs, which  then would reveal  differences in  the stellar
populations, we have subtracted the normalized F1 Hess diagram (before
decontamination)  from the  F2  one.   The F1  Hess  diagram has  been
normalized by  a factor of 0.32,  which is the  ratio of the F2  to F1
(uncorrected for M31 contamination)  stars above the 80\% completeness
level.   The F2 Hess  diagram has  been also  normalized by  its total
number  of  stars above  a  completeness  level  of 80\%.  Thus,  both
diagrams  are  normalized  so  that  they  have  the  same  amount  of
stars.   Figure~\ref{fig:m31minusm32}  shows   the   result  of   this
subtraction.  Negative  values illustrate over-subtraction  (more M32)
whereas positive values represent under-subtraction (more M31).

Interestingly the subtraction exposes clear differences between the
two CMDs.  An over-subtraction would imply an excess of the relative
fraction of M32 stars in that region of the CMD. On the contrary an
under-subtracted region would imply an excess of fraction of M31
stars. It is clear from the subtraction that the RC in the M31 field
lies at and extends towards bluer colors, indicating a more metal-poor
population than in M32. In addition, the RGB of M32 is considerably
redder than that of M31, indicating higher metallicity stars in M32
not present in M31. There is not a significant difference in the BP
population.  However there appears to be an excess of F2 stars at
higher luminosities, in the BP, indicating a larger number of stars
with younger ages present in M31.  The fraction of M32 stars in the
region of apparent magnitudes F555W between $\sim 27 $ and $\sim 28$
dominates over M31 stars (see Table~\ref{table:starcounts}).  This,
together with the different morphologies of the RCs, suggests that M32
possesses more intermediate age stars than M31 in our fields.  Note in
addition that there is a magnitude difference, $\sim 0.2$ mag, between
the RCs in F1 and F2.  The brighter RC in M31, however, does not
directly indicate that M32 must be behind M31.  As stated in previous
sections (see, e.g., Sections~\ref{sec:iapop} and~\ref{sec:distance}),
the CMD location of the RC for a stellar system depends on both its
age and metallicity.  Hence, a difference between the RC's magnitudes
in F1 and F2 CMDs may be only due to differences between the stellar
populations of M32 and M31.

Overall, this subtraction reveals that the M31 stellar populations in
our field F2 are younger and more metal-poor than the M32 population
in F1\footnote{We note that this result appears to contradict the
  mass-metallicity relations found in, e.g., \citet{Tremonti_etal04};
  \citet{Gallazzi_etal05}. However, those relations have not only
  significant spreads but also refer to the centers of galaxies. Our
  M31 field is near the edge of the visible disk and thus likely has
  lower metallicity than the center, whereas it is well known that M32
  has at best a mild metallicity gradient so that the outer regions
  are nearly as metal-rich as the center \citep[see,
  e.g.,][]{Worthey04}. We are therefore comparing the (relatively)
  metal-poor outskirts of the M31 disk with the (relatively)
  metal-rich outskirts of M32, and the mass--metallicity relation as
  such does not apply.}. The latter is in agreement with the MDFs.

\subsection{ M31 inner disk and bulge stellar populations}
The  stellar populations  of the  disk  of M31  have been  extensively
studied by several means.   However, those studies were mainly focused
on the  outer disk of M31.   The present data are  the deepest optical
observations of the inner disk of M31 to date and are sensitive to the
young stellar  populations.

We summarize here our findings about the stellar populations in F2.
The broad RGB of its CMD (right panel in
Figure~\ref{fig:hessdeconvolved}) indicates a wide metallicity spread.
We derive the MDF of M31, using the method of Section~\ref{sec:rgb},
and it is shown in the bottom panel of Figure~\ref{fig:mdf} as a gray
line representing a 10 Gyr-old population.  The F2 histogram has been
multiplied by a factor of 1.9, the ratio of the decontaminated M32 to
field F2 stars with completeness level above 80\%.  The selection of
stars to compute the MDF has the same constraints as in the M32 MDF
computation but only the results for stars below the RC are shown here
since there was no difference with those above the RC.  We have
selected 611 stars below the RC in F2, having absolute magnitudes
$0.95<M_{F555W}<1.60$ and dereddened colors $0.75<(F435W-F555W)_0 <
1.40$. We have assumed a Galactic reddening of $E(B-V) = 0.08$
\citep{burstein_heiles82} and a distance modulus for M31 of
$\mu_0(\rm{M31})= 24.45$ (this paper).  The peak of this distribution
is at $[\mathrm{Fe/H}]\sim -0.4 \, \mathrm{dex}$, indicating a more
metal-poor population of stars in our background field than in F1.
Note that almost no stars more metal-poor than $[\mathrm{Fe/H}]\sim
-1.2 \, \mathrm{dex}$ are present in our data.  We detect blue stars
above the HB level, indicating the presence of stars as young as 0.5
Gyr (see argument in \citealt{Brown_etal06} and in
Section~\ref{sec:young} above).  The detection of AGB stars indicates
the presence of an intermediate-age population.  However, the Hess
diagram subtraction suggests that this population in our M31 field is
less significant than that in M32 (see Figure~\ref{fig:m31minusm32}).

\citet{Williams_02} studied 27 fields in the disk of M31, using
HST/WFPC2 archival data.  He performed photometry on these fields and
statistically derived the star formation history of each of them.  The
random distribution of most of these fields as well as the comparison
of their SFHs allow an overall understanding of the formation and
evolution of the M31 disk.  Note, however, that these data are
strictly limited to the giant branch and, therefore, an accurate
measurement of the age cannot be achieved.  Williams concluded that
the SFR of the disk as a whole has been very active until about 1 Gyr
ago, when the overall SFR declined with the exceptions of some areas
of the spiral arms.  He also showed that the disk of M31 is deficient
in old metal-poor stars and that an intermediate-age, metal-rich
population is present in most of the fields.  A young population,
however, is only significant in fields within the spiral arms.  From
the 27 fields, three are located in the vicinity of M32 which show
nearly identical SFRs. Their SFRs indicate a moderate decline from 10
Gyr to 1 Gyr followed by a steep decline after that.  The metallicity
of these three fields is quite high, $[\mathrm{Fe/H}] >
-0.5~\text{dex}$, with only a very small fraction of stars with
$[\mathrm{Fe/H}] \sim -1.0~\text{dex}$ in one of the fields.
\citet{Worthey_etal05} have also studied archival observations of
HST/WFPC2 fields in the disk of M31.  They however restricted their
interest to the metal abundance distribution and the efficacy of the
closed-box model.  They find a robust metal abundance distribution
that appears to hold in the vicinity of our fields but also further
out, with a FWHM of $\sim 0.6 \, \mathrm{dex}$.  The peak of the
distribution at $[\mathrm{Fe/H}] =-0.25 \, \mathrm{dex}$ is quite
metal-rich, which agrees with Williams' results.
\citet{Olsen_etal06}, on the other hand, studied high resolution, deep
near-IR images of the inner disk and bulge of M31 obtained with Gemini
North telescope.  They conclude that most of the inner disk and
spheroid are indistinguishably old, with stellar populations dominated
by median ages larger than 6 Gyr.  They find a metallicity of
$[\mathrm{Fe/H}] \sim -0.7 \, \mathrm{dex}$ in the disk, in agreement
with \citet{ferguson_johnson01}.  Interestingly, their outermost disk
fields, which are comparable in radial distance to our field F2, have
a 10\% population of 1 Gyr or younger stars, more young stars than in
their inner fields.

The most extensive study about the disk of M31 was done by
\citet{Brown_etal06}.  They used deep HST ACS observations of three
fields in M31, one located on the disk at $\sim 25$ kpc from the
center of the galaxy, one in the spheroid, and one tidal stream field.
Their data reach stars well below the oldest MSTO, which allowed them
to derive the complete star formation history in those fields.
Regarding the outer disk field, they found a broad RGB, which suggests
a wide range of metallicity.  They failed to find a significant old
metal-poor population in the disk of M31, due to the small amount of
blue HB stars.  The detection of a BP indicates that there is at least
some very young population, with ages of 0.2 -- 1 Gyr, in agreement
with \citet{Olsen_etal06}.  After quantitatively fitting their CMDs,
Brown et al.  concluded that the outer disk population is dominated by
stars of 4--8 Gyr, with a mean age of 7.5 Gyr and a mean metallicity
of $\text{[Fe/H]} = -0.2~\text{dex}$.  If the contamination from their
spheroid field population is taken into account, they found a younger
mean age (6.6 Gyr) and a higher mean metallicity ($\text{[Fe/H]} =
+0.1~\text{dex}$).  Thus the spheroid contamination almost completely
accounts for the old and metal-poor stars in the outer disk.  Note,
however, that given the wide separation between their fields, they
cannot be assured that their spheroidal field is representative of the
underlying spheroid in the disk field.

In general, the stellar populations we find in F2 agree with previous
studies of M31. In particular, the MDF that we have obtained for M31
agrees very well with the findings by \citet{Williams_02}, with a peak
at $[\mathrm{Fe/H}] > -0.5~\text{dex}$ and devoid of stars with
metallicities lower than $[\mathrm{Fe/H}] \sim -1.0~\text{dex}$.  The
lack of a significant amount of metal-poor stars is a common property
found in all the studies just discussed. Our metallicity peak is,
however, slightly more metal-poor than that obtained by
\citet{Worthey_etal05} and \citet{Brown_etal06}, and slightly more
metal-rich than the value obtained by \citet{Olsen_etal06}. Hence, the
possibility raised above that field F2 contains stripped M32 stars is
unlikely.  A more quantitative study of the stellar populations in F2
will be presented in a follow-up paper.

\section{Summary}

Deep $F435W$ ($\sim B$), $F555W$ ($\sim V$) observations of two fields
located at $\sim 110 \arcsec$ (M32 field F1) and $\sim 320 \arcsec$
(M31 background field F2) from the nucleus of M32 were made using
ACS/HRC on board HST. The location of the M32 field was chosen so that
both the image crowding and contamination from M31 disk were as
minimal as possible. The M31 background field was located at the same
M31 isophotal level of our M32 field, in order to allow the photometry
of M32 to be properly corrected for the M31 background. Photometry of
these images was performed using two different techniques: aperture
photometry applied to the images deconvolved by a PSF and the DAOPHOT
II stand-alone packages.  Photometry on deconvolved images has been
shown to do a significantly better job in resolving blends and thus we
used its results for analyzing the data. Before any analysis was done,
extensive ASTs were performed to understand completeness and crowding
in our fields. The results from the completeness and error analysis
were used to statistically decontaminate M32 for the contribution of
M31 light and thus the different crowding between the fields were
taken into account when decontaminating.  Unfortunately, the severe
crowding prohibits us from reaching the oldest MSTOs of M32, and we
were forced to assume that all stars fainter than $F555W \sim 28$ are
products of blends. Nevertheless the photometry here is significantly
deeper (by at least 2 mag) than any other obtained so far for similar
field positions and we are able to analyze the stellar populations of
M32. Based on the CMD of M32 we conclude the following:
\begin{itemize}

\item We find that the core-helium burning stars are concentrated in a
  red clump, as expected for a metal-rich system and consistent with
  results from both spectral analysis and photometric studies.  By
  using the mean color and magnitude of the RC, we obtain a mean age
  of 8--10 Gyr for a metallicity of $[\mathrm{M/H}]\sim -0.2$ in M32.

\item We report the detection of the RGB bump and the AGB bump in M32
  for the first time.  We use their positions in the CMD relative to
  the position in color and magnitude of the RC to constrain the mean
  age and metallicity of the population.  We find that the mean
  metallicity of M32 is higher than $\mathrm{[M/H]} \sim -0.4$ dex, in
  agreement with the RGB results, and that the mean age from this
  method is between 5 and 10 Gyr.

\item The metallicity distribution of M32 inferred from the CMD has
  its peak at $[\mathrm{M/H}]\sim -0.2 \, \mathrm{dex}$. Overall, the
  metallicity distribution function implies that there are more
  metal-rich stars than metal-poor ones.  We find that metal-poor
  stars with $[\mathrm{M/H}] <-1.2$ contribute very little, \emph{at
    most} 6\% of the total $V$-light or 4.5\% of the total mass in our
  M32 field, implying that the enrichment process largely avoided the
  metal poor stage.

\item Bright AGB stars at $F555W < 24$, i.e.  above the TRGB, confirm
  the presence of an intermediate-age population in M32 (ages of 4$\pm
  3$ Gyr).

\item The observed blue plume is a genuine blue plume and not an
  artifact of crowding.  It contains stars as young as $\sim$ 0.5 Gyr.
  The detected blue loop having stars with masses of $\sim 2-3$
  $M_{\odot}$ and ages between $\sim 0.3$ and $\sim$ 1 Gyr, as well as
  the possible presence of a bright SGB are different manifestations
  of the presence of a young population.  However, it is likely that
  this young population belongs to the disk of M31 rather than to
  M32. The fainter portion of the blue plume ($F555W > 26$)
  \emph{does} belong to M32 and it indicates the presence of stars
  between 1 and 2 Gyr or/and the first direct evidence of blue
  straggler stars in M32.

\item We do  not observe either a significant BHB  or the oldest MSTO,
  but studies  in our fields have  found RR Lyrae stars  in M32, hence
  confirming the presence of an ancient metal-poor population with our
  data. 
 
\item  We note that  in general  the CMDs  of both  fields F1  and F2
  present  an  unexpectedly  similar  morphology. By  subtracting  the
  normalized F1  CMD from  the F2  one, we see  that there  are subtle
  differences,  such that  M31  appears  to have  a  younger and  more
  metal-poor  population than  M32, and  M32  appears to  have a  more
  predominant intermediate age  population. 

\item The CMD of our background field F2 exhibits a wide RGB,
  indicative of a metallicity spread with its peak at
  $[\mathrm{M/H}]\sim -0.4 \, \mathrm{dex}$. The presence of a blue
  plume indicates the presence of stars as young as 0.3 Gyr. We have
  also detected bright AGB stars which reveal the presence of
  intermediate-age population in M31.

\item We have calculated the distance to M32 using the RC stars.  We
  obtained a distance modulus of $\mu_0 =24.53 \pm 0.12$ which agrees
  with previous results. This estimate is comparable to the distance
  modulus obtained for M31, $\mu_0 =24.45 \pm 0.12$, also using the RC
  stars.  Hence, within the errors we cannot conclude whether M32 is
  situated in front of or behind M31.
\end{itemize}

M32 clearly has a complex SFH: it is dominated by metal-rich
intermediate-age stars and it contains some, but few, old, metal-poor
stars as well as possible young populations.  M32 appears to be a
normal, low-luminosity elliptical \citep{kormendy_etal09}. However,
the model proposed by \citet{Bekki_etal01} of M32 as a
tidally-stripped, low-luminosity, early-type spiral galaxy cannot be
ruled out with our data.  Whatever its formation history, we can still
attempt to place M32's SFH into a cosmological context. F1's dominant
population has an age of 8--10 Gyr, corresponding to a formation
redshift of $1\la z_f\la2$ \citep[for the cosmology of][]{Komatsu10}.
M32's total stellar mass can be inferred from the values in Table 1 of
\citet{Cappellari06} to be $M_*=10^{8.66}\,M_{\odot}$.  If we assume
that, say, 90\% of M32's stellar mass was formed uniformly in the
interval of 8--10 Gyr ago (assumptions that will be tested when we
complete our detailed analysis of the SFH from the CMD), we derive a
specific star formation rate
$SSFR=SFR/M_*\sim0.45\,\mathrm{Gyr^{-1}}$---typical of or perhaps even
slightly lower than the \emph{observed} specific star formation rates
in galaxies with mass $\sim10^9\,M_{\odot}$ at $1\la z\la2$
\citep[e.g.][]{Noeske07,Santini09}.  Our observations of M32 thus
represent a ``ground-truth'' of the downsizing of star formation in
galaxies \citep[see, e.g.,][]{Cowie96,Fontanot09}: a low-mass
early-type galaxy forming the majority of stars at lower redshifts
than the peak of the Universe's star formation rate density
\citep[e.g.,][]{Hopkins06}.

In spite of the fact that we do not reach the oldest MSTOs, our new
photometric results dramatically improve our understanding of the
stellar composition of M32.  Synthetic stellar population models ought
to be able to reproduce all the features we presented in this work.
We note once again that M32 is the only system with properties similar
to normal elliptical galaxies close enough for which a direct
comparison between resolved intrinsically faint individual stars and
stellar population models of integrated light can currently be
obtained.  A follow-up paper will present a detailed analysis of the
recent and intermediate SFH obtained from these data. Its subsequent
implications on the formation and evolution of M32, and the different
proposed models about its origins, will be further tested and
discussed there.

\acknowledgments AM is grateful to Carme Gallart for very valuable
discussions.  Special thanks to Eline Tolstoy, Giuliana Fiorentino,
Annette Ferguson, and Kathryn Johnston for useful comments and
suggestions, and also to Eva Busekool for bringing to our attention
the BP in the archival WFPC2 observations.  We are grateful to Kirsten
Howley for providing us with the values of the contributions of M31
and M32 light to our fields before publication. We thank Peter Stetson
for providing his stand-alone DAOPHOT II software and guidance on its
use.  We thank the referee, Ivo Saviane, for his careful reading of
the manuscript and comments that helped to improve this paper. This
work has made use of the IAC-STAR synthetic CMD computation
code. IAC-STAR is supported and maintained by the computer division of
the Instituto de Astrof\'isica de Canarias.  NOVA is acknowledged for
financial support.  Support for program GO-10572 was provided by NASA
through a grant from the Space Telescope Science Institute, which is
operated by the Association of Universities for Research in Astronomy,
Inc., under NASA contract NAS 5-26555.

\emph{Facility:} \facility{HST (ACS)}

\bibliographystyle{apj}
\bibliography{references}

\appendix

\section{BP in  WFPC2 archival observations}
To investigate  the presence of  a BP in  M32, we retrieved  HST WFPC2
observations taken  at four different  positions around the  center of
M32 using the  $F606W$ (wide $V$) and $F814W$  (wide $I$) filters from
the ESO/STECF Science Archive. The images are already calibrated using
the  standard  pipeline  data  processing.   Figure~\ref{fig:location}
shows the  location of the F3, F4,  F5 and F6 fields  we have analyzed
for this purpose and Table ~\ref{table:wfpc2fieldsdata} summarizes the
information regarding to these  archival observations. Note that these
fields, except  F3, should have a small  fractional light contribution
of M32 compared to that of M31.   In fact F4, the nearest field to the
center of  M32 after  F3, contains our  F2 background  field, implying
that M31 dominates over M32 at that
position. 

For   each  of   these  fields,   stellar  photometry   was  performed
simultaneously    on    all    available    images    using    HSTphot
\citep{dolphin00},  a package  specifically  designed for  use on  HST
WFPC2  images.  Before  running HSTphot  several  pre-processing steps
were done  on each individual image  for each of the  fields.  All the
routines are described in  detail in \citet{dolphin00} and the HSTphot
manual. Briefly, each  image was first masked, using  the data quality
image included in  the HSTphot \texttt{mask} routine; we  next made an
initial sky estimate using  the routine \texttt{getsky}; after that we
removed  cosmic  rays  using  \texttt{crmask}  and  used  the  routine
\texttt{hotpixels}  to mask  hot pixels  that differ  significantly in
value from  their neighbors. Finally,  all the images per  field were
run together in the \texttt{hstphot} photometry routine. The output of
this  routine  was  a  list  of  magnitudes  and  positions  for  each
stellar-like  object as  well as  several some  global  and individual
frame solution information  that includes a goodness-of-fit parameter,
sharpness,  roundness, and  object  type. We  kept  objects that  were
classified  as   good  stars,  having   $\chi  <  4$  and   $-0.5  \la
\mathrm{sharpness}  \la 0.5$.  The  magnitudes are  given in  both the
WFPC2   photometric   system   with   the  applied   calibrations   of
\citet{dolphin00} and the UBVRI system using the color transformations
of \citet{holtzman_etal95}.  To  understand completeness in these data
we  have run  ASTs on  each  of the  fields. HSTphot  contains an  AST
routine which  distributes a grid  of artificial stars generated  on a
two-dimensional CMD  according the flux  of the image.   Details about
how ASTs work in HSTphot  can be found in \citet{holtzman_etal06}.  In
short, the  AST routine is  a post-processing step of  the photometry;
the artificial stars  are treated in absence of  neighbors, i.e. real
stars,  and thus  crowding  is not  taken  into account.  In order  to
compensate for this problem, we  have compared the output magnitude of
each fake  recovered star with the  magnitudes of real  stars that are
within 2 pixels  of its position.  We have only kept  the fake star as
recovered if it contributes most of the light \citep{holtzman_etal06}.
We have performed 100 ASTs  on each image, injecting $\sim 5000$ stars
per AST. We have located the artificial  stars on the CMD with $20 < V
< 29$ and  $-1 < (V-I) <  4$ distributed according to the  flux of the
image.

\begin{figure*}\centering
%\hspace{0.8cm}
\includegraphics[width=140mm]{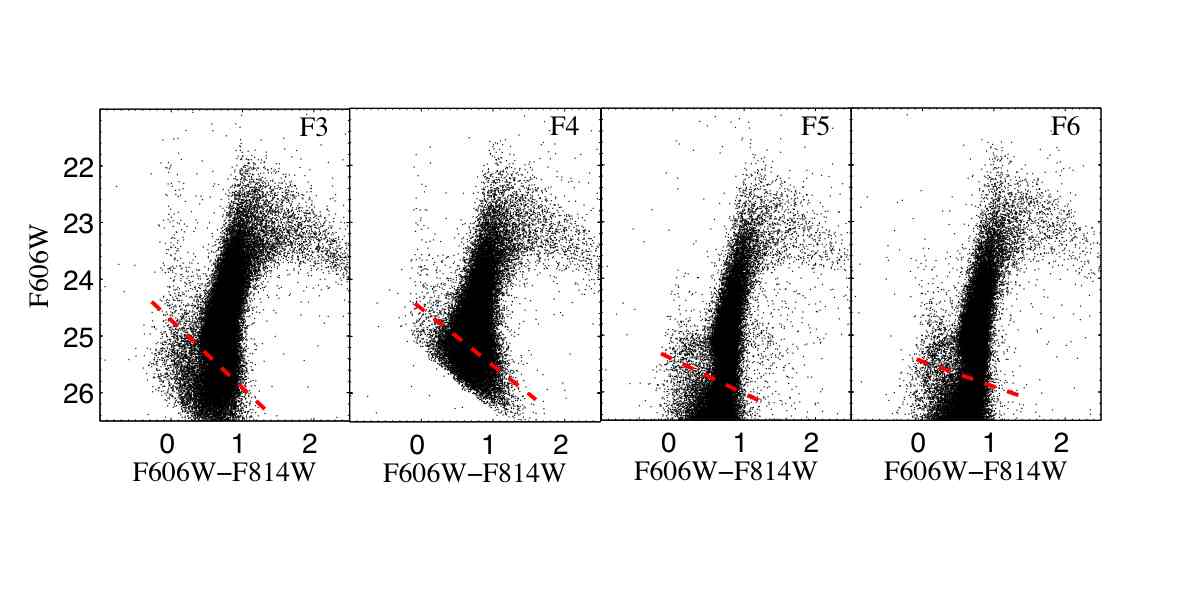}
\caption{CMDs of the fields analyzed to investigate the presence of a
  blue plume in M32. The magnitudes are calibrated onto the Vega mag
  WFPC2 system.  The dashed red line indicates the 50\% completeness
  level of the data. We can see a blue plume above the 50\%
  completeness level in \emph{all} fields.  More information about
  these fields and CMDs can be found in
  Table~\ref{table:wfpc2fieldsdata}.}
\label{fig:wfpc2cmds}
\end{figure*}

Figure~\ref{fig:wfpc2cmds} shows the ($F606W-F814W$, $F606W$) CMDs for
each of the outlying fields. The main features in these CMDs are the
RGB and the TRGB. We can also see that a blue plume is present in {\it
  all} the fields above the 50\% completeness level, which is
indicated as a dashed line in each of the CMD.  Note that F5 and F6
have a 50\% completeness level at a fainter magnitude than F3 and F4.
This is probably due to the fact that there is less crowding, as
  expected for these fields which are located further away from the
  center of M32 than F3 and F4, and their exposure times were longer.

\begin{figure}
\centering \includegraphics[width=70mm,clip]{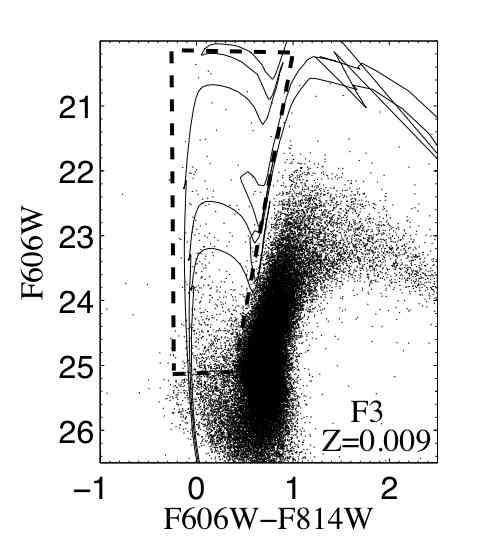}
\caption{ CMD of F3 with young Padova isochrones \citep{marigo_etal08,
    girardi_etal08} of ages 0.08, 0.2, and 0.3 Gyr for a metallicity
  of $Z=0.009$ superimposed. We have assumed a reddening of $E(B-V) =
  0.08$ \citep{burstein_heiles82} and a true distance modulus of
  $\mu_0=24.53$ (this paper). The dashed-box represents the region
  from which we counted blue plume stars in this field.}
\label{fig:wfpc2_1cmd}
\end{figure}

To analyze the relative significance of the BP between the fields, we
have counted their blue plume stars as follows.  For each field, we
defined a box in such a way that only BP stars that are over the 50\%
completeness level are inside.  This box is different in each field,
given that the completeness levels are not exactly always the same. We
then divided the number of stars in this box by the RGB stars that are
also over the 50\% completeness level, obtaining thus a specific
frequency of blue plume stars per field.  Both the specific frequency
of blue plume stars and the angular distances of each field to M32
center are indicated in Table~\ref{table:wfpc2fieldsdata} as ``f'' and
``d'' respectively. An example of how the BP stars box was defined is
shown for F3 in the Figure~\ref{fig:wfpc2_1cmd}.  Padova isochrones
\citep{marigo_etal08, girardi_etal08} having ages of 0.08, 0.2 and 0.3
Gyr and a metallicity of $Z=0.009$ are superimposed on the CMD
corrected for reddening $E(B-V) = 0.08$ \citep{burstein_heiles82} and
a true distance modulus of $\mu_0=24.53$ (this paper). We can see that
they fit the data very well suggesting that a young population is the
origin of the blue plume.  Finally, we compared the specific frequency
of blue plume stars between the fields and we found that the value
increases for the fields closer to the center of M32, where the
contribution from M32 becomes significant.  This suggests that the
blue plume truly belongs to M32 and is not due solely to the blue
plume seen in M31 fields.
\begin{deluxetable*}{@{}cccccccccc}
\tablecaption{Archival observations of fields in M32 \label{table:wfpc2fieldsdata}}
\tablewidth{0pt} 
\tablecolumns{10}
\tablehead{\colhead{ Field} &\colhead{Proposal ID}&\colhead{P.I.} &
\colhead{$\alpha_{J2000.0}$}&\colhead{$\delta_{J2000.0}$}&\colhead{$t_{F606W}$(s) \tablenotemark{a}} & \colhead{$t_{F814W}$(s) \tablenotemark{b}} &\colhead{$\rm{N_{stars}}$} & \colhead{d($\arcmin$)\tablenotemark{c}}&\colhead{f\tablenotemark{d}}}
\startdata
 F3&
  GO6664&G. Worthey &00 42 58.80 &+40 50 34.6 &8800 &4400 &56,220 & 3.48&0.008\\
%P2 & &G.Worthey & & & & & & 0.2497\\ 
  F4&GTO7566 &R. Green &00 43 07.85 &+40 53 32.8 &1700 &1100&48,210 &5.16 & 0.006 \\
  F5 &GO9392 &M.Mateo &00 43 16.93 &+40 46 31.2
  &34700 &10000&47,821 &8.56 & 0.005\\  F6 &GO9392
  &M.Mateo &00 43 50.75 &+40 59 35.7 &29700 &15000&47,289 &15.09 & 0.005
\enddata
\tablenotetext{a}{Total exposure time in the F606W filter}
\tablenotetext{b}{Total exposure time in the F814W filter}
\tablenotetext{c}{Angular distance to the center of M32}
\tablenotetext{d}{Ratio of BP stars to RGB stars}
\end{deluxetable*}

Thus, it remains as  an open question as to why the  BP in M32 was not
seen in earlier optical data.  The fact that the $B$ band has not been
previously observed but just the  $V$ and $I$ cannot be reason because
the BP is  clearly visible in the $VI$  archival observations shown in
Figure~\ref{fig:wfpc2cmds} and  in \citet{alonsogarcia_etal04}. The BP
was  not  however  seen in  G96,  who  observed  with the  same  WFPC2
camera. However,  the G96 data  are closer to  the center of  M32 than
those of Alonso-Garc\'ia et al. It is likely that the extreme crowding
affecting the region of interest  made the detection of the blue plume
impossible in  previous works.  In  this context it is  interesting to
compare M32 with the dwarf  elliptical (dE) NGC 205, also satellite of
M31,  which  has   been  known  to  have  blue   OB-type  stars  since
\citet[][see also  \citealt{Hodge73} and references therein]{Baade51}.
Only recently has  there been a significant and  genuine population of
young  blue   stars  observed  in   its  center  using   ACS/HRC  data
\citep{Monaco_etal09}.  Previous works of this galaxy have shown no BP
stars,    or   some   blue    stars   in    the   WFPC2    $VI$   data
\citep[see][]{butler_martinezdelgado05} although, curiously, few.  The
fact that in NGC 205, a genuine BP was only observed with ACS/HRC data
whereas it was not (significantly) seen in previous works, favors our
hypothesis  that the  previous  non-detection of  a  BP in  M32 was  a
problem of crowding and resolution.

\end{document}